%% file: main.tex
\documentclass{article}

\usepackage{arxiv}

\usepackage[utf8]{inputenc} % allow utf-8 input
\usepackage[T1]{fontenc}    % use 8-bit T1 fonts
\usepackage{hyperref}       % hyperlinks
\usepackage{url}            % simple URL typesetting
\usepackage{booktabs}       % professional-quality tables
\usepackage{amsfonts}       % blackboard math symbols
\usepackage{nicefrac}       % compact symbols for 1/2, etc.
\usepackage{microtype}      % microtypography
\usepackage{lipsum}
\usepackage{graphicx}
\usepackage{tikz,lipsum,lmodern}
\usepackage[most]{tcolorbox}
\usepackage{xcolor}
\usepackage{color}
\usepackage{longtable}

\usepackage{tabularx}
\usepackage{booktabs}
\usepackage{array}

\usepackage{booktabs}
\usepackage{graphicx}       % for \resizebox
\usepackage{caption}
\usepackage{comment}

\definecolor{dark-gray}{gray}{0.25}

\graphicspath{ {./images/} }

\newcommand{\code}[1]{\textcolor{dark-gray}{\texttt{#1}}}

\title{Exploring the Capabilities of the Frontier Large Language Models for Nuclear Energy Research}

\author{%
  \parbox{\textwidth}{%
    \centering
    Ahmed Almeldein, Mohammed Alnaggar, Rick Archibald, Tom Beck, Arpan Biswas, Rike Bostelmann, Wes Brewer, Chris Bryan, Christopher Calle, Cihangir Celik, Rajni Chahal, Jong Youl Choi, Arindam Chowdhury, Mark Cianciosa, Franklin Curtis, Gregory Davidson, Sebastian De Pascuale, Lisa Fassino, Ana Gainaru, Yashika Ghai, Luke Gibson, Qian Gong, Christopher Greulich, Scott Greenwood, Cory Hauck, Ehab Hassan, Rinkle Juneja, Soyoung Kang, Scott Klasky, Atul Kumar, Vineet Kumar, Paul Laiu, Calvin Lear, Yan-Ru Lin, Jono McConnell, Furkan Oz, Rishi Pillai, Anant Raj, Pradeep Ramuhalli, Marie Romedenne, Samantha Sabatino, José Salcedo-Pérez, Nathan D. See, Arpan Sircar, Punam Thankur, Tim Younkin, Xiao-Ying Yu, Prashant Jain, Tom Evans, Prasanna Balaprakash\thanks{Contact Prasanna Balaprakash (pbalapra@ornl.gov) for workshop-related questions; the authors of each section related to the problem are explicitly identified in the report.}
    \\[1ex]
    {Oak Ridge National Laboratory, Oak Ridge, TN 37831, USA}\\ [1ex]
    {\small\texttt{\{almeldeinam, alnaggarmg, archibaldrk, becktl, biswasa, bostelmannf, brewerwh, bryancd, calleci, celikc, chahalr, choij, chowdhurya1, cianciosamr, curtisfg, davidsongg, depascuales, reedke, gaianrua, ghaiy, gibsonld, gongq, greulichcr, greenwoodms, hauckc, hassanem, junejar, kangs1, klasky, kumara, kumarv, laiump, learcr, liny, mcconnelljh, ozf1, pillairr, raja, ramuhallip, romedennem, sabatinosm, salcedoperjl, seend, sircara, thakurp, younkintr, yuxiaoying, jainpk, evanstm, pbalapra\}@ornl.gov}}
  }%
}

\begin{document}
\maketitle

\begin{abstract}
The AI for Nuclear Energy workshop at Oak Ridge National Laboratory evaluated the potential of Large Language Models (LLMs) to accelerate fusion and fission research. Fourteen interdisciplinary teams explored diverse nuclear science challenges using ChatGPT, Gemini, Claude, and other AI models over a single day. Applications ranged from developing foundation models for fusion reactor control to automating Monte Carlo simulations, predicting material degradation, and designing experimental programs for advanced reactors. Teams employed structured workflows combining prompt engineering, deep research capabilities, and iterative refinement to generate hypotheses, prototype code, and research strategies.
Key findings demonstrate that LLMs excel at early-stage exploration, literature synthesis, and workflow design, successfully identifying research gaps and generating plausible experimental frameworks. However, significant limitations emerged, including difficulties with novel materials designs, advanced code generation for modeling and simulation, and domain-specific details requiring expert validation. The successful outcomes resulted from expert-driven prompt engineering and treating AI as a complementary tool rather than a replacement for physics-based methods.
The workshop validated AI's potential to accelerate nuclear energy research through rapid iteration and cross-disciplinary synthesis while highlighting the need for curated nuclear-specific datasets, workflow automation, and specialized model development. These results provide a roadmap for integrating AI tools into nuclear science workflows, potentially reducing development cycles for safer, more efficient nuclear energy systems while maintaining rigorous scientific standards.
\end{abstract}

\section{Introduction}
The intersection of artificial intelligence and nuclear energy represents a frontier of immense potential for accelerating scientific discovery and engineering innovation. Both fusion and fission technologies stand as critical pillars for achieving energy abundance and economic prosperity. The promise of virtually limitless, reliable baseload power from nuclear sources could fundamentally transform global energy markets and enable new scales of industrial capability. However, the complexity of nuclear systems—spanning plasma physics, materials science, neutronics, thermal hydraulics, and structural mechanics—presents formidable challenges that traditionally require decades of research and development. The emergence of Large Language Models (LLMs) and advanced AI reasoning systems offers an opportunity to compress these timelines while maintaining the rigorous safety and reliability standards essential to nuclear applications.

The ``AI for Nuclear Energy" workshop, convened at Oak Ridge National Laboratory on May 1, 2025, was designed to systematically evaluate how state-of-the-art LLMs  could enhance nuclear science research workflows. This one-day workshop brought together experts from diverse disciplines—nuclear engineering, computational science, materials science, and machine learning—to explore practical applications of LLMs in addressing  nuclear energy research challenges. The timing of this workshop was particularly significant, as it coincided with rapid advances in AI capabilities, including enhanced reasoning models, deep research functionalities, and improved code generation tools.
The workshop's primary objective was to assess whether AI could meaningfully accelerate early-stage research in nuclear energy applications while identifying both the opportunities and limitations of current technology. Unlike traditional computational tools that excel at numerical simulation, LLMs offer unique capabilities in synthesizing vast bodies of literature, generating initial hypotheses, identifying research gaps, and facilitating cross-disciplinary communication—all critical activities in the complex landscape of nuclear research.

Fourteen teams tackled diverse challenges spanning the nuclear energy spectrum: from developing foundation models for real-time fusion reactor control to predicting radiation-induced material degradation, from automating complex Monte Carlo simulations to designing experimental programs for advanced reactor concepts. Each team employed structured workflows that combined prompt engineering, iterative refinement, and careful validation to explore how AI tools could complement traditional physics-based approaches.
This report documents the methodologies, findings, and insights from this collaborative effort. We present detailed case studies from each team, highlighting successful applications where AI demonstrated clear value—such as literature synthesis, workflow design, and initial code prototyping—as well as areas where current limitations became apparent, including challenges with novelty, technical accuracy, and the need for expert validation. The synthesis of these experiences provides an advanced  understanding of AI's current role in nuclear science: not as a replacement for domain expertise or rigorous physical modeling, but as an accelerant for scientific exploration.

The implications of this work extend beyond the immediate findings. As U.S. race to achieve energy dominance through advanced nuclear technologies, the ability to rapidly explore design spaces, identify promising research directions, and integrate knowledge across disciplines becomes a strategic imperative. The potential for abundant, affordable nuclear energy to power massive data centers, advanced manufacturing, and even space exploration depends on our ability to accelerate the development cycle. This report serves as both a snapshot of current capabilities and a roadmap for future development, outlining how the nuclear community can harness AI tools while maintaining the exceptional standards of safety and scientific rigor that define this field.

Through careful documentation of successes, failures, and lessons learned, we aim to provide the nuclear science community with practical guidance for integrating AI into research workflows, ultimately contributing to the acceleration of energy abundance that will define the next era of human prosperity and technological advancement.

%\section*{Summary of Workshop Team Projects}

\section{Executive summary}

The workshop showed LLM's versatility across 14 distinct fusion and fission nuclear energy challenges, from fundamental fusion physics to practical reactor engineering. Teams explored applications spanning materials science, plasma control, neutronics modeling, and data management—representing the full spectrum of nuclear energy research and development.

\subsection*{Success Patterns}

\textbf{Workflow development and integration:} Nearly all teams successfully used AI to design complex multiphysics workflows. Notable examples include integrating multiple simulation codes (XGC, NekRS, M3D-C1) for fusion modeling and coupling neutronics–thermal hydraulics–structural mechanics for PWR analysis. AI excelled at identifying data streams, coupling strategies, and process bottlenecks.

\textbf{Knowledge synthesis and gap analysis:} AI models demonstrated exceptional capability in literature review and research gap identification. Teams successfully identified missing experimental data for reactor corrosion, proposed experimental programs with detailed budgets, and ranked candidate materials based on multiple criteria. The medical isotope team correctly identified HFIR production capabilities from dispersed sources.

\textbf{Hypothesis generation and initial design:} Multiple teams generated novel research directions, including alpha particle channeling for fusion optimization, isotopic tailoring for reduced activation, and RF sheath mitigation strategies. The material degradation team achieved high predictive accuracy using AI-generated models.

\textbf{Visual and structural generation:} AI successfully created realistic microscopy images of material defects, atomic void structures, and STEM simulations that aligned with published literature, demonstrating potential for synthetic training data generation.

\subsection*{Limitations}

\textbf{Data and code access barriers:} The most significant limitation across projects was the inability to access or process in-house nuclear data. Teams working with SCALE, MCNP, and other restricted codes could not generate functional inputs despite extensive prompting.

\textbf{Implementation depth:} While AI excelled at high-level design and framework development, detailed implementation consistently required human expertise. No team achieved production-ready code, with particular struggles in Python generation and mathematical solver development.

\textbf{Data access and validation:} Projects were hampered by paywalled literature, missing datasets, and the need for extensive expert validation. The fusion data annotation team could not access TRANSP datasets, while the corrosion team missed state-of-the-art studies behind paywalls.

\textbf{Technical accuracy:} Several projects encountered accuracy issues—incorrect parameter values for heat transfer, failed dislocation loop generation, problematic citations, and overly general responses that lacked domain specificity.

\subsubsection*{Implications}

The workshop results reveal LLMs as an accelerator for early-stage nuclear research, particularly in:
Rapid literature synthesis and knowledge integration;  
Cross-disciplinary workflow design; 
Experimental program planning; 
Initial hypothesis generation. 
However, the technology requires careful integration with domain expertise and cannot replace detailed technical implementation or rigorous validation. Success depends on developing secure AI infrastructure for classified data, creating nuclear-specific training datasets, and establishing validation protocols.

The diversity of successful applications, from fusion plasma control to fission material degradation, suggests that LLM integration could provide competitive advantages across the entire nuclear energy sector, potentially accelerating development timelines by effectively harness these capabilities.

\begingroup
\footnotesize
\renewcommand{\arraystretch}{1.2}
\setlength{\tabcolsep}{3pt} % Slightly reduce column padding

% Adjust column widths to fit within \textwidth (e.g., 15.9cm max for 1in margins)
\begin{longtable}{>{\raggedright\arraybackslash}p{3.5cm} 
                  >{\raggedright\arraybackslash}p{3.7cm} 
                  >{\raggedright\arraybackslash}p{4.2cm} 
                  >{\raggedright\arraybackslash}p{4.2cm}}
\toprule
\textbf{Project Title} & \textbf{Key Challenge} & \textbf{Main Outcomes} & \textbf{Key Limitations} \\
\midrule
\endfirsthead

\toprule
\textbf{Project Title} & \textbf{Key Challenge} & \textbf{Main Outcomes} & \textbf{Key Limitations} \\
\midrule
\endhead
\bottomrule
\multicolumn{4}{c}{\bf{Domain: Fusion}}\\
\bottomrule
Foundation Model for Fusion & Real-time fusion reactor control and simulation & Structured workflow for integrating XGC, NekRS, M3D-C1 simulations; identified data pipeline needs & Could not delve into implementation details; requires extensive compute resources \\
\addlinespace
Structural Materials \& Molten Salt Compatibility & Material selection for extreme fusion environments & Ranked 5 candidate materials (W, RAFM steels, SiC/SiC); identified MHD interface challenges & No truly novel alloy proposals; limited to known materials \\
\addlinespace
Isotopically Tailored Materials & Reducing radioactive activation in fusion reactors & Identified materials benefiting from isotope engineering; proposed experimental roadmap & Incomplete analysis; AI provided incorrect values for some parameters \\
\addlinespace
Alpha Particle Confinement & Optimizing fusion self-heating from alpha particles & Proposed alpha channeling and divertor optimization strategies; identified experimental gaps & Model struggled with device-specific synthesis; some incorrect citations \\
\addlinespace
Fusion Data Annotation & Semantic enrichment of fusion datasets & Developed RAG pipeline for TRANSP data; created annotation workflow & Could not access needed datasets; incomplete implementation \\
\addlinespace
Multiphysics Modeling for MPEX & Plasma-material interaction modeling & Identified RF sheath erosion as key issue; developed inverse solver for diffusion coefficients & Limited to simplified 1D models; adjoint solver generation failed \\
\addlinespace
\bottomrule
\multicolumn{4}{c}{\bf{Domain: Fission}}\\
\bottomrule
Monte Carlo Surrogate Models & Accelerating MCNP simulations & Proposed neural network surrogate architecture; identified data gathering strategies & No existing Shift-based ML surrogates; lacks public training data \\
\addlinespace
Molten Salt Corrosion & Understanding corrosion in nuclear reactor salts & Literature review of MD/AIMD studies; identified need for MLIPs & State-of-the-art studies not captured; limited by paywall access \\
\addlinespace
Material Degradation Prediction & Inverse modeling from acoustic/strain data & Achieved R² $\geq$ 0.90 for degradation prediction; generated Mathematica ML code & Python code generation problematic; requires extensive validation \\
\addlinespace
Experimental Program Design & AI-directed experiments for advanced reactors & Generated complete experimental program with budget; identified corrosion knowledge gaps & High-level only; lacks technical implementation details \\
\addlinespace
Atomic Structure \& S/TEM Imaging & 3D reconstruction from electron microscopy & Generated atomic void structures; simulated STEM images & Failed to create accurate dislocation loops; limited to $<$10nm structures \\
\addlinespace
Pressurized Water Reactor Power Uprate Analysis & Multiphysics framework for reactor uprates & Outlined neutronics-TH-fuel-structural coupling strategy; identified relevant tools & Failed with export-controlled software; could not generate working inputs \\
\addlinespace
Microscopy Defect Imaging & Simulating defect images for SS304 & Generated realistic sensitization and crack defect images & No source transparency; unclear if SS304-specific \\
\addlinespace
Nuclear Isotopes \& MCNP Automation & Medical isotope ranking and LiDAR-to-MCNP geometry & Correctly identified HFIR isotope capabilities; proposed geometry workflow & Missed latest research; overly general responses \\
\bottomrule
\bottomrule
\end{longtable}
\endgroup

\section{Challenge problems explored at the workshop} 
This section present detailed case studies from 14 interdisciplinary teams that explored diverse applications of AI models across the nuclear energy spectrum. Each team tackled a specific challenge, employing structured workflows that combined prompt engineering, iterative refinement, and domain expertise validation. The projects span both fusion and fission applications, from fundamental physics to practical engineering solutions.

\input{team01.tex}
\input{team03.tex}
\input{team04.tex}
\input{team05.tex}
\input{team19.tex}
\input{team06.tex}
%\input{team06_2.tex} 
\input{team07.tex}
\input{team08.tex}
%\input{team09.tex} %remove
\input{team10.tex}
\input{team11.tex}
\input{team12.tex}
\input{team13.tex}

%\input{team14.tex} %remove
%\input{team15.tex} %remove
\input{team16.tex}
\input{team17.tex}
%\input{team18.tex} %remove

\section{Conclusion}

The workshop demonstrated that large language models (LLMs) can meaningfully contribute to nuclear energy research by accelerating hypothesis generation, facilitating multidisciplinary integration, and enhancing early-stage design workflows. Across 14 diverse projects, AI consistently proved useful for synthesizing literature, identifying knowledge gaps, generating structured workflows, and producing  scientific artifacts such as defect images and simulation strategies.

At the same time, the workshop underscored clear boundaries to current LLM capabilities. Technical depth, data access limitations, and constraints such as export control remain significant barriers. These challenges highlight the necessity of close integration between AI systems and domain experts, as well as the development of nuclear-specific data infrastructure and validation protocols.

Going forward, the path to impactful AI deployment in nuclear science lies in building secure, domain-aware AI platforms, expanding curated training datasets, and establishing hybrid human–AI workflows. The results of this workshop provide a compelling foundation for continued national investment in AI-augmented nuclear R\&D, with the potential to reduce development timelines, improve decision-making, and unlock new capabilities across the nuclear enterprise.

%\section*{Conclusions}
%The “AI for Nuclear Energy” workshop at ORNL demonstrated that AI reasoning models can meaningfully accelerate early‐stage exploration of fusion and fission science problems. Across six interdisciplinary teams, participants successfully formulated non-sensitive science challenges, applied structured workflows—encompassing telescoping, hypothesis generation, and code prototyping—and produced preliminary results within a single day. Key outcomes included novel domain hypotheses (e.g., surrogate models for plasma filament dynamics), proof-of-concept code prototypes for materials screening, and clear visibility into model performance constraints such as data sparsity and interpretability. The cross-team synthesis underscored common technical bottlenecks (limited labeled data, integration overhead) as well as process lessons (the value of precise prompting, the importance of in-session collaboration tools). Overall, the workshop validated AI’s role as a complementary tool to physics-based methods, enabling rapid iteration on complex nuclear science questions and fostering closer collaboration between domain experts and AI practitioners.

\section{Future Research Directions}

The workshop revealed significant potential for advanced  Large Language Models in nuclear energy research while exposing critical limitations. To that end, we identified  key research directions that will advance the nucelar energy research with advanced, physics-informed AI systems capable of accelerating nuclear energy development.

{\bf Multimodal Graph-Enhanced Models for Nuclear Materials Property Prediction} will advance materials discovery by combining atomic-scale graph representations with nuclear-specific property prediction capabilities, enabling the design of novel alloys and ceramics optimized for extreme nuclear environments. These models will integrate crystal structures, defect configurations, and radiation damage evolution to predict material behavior under neutron irradiation, high temperatures, and corrosive conditions, dramatically accelerating materials qualification from decades to years while ensuring safety-critical reliability standards.

{\bf Large-Scale Multimodal Foundation Models for Nuclear Energy Research} will establish comprehensive AI systems trained on integrated nuclear datasets spanning simulation outputs, experimental data, technical literature, and regulatory documentation, while maintaining security and export control compliance. These foundation models will process diverse data modalities including microscopy images, sensor measurements, simulation results, and engineering drawings to enable seamless integration of theoretical knowledge with experimental observations, supporting real-time decision making and cross-disciplinary research acceleration.

{\bf AI Surrogates for Exascale Nuclear Simulations} will create physics-informed neural network surrogates for leadership-class nuclear simulation codes, enabling orders-of-magnitude speedup while maintaining high-fidelity accuracy for reactor design optimization. These surrogates will leverage exascale training data from codes such as  Vertex CFD to support reactor design exploration, safety analysis, and optimization workflows that currently require weeks of supercomputing time, fundamentally transforming nuclear engineering design cycles.

{\bf Advanced Reasoning Models for Nuclear Energy Applications} will develop domain-specialized AI systems capable of advanced technical reasoning about complex nuclear phenomena, moving beyond current models' responses to provide deep, multi-step analysis grounded in nuclear physics principles. These reasoning models will incorporate conservation laws, thermodynamic principles, and nuclear data libraries to enable automated problem-solving, theoretical analysis, and educational support that matches or exceeds human expert-level technical reasoning capabilities.

{\bf Assured AI with Uncertainty Quantification and Causal Reasoning} will create nuclear-grade AI systems that meet the rigorous validation and safety standards required for nuclear applications, providing calibrated uncertainty estimates, explainable decisions, and regulatory-compliant documentation. These systems will incorporate Bayesian approaches, causal discovery algorithms, and formal verification methods to ensure AI predictions are trustworthy, interpretable, and suitable for safety-critical nuclear applications, enabling AI deployment in operational nuclear facilities with regulatory approval.

{\bf Agentic Workflows for Theory Development, Modeling and Simulation, and Experimental Synthesis} will establish autonomous AI research ecosystems capable of orchestrating complex nuclear research workflows spanning hypothesis generation, computational modeling, experimental design, and data synthesis with minimal human intervention. These agent networks will coordinate theory development, simulation execution, and experimental validation in continuous feedback loops, enabling autonomous research capabilities that can systematically explore nuclear design spaces and discover novel phenomena beyond current human research capacity.

{\bf Convergence of AI, Exascale, Nuclear Energy Research} will establish co-designed computational ecosystems that seamlessly integrate artificial intelligence algorithms with exascale high-performance computing infrastructure specifically optimized for nuclear energy applications. This convergence will enable AI training and inference on leadership-class supercomputers and distributed AI training across nuclear facility networks, creating unprecedented computational capabilities that leverage the synergies between AI algorithms, advanced computing architectures, and nuclear domain requirements to solve previously intractable problems in reactor design, safety analysis, and materials discovery.

\begin{comment}
\section{Future Work}
Building on these promising results, several avenues warrant further investigation:
\begin{enumerate}
  \item \textbf{Data Expansion and Curation:} Establish curated, domain-specific datasets—annotated experimental and simulation outputs—to mitigate data sparsity. Creating shared repositories will allow benchmarking and continuous improvement of AI models.
  \item \textbf{Workflow Automation and Scaling:} Develop end-to-end pipelines that automate key workflow stages (data preprocessing, hypothesis ranking, code template generation) and support larger teams or multi-day workshops. Containerized AI tools and collaborative notebooks could streamline onboarding and iteration.
  \item \textbf{Interdisciplinary Training and Best Practices:} Create training modules on AI-for-science best practices—covering prompt engineering, data governance, and ethical considerations—to empower nuclear scientists with AI fluency and to foster reproducible workflows.
  \item \textbf{Extended Pilot Studies:} Launch multi-week pilot projects where teams iteratively refine models against live experimental or simulation data, enabling deeper validation of AI-derived insights and stronger integration into ongoing nuclear R\&D programs.
\end{enumerate}
\end{comment}

\section*{Acknowledgements}
This workshop is sponsored by the Artificial Intelligence Initiative as part of the Laboratory Directed Research and Development (LDRD) Program of Oak Ridge National Laboratory, managed by UT-Battelle, LLC, for the US Department of Energy under contract DE-AC05-00OR22725.

\bibliographystyle{unsrt} 
\bibliography{refernces}

\end{document}

%% file: team01.tex
\subsection{Towards a Foundation Model for Fusion}

\begin{tcolorbox}[colback=gray!10, colframe=gray!80, title=\textbf{Team Members}, fonttitle=\bfseries]
    \textbf{Furkan Oz}, \href{mailto:ozf1@ornl.gov}{ozf1@ornl.gov}\\
    \textbf{Wes Brewer}, \href{mailto:brewerwh@ornl.gov}{brewerwh@ornl.gov}\\
    \textbf{Christopher Calle}, \href{mailto:calleci@ornl.gov}{calleci@ornl.gov}\\
    \textbf{Scott Klasky}, \href{mailto:klasky@ornl.gov}{klasky@ornl.gov}
\end{tcolorbox}

% \PB{please add at least 1 figure and 5 domain related citations}

\begin{figure}[htbp]
  \centering
  \includegraphics[width=0.8\linewidth]{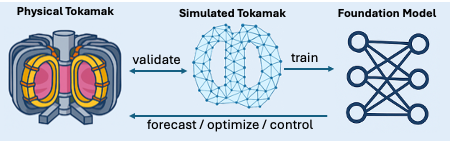}
  \caption{Conceptual overview: A foundation model (FM) trained on high-fidelity fusion simulations provides real-time control and receives validation feedback from a tokamak reactor.}
  \label{fig:team1_teaser}
\end{figure}

At the start of the workshop, our team held a brainstorming session to identify a fusion-relevant challenge that would meaningfully integrate each member’s area of expertise. We aligned on the development of a foundation model for fusion reactor simulation and control as a unifying goal, as illustrated in Figure~\ref{fig:team1_teaser}. With backgrounds spanning workflows, magnetohydrodynamics (MHD) simulations, control theory, data management, and high-performance computing (HPC), and AI, each team member assessed their strengths and contributed to the initial problem framing.

To accelerate planning and task division, we provided our brainstorming notes and domain specializations to ChatGPT-4.5, which helped decompose the project into structured tasks and suggested how responsibilities could be distributed. Each team member then worked on their assigned component independently for 2-3 hours, identifying key requirements, developing domain-specific contributions, and collaboratively integrating these elements into a unified workflow. Iterative feedback and refinement—both among teammates and through continued interaction with ChatGPT—helped us converge on a coherent and technically sound concept paper in just a few hours. 

\begin{figure}[htbp]
  \centering
  \includegraphics[width=0.95\linewidth]{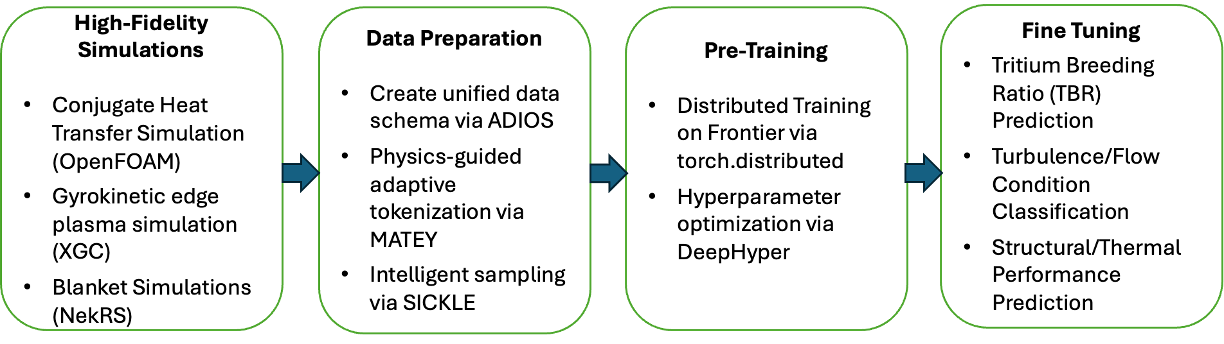}
  \caption{End-to-end workflow from high-fidelity simulations through data preparation and pre-training to fine-tuned foundation model tasks, illustrating how simulation, data, and AI components are integrated into a unified pipeline for scientific modeling.}
  % \caption{End-to-end workflow from high-fidelity simulations through data preparation and pre-training to fine-tuned foundation model tasks. This visual illustrates how each team member's contributions \PB{not sure what you mean by this; the figure does not have team members name} integrated into a unified AI pipeline.}
  \label{fig:team1_workflow}
\end{figure}

The overall resulting workflow—from high-fidelity simulations through data preparation, pre-training, and fine-tuning—is summarized in Figure~\ref{fig:team1_workflow}.

\subsubsection{Problem Statement}
\textbf{Description:}  
Current fusion simulation workflows are fragmented, relying on isolated high-fidelity codes (e.g., XGC~\cite{Chang2008XGC}, NekRS~\cite{Fischer2021NekRS}, M3D-C1~\cite{Jardin2012M3DC1}, OpenFOAM~\cite{Jasak2007OpenFOAM}) with incompatible data formats and inefficient communication strategies.
% Current fusion simulation workflows are fragmented, relying on isolated high-fidelity codes (e.g., XGC, NekRS, M3D-C1, OpenFOAM) \PB{add references} with incompatible data formats and inefficient communication strategies. 
This fragmentation leads to significant delays in data integration, limiting real-time prediction capabilities and slowing reactor design iteration cycles.

\textbf{Motivation:}  
Rapid and reliable integration of multiphysics simulation data is critical to meet DOE's strategic goal of developing fusion pilot plants by the early 2030s. By creating a unified data pipeline, we facilitate fast, iterative design optimization, drastically reducing simulation-to-insight latency and accelerating fusion energy research. The data pipeline will be constructured with a simliar schema which is mostly compliant with the International Thermonuclear Experimental Reactor (ITER) Integrated Modelling \& Analysis Suite (IMAS) standard~\cite{Imbeaux2015IMAS} since IMAS is not readily suited for large data. Our extensions will be the inclusion of ADIOS into IMAS which will work for both stellarators and tokamak physics.

% Rapid and reliable integration of multiphysics simulation data is critical to meet DOE's strategic goal of developing fusion pilot plants by the early 2030s. By creating a unified data pipeline that is compliant with the International Thermonuclear Experimental Reactor (ITER) Integrated Modelling \& Analysis Suite (IMAS) standard~\cite{Imbeaux2015IMAS} and capable of handling exascale-scale data and compute workflows, we facilitate fast, iterative design optimization, drastically reducing simulation-to-insight latency and accelerating fusion energy research.
% Rapid and reliable integration of multiphysics simulation data is critical to meet DOE's strategic goal of developing fusion pilot plants by the early 2030s. By creating a unified, IMAS-compliant \PB{IMAS defined?}, exascale data pipeline \PB{what is exascale data pipeline? large scale data pipeline or large scale data?}, we facilitate fast, iterative design optimization, drastically reducing simulation-to-insight latency and accelerating fusion energy research.

\subsubsection{Methodology}
\begin{itemize}
 \item \textbf{AI Models \& Tools:} Our team extensively utilized OpenAI’s ChatGPT-4.5, providing initial notes and concepts about our workflow, simulations (XGC, NekRS, OpenFOAM, M3D-C1), and our goals around the IMAS-compliant data schema and HPC integration. ChatGPT assisted in structuring our ideas clearly into a goal statement, objectives, specific tasks, subtasks, and success metrics.
  % \item \textbf{AI Models \& Tools:}  
  % We will leverage ADIOS2 for high-performance data transfer, IMAS/OMAS for schema standardization, and advanced orchestration tools (Airflow, RADICAL-EnTK) to automate data pipelines. Simulation data from XGC, NekRS, M3D-C1, and OpenFOAM were harmonized into a common schema for efficient downstream AI model ingestion.

  \item \textbf{Workflow:}  

Data Inventory and Notes Gathering $\rightarrow$ Problem Formulation $\rightarrow$ Prompting and Clarification with ChatGPT $\rightarrow$ Iterative Refinement and Telescoping (in parallel) $\rightarrow$ Task and Objective Structuring $\rightarrow$ Success Metric Definition $\rightarrow$ LaTeX Integration and Formatting $\rightarrow$ Final Concept Paper Draft 

The team began with Data Inventory and Notes Gathering, compiling individual expertise, relevant simulation tools, and domain-specific challenges. This was followed by Problem Formulation, where we defined the high-level objective of developing a foundation model for fusion reactor simulations.

We then engaged in Prompting and Clarification with ChatGPT to refine our ideas, decompose the problem into subcomponents, and explore alternative directions. Through Iterative Refinement and Telescoping, we continuously revised our project scope and structure based on AI-assisted feedback and peer discussion.

Next, Task and Objective Structuring helped us assign responsibilities aligned with each member's domain, enabling parallel development. We then articulated clear Success Metric Definitions, outlining how to evaluate each component's contribution to the overall concept.

LaTeX Integration and Formatting ensured our concept paper adhered to professional standards and supported seamless collaboration. Finally, we converged on a Final Concept Paper Draft, synthesizing all components into a cohesive document.
 
  % Each member contributed to building data pipelines (via IMAS and ADIOS), refining MHD-turbulence simulations, formulating control strategies using Lyapunov analysis, and developing physics-informed AI surrogates optimized for exascale systems like Frontier.
  % Simulation Outputs $\rightarrow$ ADIOS Data Management $\rightarrow$ IMAS Schema Harmonization $\rightarrow$ Automated Data Ingestion \& Validation $\rightarrow$ AI Readiness (feature extraction) $\rightarrow$ Downstream Tasks (foundation model pre-training and fine-tuning).

  \item \textbf{Team Roles:}  
  - Furkan Oz: Magnetohydrodynamics simulations and Turbulence Modeling  
  - Wes Brewer: AI Model Development and HPC Optimization  
  - Christopher Calle: Plasma Control and Stability  
  - Scott Klasky: Data and Workflow Integration, Project oversight, strategic guidance on exascale data management 
\end{itemize}

\subsubsection{Key Findings}
\begin{itemize}
  \item \textbf{Results:}  
  Through structured interactions with ChatGPT, our team successfully produced a detailed, coherent concept paper outlining clear project goals, objectives, tasks, and relevant success metrics. ChatGPT facilitated generating concise, accurate language to communicate complex technical concepts clearly, including a detailed LaTeX-formatted workflow figure caption and integration guidance.
  % The result of this study was to develop a concept paper outlining how we could create a foundation model for fusion. 
    % Developed an IMAS-compliant unified data pipeline capable of ingesting and validating high-volume simulation outputs at Frontier scale. Demonstrated seamless coupling between plasma, blanket, and turbulence codes via ADIOS zero-copy transfers, reducing data integration overhead from hours to minutes.

  \item \textbf{Insights:}  
  We found that by having a talented and diverse team, along with a structured layout for how each team member would interact with ChatGPT, and a well-defined concept paper template, we could produce a concept paper in just a few hours. ChatGPT helped to synthesize ideas and bring them together in a unified manner. 
  
  % ChatGPT significantly accelerated conceptual clarity and content generation, particularly helping to pinpoint the most effective language and formatting strategies for presenting complex scientific workflows and ensuring consistency in technical communication.
  % Integrating GPU-resident zero-copy middleware (ADIOS2) dramatically reduced simulation I/O bottlenecks, proving essential for real-time AI inference readiness. The developed pipeline provided a robust, extensible framework easily adaptable to new codes and datasets.
\end{itemize}

\subsubsection{Limitations \& Challenges}
\begin{itemize}
  \item \textbf{Technical Constraints:}  
  Due to the complexity of what was being proposed, we were not able to delve into any of the implementation details, but rather focused on the high-level conceptual strategy. Training a foundation model will require running large-scale simulations on Frontier, and then curating training datasets, etc., which involve minimally tens of thousands of node hours. 
  % Diverse data formats and differing simulation timestep frequencies complicated real-time data integration. Limited historical blanket data constrained validation efforts.

  \item \textbf{Process Hurdles:}  
  One challenge we encountered was in using ChatGPT-4.5 to generate a visual workflow diagram illustrating the pipeline from simulation to data preparation, pre-training, and fine-tuning Figs. \ref{fig:team1_teaser} and \ref{fig:team1_workflow}. While the model initially produced a relevant flowchart, the image was cropped and partially cut off. Subsequent attempts resulted in inconsistent diagrams with similar formatting issues. We found it much easier to generate the separate components of the diagrams and then piece them together in PowerPoint. This experience highlighted a clear opportunity for improving OpenAI’s diagram generation capabilities—particularly for producing consistent, high-quality visual content that would greatly benefit concept papers like ours.
  % One issue we came across was that we wanted ChatGPT-4.5 to create the workflow from simulations, to data preparation, to pre-training, and fine-tuning. However, while it initially generated a flow chart that had this content the flowchart was cropped in such a way that the image was cutoff. When we tried to regenerate it again, it came up with a completely different image, which was also cutoff. Clearly there are opportunities for improving how OpenAI could be used to generate flow charts, which would be very helpful for these kinds of concept papers. 
  % Ensuring timely metadata alignment across multiple simulation teams initially slowed workflow setup. Achieving consistent schema compliance among legacy simulation tools required considerable coordination.

  \item \textbf{Mitigations Tried:}  
  To address the technical scope limitations, we prioritized defining a clear, modular task structure that could guide future implementation efforts, and focused on aligning high-level objectives with feasible HPC workflows. In response to the diagram generation issues, we created a textual outline of the workflow with ChatGPT’s help and used it as a guide to manually construct the final flowchart using external tools. This ensured visual consistency while retaining the semantic structure generated by the model.
  % \textit{Need something here... Brief note on what was done to address these issues.}

  % \item \textbf{Mitigations Tried:}  
  % Developed intermediate metadata adapters and employed automated continuous-integration tests to quickly identify and rectify schema compliance issues. Implemented asynchronous data staging layers to accommodate variable simulation frequencies.
\end{itemize}

\subsubsection{Future Directions}
\begin{itemize}
  \item \textbf{Next Steps:}  
  Our next goal is to move from the concept paper to a working prototype by developing a small-scale foundation model using existing simulation datasets. This initial effort will lay the groundwork for a future publication related to the topic: ``Towards a Foundation Model for Fusion.''
  % Expand the unified schema and integration pipeline to incorporate additional fusion codes (e.g., JOREK, GENRAY). Conduct large-scale, continuous validation using experimental data from DIII-D and ITER.

  \item \textbf{Potential Extensions:}  
  There are several potential extensions that focus on enhancing our pipeline capabilities. One key area for exploration is the development of real-time visualization and monitoring systems to improve our ability to track and manage data flows more efficiently. Additionally, we see significant promise in AI-driven adaptive workflow optimization, which would enable automatic adjustments to data processing based on real-time HPC performance metrics. We also envision implementing AI-driven topology optimization to better align with our target objective function, alongside integrating neutronics into our framework.
  % Extend pipeline capabilities for real-time visualization and monitoring. Explore AI-driven adaptive workflow optimization, automatically adjusting data processing based on real-time HPC performance metrics. AI-driven topology optimization for target objective function. Neutronics integration. \textit{This one needs work}
\end{itemize}

%% file: team03.tex
\subsection{Compatibility of Structural Materials and Molten Salt Blanket in Fusion Reactors}

%\PB{please add at least 1 figure and 5 domain related citations}

\begin{tcolorbox}[colback=gray!10, colframe=gray!80, title=Team Members, fonttitle=\bfseries]
\textbf{Rinkle Juneja}, \href{mailto:junejar@ornl.gov}{junejar@ornl.gov}\\
\textbf{Arpan Sircar}, \href{mailto:sircara@ornl.gov}{sircara@ornl.gov}\\
\textbf{Luke Gibson}, \href{mailto:gibsonld@ornl.gov}{gibsonld@ornl.gov}
\end{tcolorbox}

%\PB{
%add name and emails in the following format:\\
%name1, email\\
%name2, email\\
%}

\begin{figure}[htbp]
  \centering
  \includegraphics[width=0.92\linewidth]{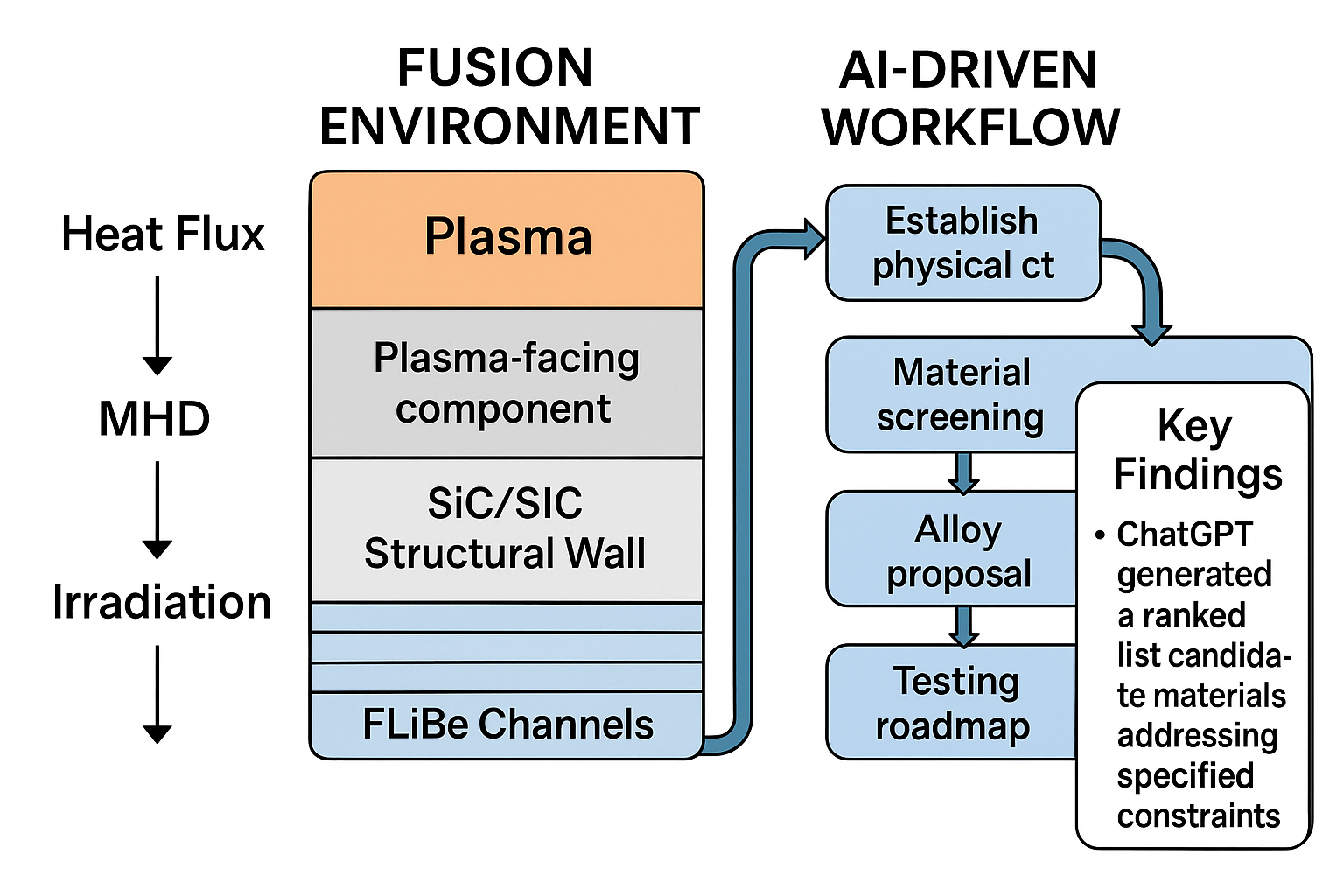}
  \caption{AI-generated modeling workflow highlighting the specified constraints on building a resilient material systems of structural materials and blankets and providing the AI-driven workflow.}
  \label{fig:team03_workflow}
\end{figure}

Our multidisciplinary team, comprising experts in fusion materials, machine learning, computational modeling, and blanket engineering, brainstormed together to bridge two critical research areas: high‐performance fusion structural materials and magnetohydrodynamic (MHD) behavior of molten salt blankets. Through iterative refinements, ChatGPT was able to outline the key drivers for coupling these domains and it mapped out a streamlined research roadmap (as illustrated in Figure ~\ref{fig:team03_workflow}.)that addresses material–blanket interface challenges.

\subsubsection{Problem Statement}
\textbf{Description:} It's crucial to simultaneously select a robust structural material \cite{MUROGA200213} and a blanket material \cite{Prato2019} suitable for fusion reactor %\PB{is challenging or crucial?} 
because the isolated selection does not guarantee their mutual compatibility under fusion‐relevant conditions. One of the major metrics for compatibility investigated here is electro-magnetic compatibility since large jumps in electrical conductivity in molten salt blankets are difficult to handle numerically \cite{siriano2024}.

\textbf{Motivation:} Accelerating fusion pilot‐plant deployment requires materials that can withstand extreme heat fluxes, high neutron irradiation, particle fluxes, and MHD‐induced stresses. By systematically evaluating the interface between candidate structural materials and molten salt blanket concepts, we can de‐risk component qualification, integration and shorten the path to commercialization.

\subsubsection{Methodology}
\begin{itemize}
  \item \textbf{AI Models \& Tools:} We used ChatGPT-4.5 and o4-mini to iteratively refine our problem statement. ChatGPT was able to list out state-of-the-art structural alloys being explored for fusion reactors, and outline potential coupling strategies with the molten salt blankets. We also used Anthropic's Claude v3.5 Sonnet v2 to explore viable candidates for structural materials in parallel with ChatGPT.
  \item \textbf{Workflow:} %Defining material and interface requirements $\rightarrow$ screen candidate alloys based on pros and cons $\rightarrow$ telescoping by going beyond state-of-the-art materials $\rightarrow$ generating a white paper draft based on proposed roadmap. \PB{expand and provide a bit more details for each step; how did you appraoch the problem} - Resolved
  \begin{enumerate}
      \item Establish physical conditions that structural materials must withstand in fusion reactor %Defining material and interface requirements
      \item Identify current state-of-the-art materials and list their advantages and disadvantages %Screen candidate alloys based on pros and cons
      \item Based on deficiencies in state-of-the-art materials, screen candidate alloys that satisfy the constraints in performance in extreme environments
      \item Generate road map for synthesizing and testing proposed materials
      \item Generate draft of white paper based on proposed road map
  \end{enumerate}
   \item \textbf{Team Roles:} Our team members had expertise in fusion materials, machine learning, fusion blanket modeling and MHD modeling. This allowed us to work together on addressing the interface compatibility of fusion structural materials and FLiBe blankets. 
  %\PB{Say who did what at a high level} - Resolved
   \begin{itemize}
        \item \emph{Juneja:} Investigated the search for fusion structural materials which can be compatible with the molten salt blankets ChatGPT-4.5 refined the search to design a workflow addressing the material search satistying most of the specified constraints using ChatGPT 04-mini.
        \item \emph{Sircar:} Used ChatGPT-4.5 to research how to handle the large change in electrical conductivity of molten salts and structural materials at solid-liquid interfaces in fusion blankets.
        \item \emph{Gibson:} Used Claude v3.5 Sonnet v2 from Anthropic to first identify the existing state-of-the-art materials proposed for structural materials and then explored promising candidate materials for comparison with output from ChatGPT
   \end{itemize}
\end{itemize}

%\PB{Please add a releated figure} - Resolved

\subsubsection{Key Findings}
\begin{itemize}
  \item \textbf{Results:} ChatGPT generated a ranked list of five candidate structural materials (e.g. Tungsten (pure and Wf/W), RAFM steels, ODS-RAFM, SiC/SiC, and High-Entropy alloy V-4Cr-4Ti), with a summary how each of these can tackle specified four stressors such as heat-flux, neutron irradiation, particle erosion, and MHD disruptions.
  \item \textbf{Coding strategies:} ChatGPT was able to come up with some strategies to tackle the large electrical conductivity jump at interfaces. This included - using a monolithic solver for the solid-liquid coupling, trying out the thin-wall approximation, and fine-tuning the linear solvers and preconditioners for the specific problem \cite{blishchik2021}.
  \item \textbf{Insights:} ChatGPT acknowledged that among the known candidates, no material system yet give performance tackling all the specified sources of damage. However, it ranked these systems based on what should be the best candidates if any of the specified constraints are our top constraints. We also compared the search with Google's Gemini platform. It did not list state-of-the-art materials but ChatGPT was able to do so.
\end{itemize}

\subsubsection{Limitations \& Challenges}
\begin{itemize}
  \item \textbf{Technical Constraints:} ChatGPT reliably summarized known materials \cite{MUROGA200213, GORLEY2021112513, SMITH1985125, Tavassoli2014} but could not propose truly novel alloys; it did, however, highlight SiC for its extensive validation data \cite{KATOH2019151849}. 
  \item \textbf{Process Hurdles:} ChatGPT was not able to provide high quality summary figures when we asked to generate a visual of high level figure for the proposed research summary. Also, ChatGPT was not able to exactly identify the issue in the simulations for MHD in solid-liquid domains.
  \item \textbf{Mitigation Tried:} We tried different reasoning models to go beyond state-of-the-art materials and it suggested "best-of-all-worlds" systems. We experimented with different prompt strategies, but manual figure curation remains necessary.
\end{itemize}

\subsubsection{Future Directions}
\begin{itemize}
  \item \textbf{Next Steps:} ChatGPT suggested systems beyond single materials such as stacking a layer of functionally-graded SiC / Refractory Low-Activation HEA / ODS-RAFM laminate (SiC-HEA-ODS), which might be worth exploring.
  \item \textbf{Potential Extensions:} Discussing the ideas with experimental colleagues.
\end{itemize}

%\PB{please add at least 5 domain related citations} - Resolved

% === End LaTeX snippet ===

%% file: team04.tex
\subsection{Isotopically Tailored Materials for Improvements to Fusion Reactor Design}

%\PB{please add at least 1 figure and 5 domain related citations}

%\PB{Cory will review} For each team, use the following headings:

\begin{tcolorbox}[colback=gray!10, colframe=gray!80, title=Team Members, fonttitle=\bfseries]
\textbf{Franklin Curtis},  \href{mailto:curtisfg@ornl.gov}{curtisfg@ornl.gov}\\
\textbf{Tim Younkin},  \href{mailto:younkintr@ornl.gov}{younkintr@ornl.gov}\\
\textbf{Qian Gong},  \href{mailto:gongq@ornl.gov}{gongq@ornl.gov}\\
\textbf{Ahmed Almeldein},  \href{mailto:almeldeinam@ornl.gov}{almeldeinam@ornl.gov}\\
\textbf{Ehab Hassan},  \href{mailto:hassanem@ornl.gov}{hassanem@ornl.gov}
\end{tcolorbox}

A multidisciplinary team consisting of experts from fusion, isotope separation, computational modeling, and computer science came together to evaluate the use of AI models to analyze materials within the fusion reactor that would benefit from isotopically tailored materials to improve design and efficiency \cite{duran2023}. The team used the AI tools to explore and generate ideas for materials that are affected by things like neutron bombardment that can benefit from isotopes with different cross-sections to aid with embrittlement, radioactive decay, and other issues related to neutron capture.

%\PB{the remaining section is incomplete; please fill these asap}  

\subsubsection{Problem Statement}

%\CH{ Please add at least 5 domain relevant citations.}
\textbf{Description:} 
Developing sustainable fusion energy requires addressing critical challenges in the design and operation of plasma-facing components (PFCs) \cite{linke2019challenges}, such as divertors, which endure extreme heat and particle fluxes. These components must tolerate heat loads exceeding 10 MW/m² while minimizing material erosion and impurity generation that can compromise plasma confinement. In addition to thermal stress, PFCs are exposed to intense 14 MeV neutron flux, leading to the activation of structural materials and the production of radioactive isotopes. To mitigate long-term waste and maintenance challenges, there is growing interest in isotopically tailored materials—engineered to reduce activation through selective nuclear cross-sections—offering potential benefits in safety, sustainability, and lifecycle cost.

Addressing these multiphysics challenges demands the integration of high-fidelity HPC simulation codes with machine learning (ML) models \cite{kerboua2024curriculum,wei2022perspective,giannini2025machine}. Surrogate models, trained using tools like NVIDIA Modulus, can replicate complex physics such as turbulence or material response at significantly lower computational cost. Coupling these AI models with fusion codes—either loosely as pre/post-processing tools or tightly within solver kernels—enables scalable, real-time prediction and design-space exploration. This code coupling strategy is essential to unlocking the full potential of exascale computing and accelerating innovation in fusion reactor design through data-driven, multidisciplinary workflows.

%\CDH{Break up into smaller sentences.} 
\textbf{Motivation:}  Achieving practical fusion energy requires addressing critical challenges in the design of plasma-facing components (PFCs), such as divertors, which must withstand extreme heat and neutron fluxes. Traditional workflows for exploring these challenges are time-consuming and fragmented, often requiring deep domain expertise across materials science, computational physics, and reactor engineering. Accelerating scientific discovery in this space demands tools that can support rapid information synthesis, hypothesis generation, and design iteration.

AI-driven language models like ChatGPT offer new opportunities to streamline early-stage research and cross-disciplinary exploration. By enabling rapid comparison of simulation tools, clarification of physical phenomena, and generation of code integration strategies, ChatGPT can assist scientists in evaluating design trade-offs and guiding research directions more efficiently. This is particularly valuable when exploring novel concepts, such as surrogate model integration into HPC workflows or adaptive sampling methods for physics-informed neural networks (PINNs).

A key topic of interest is the use of isotopically tailored materials—elements engineered to exploit differences in nuclear cross-sections to reduce long-term radioactive activation. These materials can lower the cost and complexity of managing activated waste when reactor components are replaced. In fusion environments, tailoring isotopes in structural and shielding materials may provide a means of improving safety and sustainability. Similar approaches have been explored in fission systems, such as the use of gadolinium-157 \cite{egle2020plasma} as a burnable poison due to its favorable neutron absorption properties \cite{seki1998impact}.

% Isotopically tailored materials, which are elements that take advantage of differing nuclear cross-sections to take advantage of nuclear activation properties of those materials. These materials have been of interest to the nuclear community with potential benefits that include a reduction in cost from the need to store radioactively activated materials when components within a fusion reactor are replaced \cite{seki1998impact}. Other benefits of tailoring isotopes can be for materials like burnable poisons for fission reactors, for example gadolinium-157 \cite{egle2020plasma}.  

%\CH{Remaining sections do not appear complete.}

\subsubsection{Methodology}
\begin{figure}[h]
      \centering
      \includegraphics[width=0.5\linewidth]{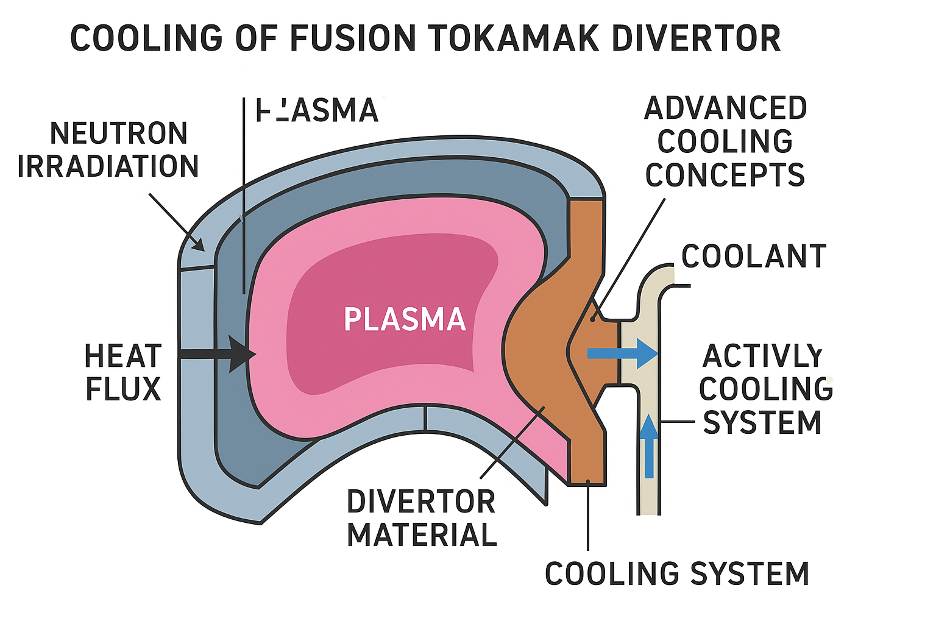}
      \caption{AI generated figure depicting the role of the divertor for a tokamak-style fusion reactor.  }
      \label{fig:team4_divertor}
  \end{figure}

\begin{itemize}
  \item \textbf{AI Models \& Tools:} We used ChatGPT iteratively during technical brainstorming meetings. The interaction involved: (1) Inputting domain-specific prompts extracted from team questions (e.g., thermal limits on tungsten coatings, neutron activation); (2) Requesting comparative assessments of computational platforms (e.g., Vertex vs. COMSOL, Trilinos); (3) Exploring integration approaches for AI surrogates with high-performance simulation codes.  
  % Each member of the team chose to utilize a specific AI tool to explore the isotopically tailored problem set described above. 
  The majority of the team did not have much experience using any of the AI tools so utilized the approaches suggested by the workshop organizers in an attempt to extract the best information from the provided platforms. 
  \item \textbf{Workflow:} The team began with a general search of what materials would have the largest impact on the fusion reactor for isotopic tailoring. Taking the approach of a naive researcher, little information was provided to the algorithm in an attempt to avoid providing bias to the results. Once this was established, the team began to dive deeper into the results and asked more probing questions with requests for computer codes and figure generation (\ref{fig:team4_divertor}).
    %\PB{Say who did what at a high level} 
    \item \textbf{Team Roles:} 
     \begin{itemize} 
     \item \emph{Qian Gong}: Used ChatGPT-4o to explore the unique features and performance of VERTEX compared against other state-of-the-art CFD simulation codes. Used ChatGPT-4.5 to brainstorm code-coupling strategies allowing for integrating ML surrogate models with VERTEX code architecture and study examples.
     \item \emph{Ehab Hassan}: Used ChatGPT-4o to explore the physical and mechanical properties of materials and designs that extend the sustainability of the divertor under extreme heat and particle fluxes, and develop a scaling-law for erosion rate of several materials and energy confinement time.
     \end{itemize}
\end{itemize}

\subsubsection{Key Findings}
\begin{itemize}
  \item \textbf{Results:} ChatGPT significantly enhanced our ability to explore and iterate on complex scientific problems related to fusion reactor design. It facilitated rapid understanding of domain-specific topics such as heat flux limits on divertor tiles, neutron activation of materials, and the viability of isotopically tailored materials. It provided valuable comparions between limits of various materials and geometic designs that can be used in the divertors.
  Advanced divertor geometries aim to control impurity transport and maximize pumping efficiency:

\begin{itemize}
    \item \textbf{Conventional Divertor}: Minimal flux expansion; high erosion; poor impurity retention.
    \item \textbf{Super-X Divertor (SXD)}: Long leg, extended connection length; excellent heat spreading and impurity screening.
    \item \textbf{Snowflake Divertor (SFD)}: Double X-point improves volumetric radiation and strike-point distribution.
    \item \textbf{X-Point Target (XPT)}: Magnetic mirror geometry reflects impurities and traps neutrals; superior performance in opacity and erosion suppression.
\end{itemize}

\begin{table}[h]
\centering
\begin{tabular}{lccc}
\toprule
Divertor Type & Impurity Recycling & W Erosion & SOL Control \\
\midrule
Conventional  & High               & High      & Poor        \\
SXD           & Low                & Low       & Moderate    \\
SFD           & Moderate           & Moderate  & High        \\
XPT           & Lowest             & Lowest    & Best        \\
\bottomrule
\end{tabular}
\end{table}

  Additionally, ChatGPT provides scaling-laws for the rate of erosion of the Tungsten material and their effect of the energy confinement time.

  Tungsten erosion rate per unit area (atoms/m\textsuperscript{2}/s):

\[
\dot{E}_W \propto \left( \frac{S}{R^2 A} \right)^{0.5} \left( \frac{I_p}{B_T} \right)^{1.2} \left( \frac{p_{ped}}{\Delta_{ped}} \right)^{0.6} (1 - \delta^2)^{0.5} \kappa^{0.3} \left( \frac{1}{\omega_\phi + \epsilon} \right)^{0.4} \left( \frac{1}{t_W} \right)
\]

\noindent Where:
\begin{itemize}
    \item $S$: neutron source rate (n/s)
    \item $R$: major radius, $A$: aspect ratio $R/a$
    \item $I_p$: plasma current, $B_T$: toroidal magnetic field
    \item $p_{ped}$: pedestal pressure, $\Delta_{ped}$: pedestal width
    \item $\delta$: triangularity, $\kappa$: elongation
    \item $\omega_\phi$: toroidal rotation, $t_W$: tungsten thickness
\end{itemize}

\[
\tau_E = \tau_E^{\text{ITER98}} \cdot \left(1 + \alpha \cdot \dot{E}_W^{\text{norm}} \right)^{-1} \cdot \left(1 + \beta \cdot \frac{\lambda_{\text{mfp}}}{\Delta_{\text{SOL}}} \right)^{-1}
\]

\noindent Where:
\begin{itemize}
    \item $\tau_E^{\text{ITER98}}$: baseline confinement time
    \item $\dot{E}_W^{\text{norm}}$: normalized erosion rate
    \item $\alpha, \beta$: empirical degradation constants
\end{itemize}
  
  ChatGPT also provided valuable summaries and comparisons of existing high-performance computing frameworks (e.g., Vertex, MOOSE, WarpX), enabling informed software selection for various simulation tasks. In addition to technical synthesis, ChatGPT helped deconstruct and explore strategies for coupling AI surrogate models—trained using tools like NVIDIA Modulus—with traditional fusion simulation codes. This included guidance on loose and tight integration schemes, performance optimization techniques, and deployment pathways on exascale platforms. As a result, ChatGPT acted as both a knowledge accelerator and a brainstorming assistant, helping us bridge the gap between computational capabilities and scientific objectives.
  \item \textbf{Insights:} We found ChatGPT to be most effective during early-stage exploration, problem scoping, and technical ideation. Its utility was greatest when framing new research directions, breaking down interdisciplinary questions, or synthesizing disparate concepts across physics, materials, and computational science. For instance, concise explanations of reaction-diffusion modeling, surrogate model deployment, and SDF-based adaptive sampling accelerated onboarding and team discussions. However, ChatGPT was not used as a replacement for technical validation or deep-domain authority. Instead, it served as a first-pass assistant to clarify ideas, generate code stubs, identify missing components, or spark alternative hypotheses. Crafting well-scoped prompts and iteratively refining questions yielded the most valuable responses. As such, ChatGPT proved particularly helpful in augmenting the ideation and conceptual design phases of scientific workflows, while downstream modeling, benchmarking, and analysis remained in the hands of domain experts.
\end{itemize}

\subsubsection{Limitations \& Challenges}
\begin{itemize}
  \item \textbf{Technical Constraints:} Responses require expert validation and sometimes follow-up searches for peer-reviewed confirmation. For example, the team asked what would be required to pull the massive heat out of the plasma and the AI correctly identified the heat transfer equation as:
  \begin{equation}
      \dot{Q}=hA\left(T-T_f\right)
  \end{equation}
where $\dot{Q}$ is the heat flux, $h$ is the heat transfer coefficient of the cooling fluid, $A$ is the area of the surface, $T$ is the surface temperature, and $T_f$ is the fluid temperature. For the heat flux coefficient, the algorithm suggested a cooling fluid velocity of 50,000 $m/s$, which is unreasonable. 

Although some of the values were unreasonable, the information gathered for relevant materials was used to develop a 'white paper' to begin a first start for funding opportunities. While the algorithm provided this paper, it was not able to provide relevant citations and instead supplied made up citations.
  \item \textbf{Process Hurdles:} Collaboration bottlenecks, tooling gaps, sensitivity issues.
  \item \textbf{Mitigations Tried:} Subject matter expertise was used to weed out irrelevant physics and guide the AI to provide better responses. For the unfamiliar, this could produce results that are misleading. 
\end{itemize}

\subsubsection{Future Directions}
\begin{itemize}
  \item \textbf{Next Steps:} 
  \begin{itemize}
      \item Develop a fine-tuned version of ChatGPT or RAG (retrieval-augmented generation) system using internal fusion simulation documents and experimental data. The team could use the white paper as a starting point to look at relevant materials for isotopically enriched materials for funding opportunities.
      \item 
  \end{itemize}
  
  \item \textbf{Potential Extensions:} New methods, datasets, or interdisciplinary angles to explore. 
\end{itemize}
% === End LaTeX snippet ===

%% file: team05.tex
\subsection{Tracing Energetic Particle Pathways in the Lifetime of Alpha Particle Confinement in Fusion Power Plant Reactors}

\begin{tcolorbox}[colback=gray!10, colframe=gray!80, title=Team Members, fonttitle=\bfseries]
\textbf{Sebastian De Pascuale},  \href{mailto:depascuales@ornl.gov}{depascuales@ornl.gov}\\
\textbf{Yashika Ghai},  \href{mailto:ghaiy@ornl.gov}{ghaiy@ornl.gov}\\
\textbf{Mark Cianciosa},  \href{mailto:cianciosamr@ornl.gov}{cianciosamr@ornl.gov}\\
\textbf{Xiao-Ying Yu},  \href{mailto:yuxiaoying@ornl.gov}{yuxiaoying@ornl.gov}
\end{tcolorbox}

Our team considered the unique challenges that future fusion reactors will face in the burning plasma regime \cite{hawryluk2019} compared to today's experimental devices. Primarily, operating in a fusion scenario will produce from each D-T reaction one neutron of 14.1 MeV energy and one helium ion (alpha particle) of 3.5 MeV energy. This means that approximately 20$\%$ of the energy produced by fusion reactions will be lost by magnetic confinement within the plasma or along plasma facing components from diffusive transport and other loss mechanisms. This power fraction from energetic particles would ideally be recirculated within the core plasma. Helium atoms would thereby deposit their energy back into the plasma to reach self-heating (or burning) conditions necessary to promote fusion. However, alpha particles preferentially transfer their energy to electrons via collisions, and these electrons subsequently transfer a proportion of their energy to thermal ions in a very inefficient process of events on the order 30-35$\%$. We investigated the domain reasoning of ChatGPT 4.5 to identify and infer key quantities and constraints that determine the lifetime of an alpha particle, born in the core of a fusion plasma and expiring on the periphery of the vessel through material interactions or pumped exhaust. 

\begin{figure}[h]
  \centering
  \includegraphics[width=0.6\linewidth]{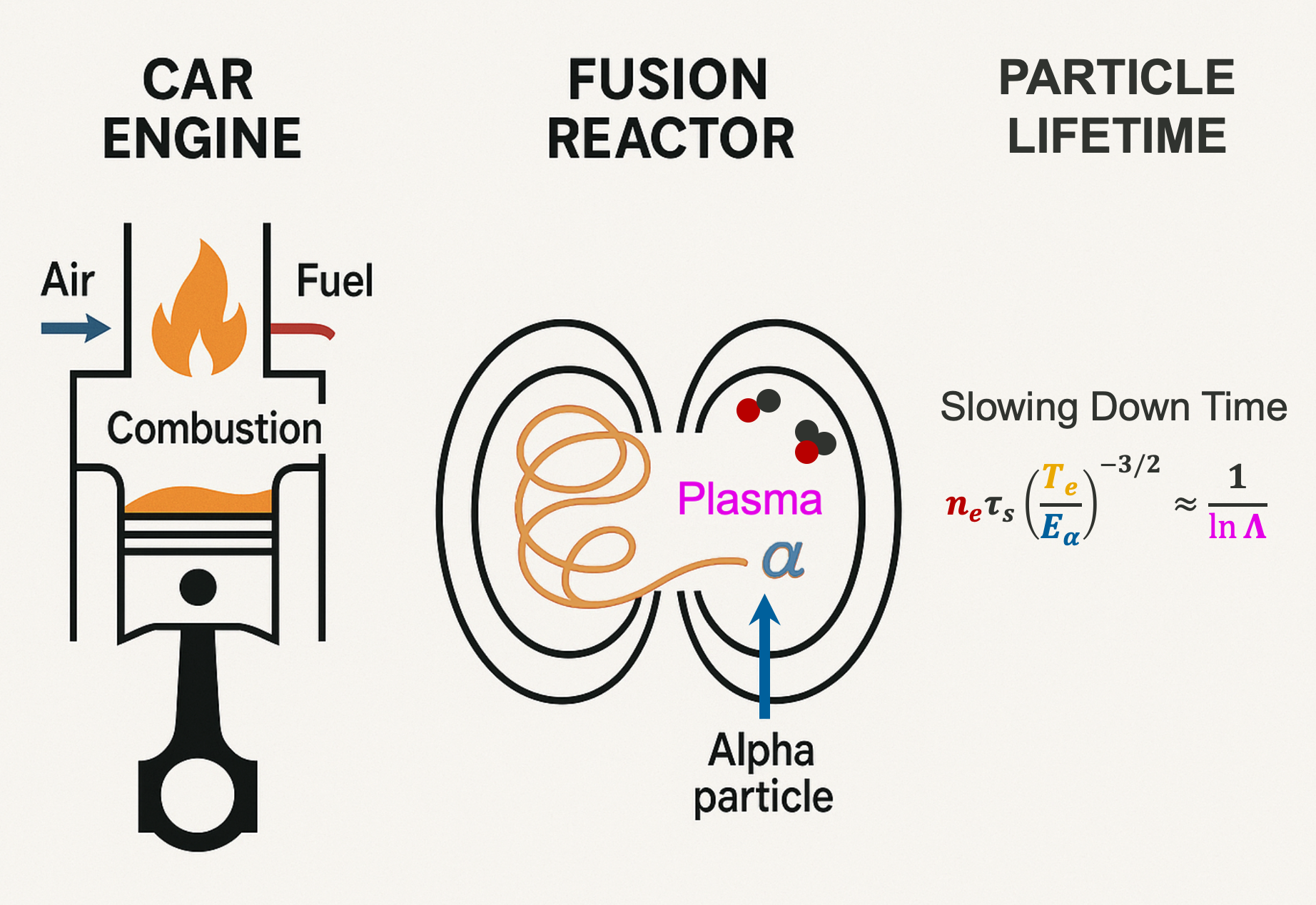}
  \caption{AI generated interpretive diagram showcasing alpha particles (of 3.5 MeV energy, $E_{\alpha}$, shown in blue) produced by fusion of D-T fuel (at high densities, $n_e$, shown in red) as a key ingredient for maintaining sustained fusion reactions through self-heating the plasma \cite{bonofiglo2024alpha} (to high temperatures, $T_e$, shown in yellow) primarily through collisions (indicated by the Coloumb logarithm, $\ln \Lambda$, shown in magenta. The alpha particle slowing down \cite{dai2023} time, $\tau_s$, is an indication of the total confinement of these particles and the triple product, $n_e\tau_eT_e$, is a key metric for obtaining fusion energy producing conditions at high enough density and temperature for long enough time.}
  \label{fig:team5_diagram}
\end{figure}

The role of energetic particles in fusion reactors is illustrated by an energy cycle, similar to the action of an engine as illustrated in Figure \ref{fig:team5_diagram}, where in the closed system of magnetic confinement the Helium ions act as a principle constituent of the plasma driving energy back into the reactions and in the open system of the plasma interactions with the vessel environment, the Helium ions are removed as "ash" pumped through exhaust pipes or deposited onto the walls of power plants \cite{hino1994reduction,hino1999helium}. Since these physical processes encompass multiple physical scales and processes of plasma, particles, materials, and thermodynamics we considered it an excellent challenge problem to test the comprehensive capabilities of ChatGPT from a theoretical level to an experimental level.

\subsubsection{Problem Statement}
\textbf{Description:} Alpha particles produced by D-T fusion reactions will carry 3.5 MeV of energy and must remain confined within the plasma long enough to transfer energy to elections and ions via collisions, thereby heating the plasma towards self-sustaining steady-state power generation.\\

\textbf{Motivation:} Helium ions produced by fusion will only remain within the plasma on the order of a typical slowing down time, which can be calculated for a reactor scale device to be of the order of 0.1 to 1 second. This amounts to approximately 200,000 orbits through a toroidal tokamak device before crossing the last closed magnetic flux surface that must carefully balance the degree of energetic particle confinement. Too short of a lifetime within the core plasma and the power plant will suffer from poor heating whereas too long of a lifetime results in fuel dilution of the primary fusion leading to dilution and inefficient energy generation. We considered detailed approaches produced by ChatGPT to optimize alpha particle lifetimes and manage power exhaust for fusion pilot plant reactors.

\subsubsection{Methodology}
\begin{itemize}
  \item \textbf{AI Models \& Tools:} We targeted the OpenAI ChatGPT o4-mini model for deployment solutions as it was the fastest performer in advanced reasoning tasks and, through the new Deep Research feature, this model was able to capture images directly from published reports with accurate and accessible citations. 
  \item \textbf{Workflow:} Our team organized its investigation following a hierarchical branching Agile framework, where each member pursued individual research queries according to their own expertise starting from a high level with the goal to produce testable code or calculations. We periodically converged on nodal queries to compare results from different conversational contexts, ensuring that collaboration could synthesize the best results from the reasoning models. Finally, we wrote descriptive assessment of model responses that allowed for changes to our line of inquiry so that more promising directions could be pursued. 
  \item \textbf{Team Roles:} Each team member brought important practical knowledge and experience to the use of ChatGPT in this exercise from the plasma theory of energetic particles, research software development, machine learning and artificial intelligence, and material science. Our breadth of expertise allowed us to better consider the cross-disciplinary challenges for power producing fusion reactors in the burning plasma regime. 
  \begin{itemize}
  \item \emph{De Pascuale:} investigated strategies for handling Helium exhaust from alpha particle production, using ChatGPT deep research to assess model extrapolation from present day devices to fusion reactor conditions.
  \item \emph{Ghai:} analysed feedback mechanisms to harness improved alpha particle heating, including proposals suggested by ChatGPT for experiments and hardware modifications to test theoretical predictions.
  \item \emph{Cianciosa:} generated input/output files from ChatGPT responses for the Integrated Plasma Simulator (IPS) software package and magnetic equilibrium solver to rapidly develop testable modeling cases.
  \item \emph{Yu:} explored potential candidate materials and novel compositions to use as plasma facing components in future devices, including expansive key word search with ChatGPT deep research to access literature and database surveys.
  \end{itemize}
\end{itemize}

\subsubsection{Key Findings}
\begin{itemize}
  \item \textbf{Results:} ChatGPT o4-mini correctly characterized the 3.5 MeV alpha particle as an important agent in the fusion process. It specified that once the energetic ion is thermalized with the plasma, it effectively becomes Helium ash  at ambient ion temperatures and can no longer sufficiently heat the plasma. In this manner, Helium doesn't participate in fusion reactors and can reduce the overall reaction rate leading to decreased fusion power output. Additionally, a population of thermal Helium can contribute to bremsstrahlung and line radiation which increases the net energy loss from the plasma.
  \item \textbf{Insights:} Initial conversations with ChatGPT led to two proposed mechanisms to manage energetic particle lifetimes in a fusion reactor. One concept is the process of $\textbf{alpha particle channeling}$, where a radio-frequency antenna generates plasma waves that can extract energy from fusion-born alpha particles $\textit{before}$ they thermalize with electrons. This allows for a more efficient pathway of energy transfer directly to the bulk ions which sustain the plasma temperature through preferential diffusion. It was suggested that Ion Cyclotron Range of Frequencies, Lower Hybrid Waves, or mode conversion from Alfv\'en to Ion Bernstein waves could facilitate in this energy transfer. The other mechanism identified was $\textbf{helium ash removal}$ through divertor design. Carefully shaping the magnetic configuration allows or increased pumping pressures and management of heatflux, necessary to mitigate excess power loss conditions along the scrapeoff layer. Turbulent diffusion and convection tend to carry thermalized helium ions away from the core towards the plasma edge, a phenomenon termed impurity screening. Additionally, in high performance plasmas transient magnetohydrodynamic events called edge-localized modes can periodically purge particles from the plasma edge down towards the divertor. ChatGPT produced a comparison of different optimized geometries that could better handle Helium exhaust, including enhanced selective pumping mechanisms and active exhaust control.
\end{itemize}

\subsubsection{Limitations \& Challenges}
\begin{itemize}
  \item \textbf{Technical Constraints:} As there are a variety of experimental tokamak research facilities, the model had a difficult time synthesizing the results across devices into one proposed fusion pilot plant design. We decided to use the international ITER experiment to base our findings on and familiarity with the literature to assess model performance. 
  \item \textbf{Process Hurdles:} We noticed that different conversation iterations could lead to several dead-ends, where the comment history adversely affected ChatGPT responses. In these cases the model was incorrectly linking different factual pieces of information together in order to produce a reasonable seaming answer, such as the name of a researcher who published a particular citation or the name of a code and its particular use cases. 
  \item \textbf{Mitigations Tried:} We decided to start new conversations from scratch in order to proceed with our investigation or to test the same query in a different reasoning model in order to cross-validate the responses.
\end{itemize}

\subsubsection{Future Directions}
\begin{itemize}
  \item \textbf{Next Steps:} Interestingly, the ChatGPT o4-mini model noted that burning plasma regimes aren't accessible in today's magnetic confinement fusion devices and instead proposed a series of experiments in order to test energetic particle physics. Some of these responses were based on previously published literature, such as the use of neutral beams to preferentially heat the plasma, however other cases indicated experimental upgrades to facilities in order to access new radio-frequency resonances or improved ion diagnostics in order to better track the particle trajectories of fast ions. 
  \item \textbf{Potential Extensions:} The Deep Research feature of ChatGPT o4-mini applied to our investigation indicates that with proper synthesis of not only the published literature from experiments, but also with unused diagnostic or operational data from comparable research facilities a fine-tuned reasoning model could propose optimization of device design in order to address important physics challenges in fusion energy science. 
\end{itemize}

%\PB{please add at least 5 domain related citations}

% === End LaTeX snippet ===

%% file: team19.tex
\subsection{AI-Assisted Semantic Annotation of Fusion Data via Retrieval-Augmented Generation }%(1--2 pages each)}
%For each team, use the following headings:

%\PB{Franklin will review}

%\PB{please add at least 1 figure and 5 domain related citations}

\begin{tcolorbox}[colback=gray!10, colframe=gray!80, title=Team Members, fonttitle=\bfseries]
\textbf{Ana Gainaru},  \href{mailto:gaianrua@ornl.gov}{gaianrua@ornl.gov}\\
\textbf{Chris Bryan}, \href{mailto:bryancd@ornl.gov}{bryancd@ornl.gov}
\end{tcolorbox}

\subsubsection{Problem Statement}
\textbf{Description:} 
Fusion research generates high-frequency, multi-modal measurements and simulation datasets~\cite{castro2012netcdf}. These datasets often lack consistent metadata or semantic annotations, limiting their reuse, interpretability, and integration across studies~\cite{anirudh20232022}. One solution of integration expert knowledge in data-driven model extractions can mitigate these challenges~\cite{10838878}, but it introduces a high overhead.
We envisioned a trained AI system based on a Retrieval-Augmented Generation (RAG) framework to automate the semantic enrichment of measurement and simulation data that can provide a schema and transformation recommendations for datasets generated with different experiments and simulation codes or versions. Such a capability would allow for fast and reliable evaluation of operation and simulation data for 
determination of causal relationships and trend evaluation for nuclear power and research reactors across unrelated datasets.

Our \textbf{main goal} can be summarized as: Given a raw dataset (e.g., NetCDF, CSV, or binary log) containing fusion measurements or simulation data, we aim to:
\begin{itemize}
    \item Extract and annotate variables with semantic metadata (units, geometry, definitions, relevance).
    \item Ground annotations in credible scientific sources (e.g., dataset readme files, publications).
    \item Provide structured outputs suitable for machine learning, visualization, or scientific interpretation.
\end{itemize}

\textbf{Motivation:}
Most nuclear power and research reactors have generated significant data over their lifetimes. Some of this data is typically used during operation by reactor control systems to adjust and control the reactor operating parameters or indicate to human reactor operators the status of specific systems~\cite{alberti2023automation}. Some data is captured but not used in operations; however this data can be mined later for better understanding of the reactor operating conditions at the time. While most of this data generated in the past 30 years is digital, some reactors started operations with analog pen plotters, so a significant amount of reactor data exists in different formats. Similar issues can be seen with simulation data~\cite{10.1007/978-3-030-63393-6_18}. Despite the vast number of datasets collected over decades from both simulations and experiments, fusion scientists often cite a scarcity of usable data—largely due to inconsistent metadata, undocumented variables, and the difficulty of integrating or interpreting legacy and heterogeneous sources.

 We investigated a system for AI-assisted ingestion and semantic annotation of raw fusion data using Retrieval-Augmented Generation (RAG). Our approach combines data chunking, scientific document indexing, and large language models (LLMs) to enrich raw data with contextually relevant metadata derived from associated publications, technical documentation, repositories and analysis codes, and domain-specific ontologies. We investigated how RAG can improve the accessibility, standardization, and utility of fusion datasets.

\subsubsection{Methodology}

\begin{itemize}
  \item \textbf{AI Models \& Tools:} ChatGPT o3 and Gemini were used as the main reasoning models and ChatGPT o4-mini-high for the initial code generation. We used the models to recommend the steps needed to build our framework, an exemplar application that we can test our methodology on and an evaluation plan. % data inputs, and software frameworks were used.
  \item \textbf{Workflow:} Our pipeline includes the following components:
  \begin{itemize}
      \item \textit{Document Corpus Construction.} For this we used the reasoning models to find ways of collecting and preprocessing scientific documents, including: i) Dataset README files;  ii) XML/JSON metadata; iii) Open-access scientific papers (e.g., IPCC, NOAA, PANGAEA); iv) Repositories with analysis codes
    \item \textit{Metadata cleaning}. The reasoning models recommanded ways of cleaning each document, chunked by semantic unit (paragraph or section), and enriched with metadata (source, section, tags). E.g. year information for papers can give context to datasets generated around that time. 
    \item \textit{Embedding and Indexing}. The cleaned chunks are encoded using domain-aware language models (e.g., SciBERT, OpenAI embeddings). We investigated ways of  fast semantic retrieval during annotation.
    \item  \textit{RAG-Based Annotation} For each data record or variable, we use a predefined API to form a query, top-k relevant chunks are retrieved, an LLM is prompted with the data and retrieved context and the  model generates structured annotations, including field definitions, units, and scientific relevance.
    \item \textit{Output Normalization} The generated annotations are converted into a unified schema (e.g., JSON-LD or CF-convention-compliant metadata) for integration with downstream tools.
  \end{itemize}
  
  \item \textbf{Team Roles} A.Gainaru was in charge of designing the entire workflow using Gemini and applying it to a commonly used dataset using ChatGPT and C. Bryan investigated ways of creating a metadata dataset with relevant publications and documentation using ChatGPT. K. Dayman has investigated the use of RAGs for the purpose of dataset annotation. All together conceived the research problem and design the workflow for adding semantics to datasets.
\end{itemize}

%Additionally, many papers have been published in journals and other media regarding reactor operations, maintenance, failures, improvements and missions. 

%Our challenge is to envision a system that leverage all data collected during reactor operations plus published papers, articles and presentations to train AI systems to ultimately enable interactive evaluation of this data to find  trends, linkages, relationships and coincidences between parameters with the goal of improving operational excellence.

\subsubsection{Key Findings}

\begin{figure}
    \centering
    \includegraphics[width=1\linewidth]{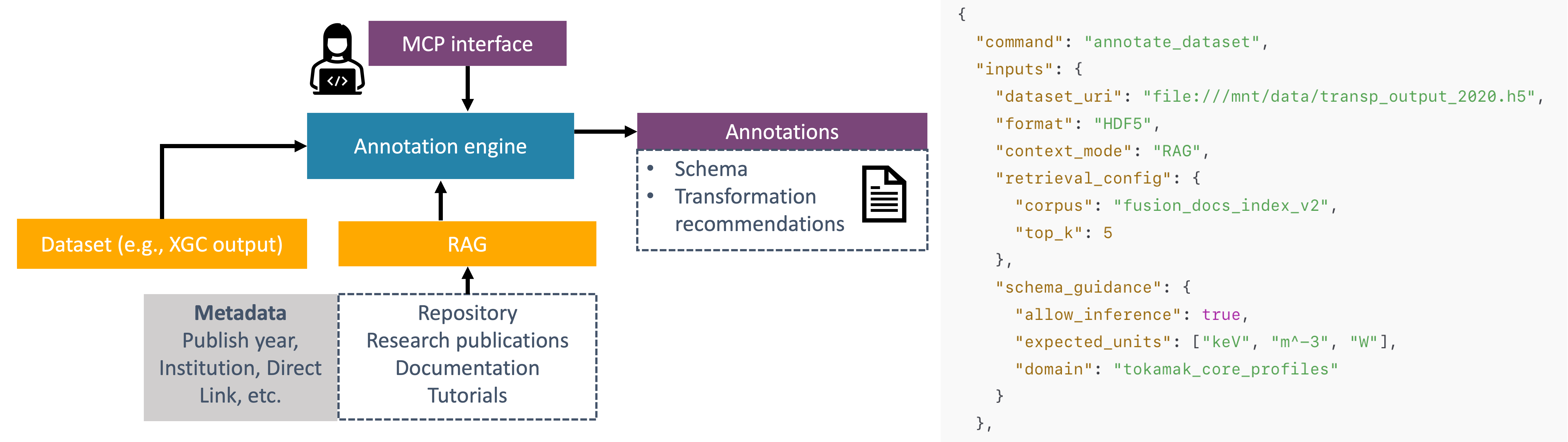}
    \caption{Methodology for users to request semantics about a fusion dataset}
    \label{fig:Team19_workflow}
\end{figure}

\begin{itemize}
  \item \textbf{Results:} The workflow used is presented in Figure~\ref{fig:Team19_workflow}. We applied the suggested methodology for Ion temperature profile in the TRANSP simulation outputs~\cite{doi:10.1177/02783649231213117} and the and we plan to generalize the methodology in the future. Our plan was to be able to read the Ion temperature in the same format for datasets generated with different versions of TRANSP without any manual intervention. Most of the time was spent developing and putting the RAG pipeline was in place. While progress has been made, we were unable to achieve our goal since we could not gather the datasets needed to test different parts of the pipeline. To overcome this, we used the MITIM framework that integrates various fusion modeling tools, including TRANSP~\cite{Rodriguez-Fernandez_2022}. While it doesn't provide the TRANSP code, it offers scripts and documentation for setting up and running simulations using TRANSP within the MITIM environment. The ChatGPT Reasoning model seemed to generate a plausible coding outline for a minimal working RAG pipeline to go through the MITIM documentation and generated datasets.
    % Summarize the primary outcomes—quantitative metrics, novel hypotheses, code prototypes.
  %\item \textbf{Insights:} %Highlight any unexpected observations or domain-specific learnings.
\end{itemize}

\subsubsection{Limitations \& Challenges}
\begin{itemize}
  \item \textbf{Technical Constraints:} Finding publicly available TRANSP simulation datasets spanning the past decade can be challenging due to access restrictions and the proprietary nature of some data. We were able to play with generated data in different formats, but we were unable to test the RAG pipeline since there was no documentation for the generated data.%Data availability, model performance ceilings, computational resources.
  \item \textbf{Process Hurdles:} Setting up the whole workflow is a bottleneck since not all the team members are able to access the data necessary for testing the hypothesis. In addition, ChatGPT and Gemini tend to give a lot of information that is not necessary useful for the given task, a lot of time was spent in extracting the needed information and double checking what was generated (this is especially true for the code generation for the RAG pipeline. The reasoning models were unable to help us track down datasets that can be used to test our pipeline. %Collaboration bottlenecks, tooling gaps, sensitivity issues. 
  \item \textbf{Mitigations Tried:} For the coding part, corrections and changes were requested until the code was usable. For the dataset discovery, many interactions were required to find datasets that are generated by codes hosted on GitHub. The TRANSP-IMAS Translator was suggested to understand how to generate TRANSP data without having access to the code.  %Brief note on what was done to address these issues.
\end{itemize}

\subsubsection{Future Directions}
\begin{itemize}
  \item \textbf{Next Steps:} We plan to try the methodology proposed to other datasets for which we have access. We plan to make experiments using XGC or GENE that are well-documented and provide comprehensive resources to understand the format of data.  %Suggest follow‐up experiments or deeper investigations.
  \item \textbf{Potential Extensions:} We plan to investigate MCP (Model Control Protocol)-style interfaces to allow users (or data management and query systems) issue structured requests to models or agents to perform tasks like annotating a dataset. %New methods, datasets, or interdisciplinary angles to explore.
\end{itemize}

% === End LaTeX snippet ===

%% file: team06.tex
\subsection{AI-assisted Multiphysics Modeling  for MPEX}

\begin{tcolorbox}[colback=gray!10, colframe=gray!80, title=Team Members, fonttitle=\bfseries]
\textbf{Atul Kumar},  \href{mailto:kumara@ornl.gov}{kumara@ornl.gov}\\
\textbf{Paul Laiu}, \href{mailto:laiump@ornl.gov}{laiump@ornl.gov}\\
\textbf{Jong Youl Choi},  \href{mailto:choij@ornl.gov}{choij@ornl.gov}
\end{tcolorbox}

\subsubsection{Problem Statement}

\textbf{Description:} The Material Plasma Exposure eXperiment (MPEX) \cite{Rapp:2016, Rapp:2024, Kumar:2024, Kumar:2023, Islam:2024} is a high-power linear plasma device designed to study plasma–material interactions under reactor-relevant conditions. A key scientific and engineering challenge is modeling the tightly coupled physical processes involved—RF wave propagation, impurity erosion and redeposition, plasma transport, and sheath dynamics—across disparate time and spatial scales.

  \begin{figure}[ht]
        \centering
        \includegraphics[width=0.95\linewidth]{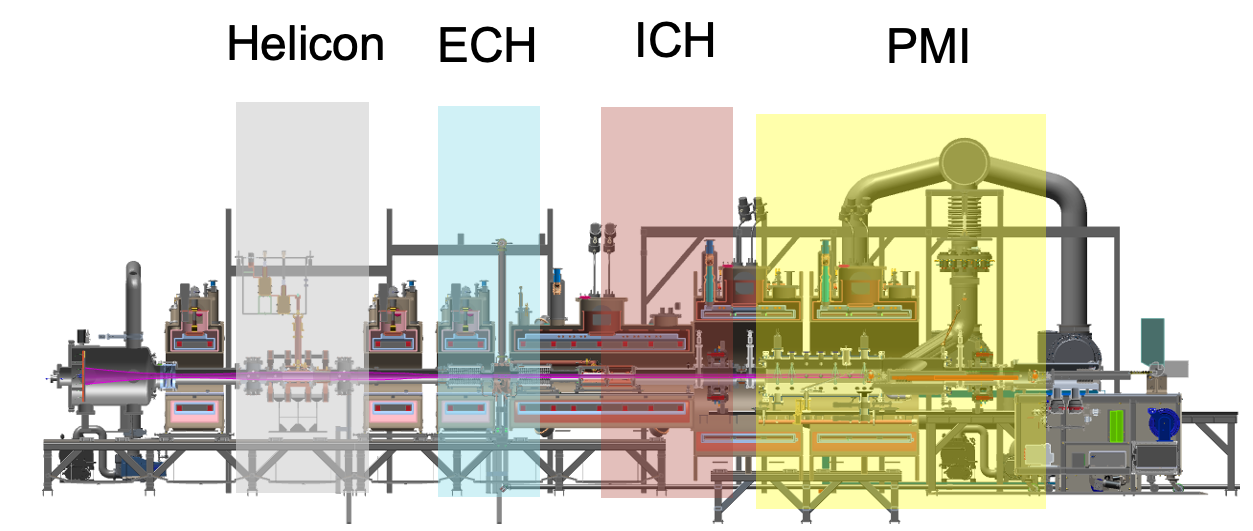}
        \caption{Geometry of the Material Plasma Exposure eXperiment (MPEX) device, highlighting the major operational regions along the axial direction. The plasma is generated and sustained through sequential heating stages: the Helicon region (gray) initiates the plasma using RF helicon waves; the ECH region (blue) applies electron cyclotron heating to increase electron temperature; the ICH region (red) uses ion cyclotron heating to further energize ions; and the PMI region (yellow) is dedicated to plasma-material interaction studies, where high-flux, high-temperature plasma is directed onto material targets for erosion, deposition, and transport analysis.}
        \label{fig:team6_code}
    \end{figure}

At the same time, achieving predictive modeling across these regimes requires not only forward modeling but also solving complex \textbf{inverse problems} to infer difficult-to-measure quantities (e.g., cross-field diffusion coefficients) from experimental or high-fidelity simulation data. This is particularly important when bridging different physical models (kinetic vs. fluid) and codes (e.g., PICOS++ \cite{Kumar:2023, Kumar:2024} vs. SOLPS\cite{Islam:2024}).

\textbf{Motivation:} The multiscale, multiphysics complexity of fusion devices like MPEX demands a robust and integrated modeling approach. Different simulation tools—such as COMSOL \cite{Beers1:2021} for RF wave propagation, SOLPS \cite{Islam:2024} for fluid plasma transport, and GITR \cite{Rapp:2024, Beers2:2021, Younkin:2021, Dhamale:2024}  for impurity dynamics—are typically developed in isolation, each tailored to specific physics regimes and dimensional constraints. This siloed structure creates inconsistencies when interpreting experimental data or designing mitigation strategies, especially for coupled phenomena like RF sheath-induced sputtering, impurity transport, and edge plasma behavior.

To address this, we explored the use of AI-assisted workflows to identify missing physics, propose model extensions, and streamline coupling strategies across simulation codes. Large language models (LLMs) like ChatGPT and Gemini were used not only to synthesize domain knowledge but also to help construct proof-of-concept workflows—including forward modeling, parameter scanning, and inverse problem-solving techniques. One such effort focused on using kinetic simulations (e.g., PICOS++) to inform radial diffusion parameters in SOLPS, with AI/ML used as a bridge between high-fidelity physics and reduced models.

Overall, the motivation behind this work is to accelerate the development of predictive, experimentally grounded, and physically consistent across scales modeling pipelines for MPEX and future fusion platforms—leveraging AI where it enhances physical reasoning, solver construction, and system integration.

\subsubsection{Methodology}

\begin{itemize}
  \item \textbf{AI Models \& Tools:}
  \begin{itemize}
    \item \textbf{ChatGPT (GPT-4o)} was used to:
    \begin{enumerate}
      \item Diagnose shortcomings in existing RF/plasma/impurity transport models.
      \item Generate conceptual and quantitative mitigation strategies (e.g., Faraday shielding, ECH screening, impurity traps).
      \item Propose multi-code modeling workflows (e.g., COMSOL + AORSA \cite{Lau:2018} + SOLPS).
      \item Draft physics-informed inverse problem frameworks involving kinetic-to-fluid translation of transport coefficients.
    \end{enumerate}
    \item \textbf{Gemini 2.0 and ChatGPT 4-mini} were used to:
    \begin{enumerate}
      \item Formulate Python-based inverse solvers.
      \item Implement synthetic tests for inferring Bohm-like diffusion coefficients from simplified 1D plasma profiles.
    \end{enumerate}
  \end{itemize}

  \item \textbf{Workflow:}
  \begin{itemize}
    \item Defined a multiphysics problem structure based on observed discrepancies in Proto-MPEX  (e.g., hollow $T_e$ profiles \cite{Kafle:2018}, target impurity deposition \cite{Beers2:2021}, sheath erosion \cite{Beers1:2021}).
    \item Used ChatGPT to analyze literature, synthesize diagnostic observations, and propose model extensions.
    \item Constructed an AI-assisted \textbf{inverse solver prototype} to derive perpendicular diffusion coefficients from 1D kinetic-like simulations (intended to mimic results from PICOS++).
    \item Broke the solver into modular Python components for easier debugging and refinement.
    \item Tested LLM-generated code using synthetic datasets to compare inferred vs. target profiles.
  \end{itemize}
  \item \textbf{Team Roles:}
  \begin{itemize}
      \item \emph{Kumar:} investigated solutions to the overall physics and operational challenges associated with the MPEX experiment, used chatGPT 4o with deep research.
      \item \emph{Laiu:} developed an AI-assisted inverse solver prototype to determine accurate diffusion coefficients for radial (cross-field) transport, used Gemini 2.0 and ChatGPT 4-mini.
      \item \emph{Choi:} explored AI-assisted pseudo turbulent flow simulator for generating synthetic turbulent data to aid the development of detector for the transition from transient to steady state.
  \end{itemize} 
\end{itemize}

\subsubsection{Key Findings}

\begin{itemize}
  \item \textbf{Results: }
  \begin{itemize}
    \item By employing ChatGPT-4o with deep research during the AI for Nuclear Workshop at ORNL, it identified RF sheath-induced sputtering from the AlN helicon window as the dominant impurity source in Proto-MPEX.  ChatGPT-4o explored relevant literatures \cite{Beers1:2021, Beers2:2021} on Proto-MPEX and correlated it with experimental deposition data to guide this interpretation.
    \item It further suggested that COMSOL cold-plasma models fail to reproduce the experimentally observed hollow $T_e$ profiles, primarily due to missing physics such as kinetic damping (e.g., Trivelpiece–Gould absorption) and RF sheath losses. ChatGPT-4o contributed by interpreting discrepancies between simulation and experiment, identifying key mechanisms omitted in the models, and referencing relevant theoretical and experimental studies.
    \item It proposed use of \textbf{AORSA} to resolve off-axis power deposition more accurately, with coupling to SOLPS for transport response.
    \item Recommended the following hardware and operational mitigations:
      \begin{enumerate}
        \item Slotted tungsten \textbf{Faraday shield} to reduce sheath-induced PMI while maintaining RF coupling.
        \item \textbf{ECH-based impurity screening} to suppress impurity flux to the target by increasing $T_e$.
        \item Magnetic geometry modifications and impurity traps near the source to intercept impurities early.
      \end{enumerate}
    \item With the help of Gemini 2.0 and ChatGPT 4-mini, developed a simplified inverse solver using LLMs to infer Bohm-like diffusion coefficients from 1D plasma profiles (density, temperature, magnetic field). The solver used gradient-free optimization and was validated on synthetic test data:

    \begin{figure}[ht]
        \centering
    \includegraphics[width=0.9\linewidth]{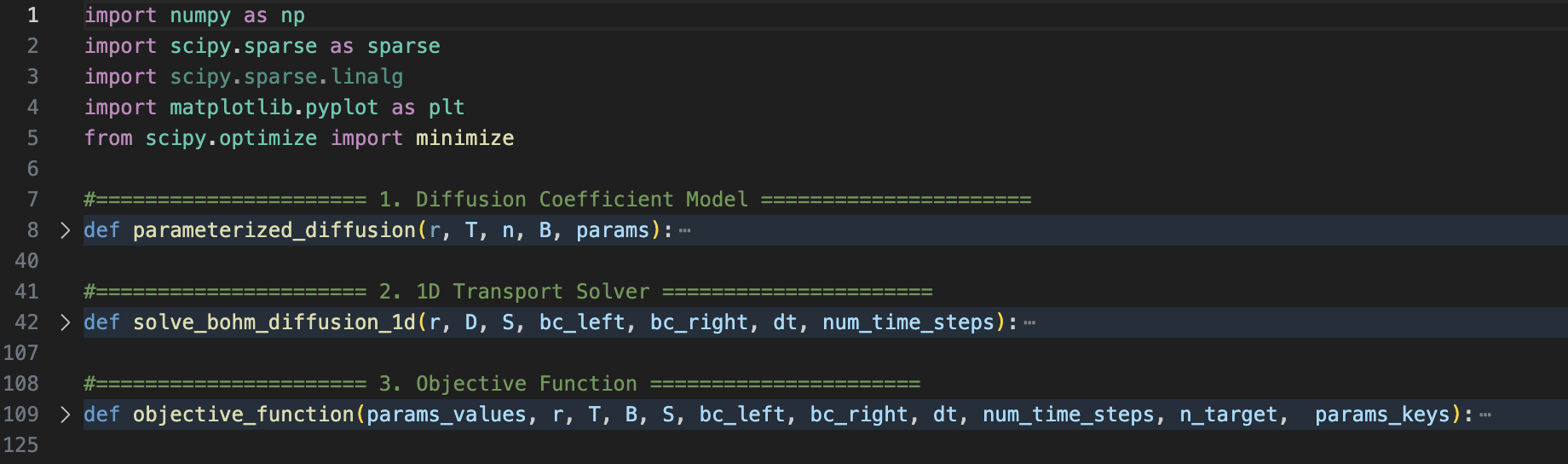}
        \caption{LLM-generated code skeleton for computing parameterized diffusion coefficients from plasma profiles.}
        \label{fig:team6_code1}
    \end{figure}

\begin{figure}[htbp]
    \centering
    \begin{minipage}[t]{0.48\textwidth}
        \centering
        \includegraphics[width=\linewidth]{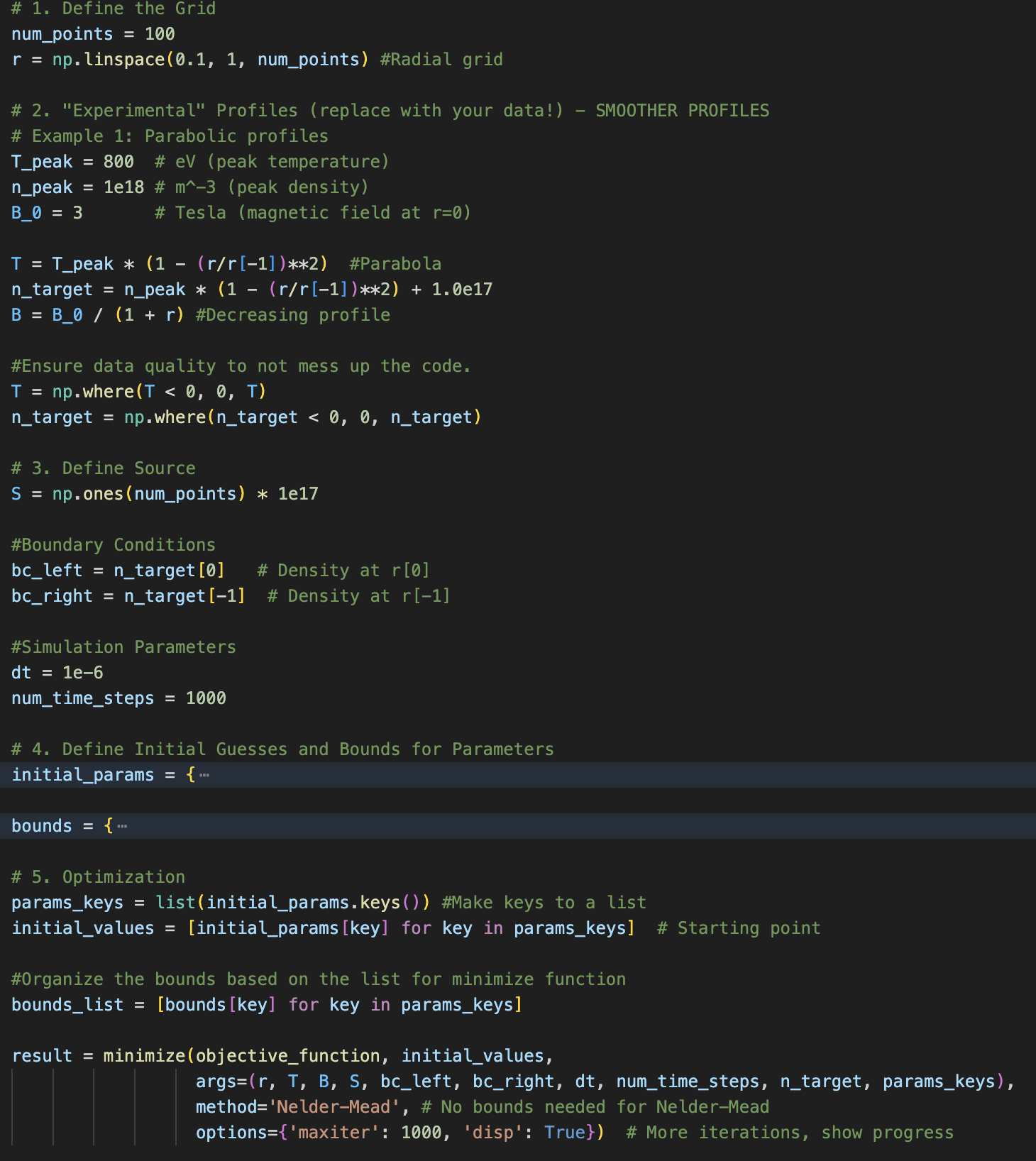}
        \caption{LLM-generated proof-of-concept script for testing the solver on synthetic plasma data.}
        \label{fig:team6_script}
    \end{minipage}%
    \hfill
    \begin{minipage}[t]{0.48\textwidth}
        \centering
        \includegraphics[width=\linewidth]{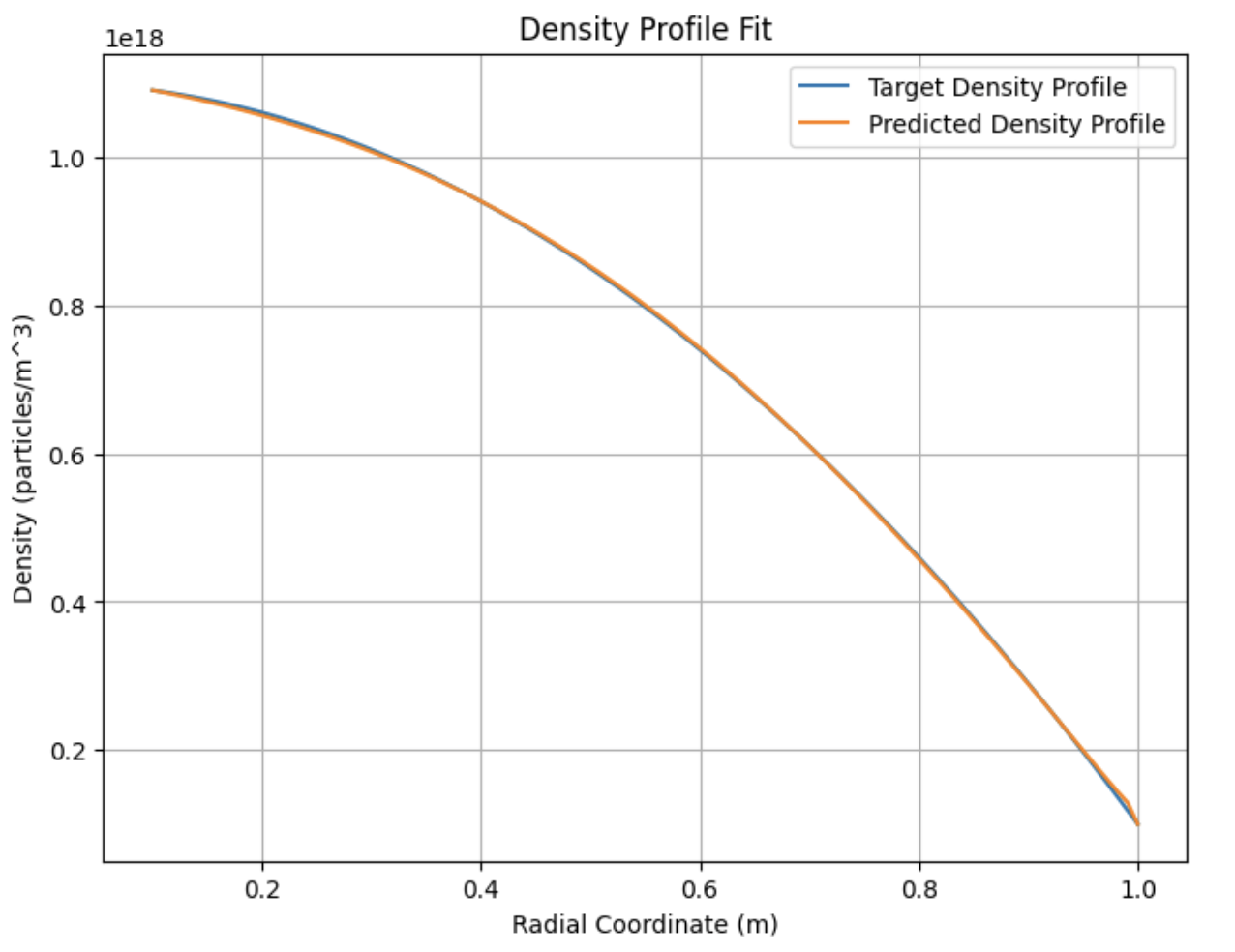}
        \caption{Comparison between target and simulated density profiles using LLM-inferred diffusion coefficients.}
        \label{fig:team6_density_profile}
    \end{minipage}
\end{figure}

    \item The toy model served as a stepping stone to a more realistic scenario: using PICOS++ to model parallel transport and using AI/ML to infer perpendicular diffusion coefficients for SOLPS. This would create a physics-informed pathway for radial transport calibration in MPEX-like regimes.
  \end{itemize}

  \item \textbf{Insights:}
  \begin{itemize}
    \item AI-assisted exploration confirmed the hypothesis that spatial RF sheath voltages (modeled via COMSOL) correlate with erosion zones on the helicon window—validating sheath-induced PMI as a dominant erosion mechanism.
    \item Comparative reasoning with PISCES-RF showed that impurity redeposition can be enhanced via magnetic or flow control, guiding MPEX design strategies.
    \item Higher azimuthal mode numbers ($m > 1$) must be included in RF simulations to explain observed off-axis power deposition and hollow heating.
    \item PICOS++ was identified as a viable kinetic tool for modeling flux-surface-resolved parallel transport. When combined with AI/ML-based inverse optimization, it provides a practical approach to infer radial diffusion coefficients in SOLPS.
    \item Faraday shielding and wall conditioning strategies emerged as essential tools to reduce source impurity loads. AI-synthesized design trade-offs between AlN, BeO, and pBN windows highlighted the importance of evaluating thermal, RF, and PMI resistance properties in tandem.
    \item In building the inverse solver, we learned that LLM-generated solutions work best when decomposed into small, modular functions, and that derivative-free optimization (though inefficient) was most reliable given the current limits of LLMs in generating adjoint-based solvers.
  \end{itemize}
\end{itemize}

\subsubsection{Limitations \& Challenges}

\begin{itemize}
  \item \textbf{Technical Constraints:} High-fidelity kinetic solvers like AORSA and PICOS++ are resource-intensive. LLMs struggled to generate robust adjoint-based solvers due to limited public code examples and gaps in memory/context.
  \item \textbf{Process Hurdles:} Cross-code parameter mapping remains a bottleneck. Code generation from LLMs is sensitive to prompt structure and benefits from step-by-step modularization. Optimization techniques used in inverse modeling were simple but not scalable.
  \item \textbf{Mitigations Tried:} Used surrogate sheath models in COMSOL. Reverted to gradient-free optimization methods for inverse solver prototyping. Introduced unit testing and modular code generation strategies to improve LLM usability.
\end{itemize}

\subsubsection{Future Directions}

\begin{itemize}
  \item \textbf{Next Steps:}
  \begin{enumerate}
    \item Integrate COMSOL, AORSA, SOLPS, and GITR in a unified framework with physics-informed feedback loops.
    \item Scale the inverse modeling approach by incorporating realistic PICOS++ outputs for full-core MPEX scenarios.
    \item Test and deploy Faraday shields and impurity traps in Proto-MPEX to evaluate operational viability.
    \item Enhance LLM usage by combining with symbolic computing and physics-informed neural networks (PINNs) for faster inversion.
  \end{enumerate}
  
  \item \textbf{Potential Extensions:}
  \begin{itemize}
    \item Develop surrogate ML models for expensive solver stages (e.g., sheath modeling, impurity redepositions).
    \item Deploy LLMs as real-time assistants during experiments for scenario adjustment and model-informed decision-making.
    \item Expand AI-generated solver frameworks to support transport coefficient inference in other edge/scrape-off layer codes (e.g., UEDGE, BOUT++).
  \end{itemize}
\end{itemize}

%\PB{please add at least 5 domain related citations: already done in throughout this section-Kumar}

%% file: team07.tex
\subsection{Surrogate Models for Monte Carlo N-Particle Transport}

\begin{tcolorbox}[colback=gray!10, colframe=gray!80, title=Team Members, fonttitle=\bfseries]
\textbf{Rick Archibald},  \href{mailto:archibaldrk@ornl.gov}{archibaldrk@ornl.gov}\\
\textbf{Cory Hauck},  \href{mailto:hauckc@ornl.gov}{hauckc@ornl.gov}\\
\textbf{Jono McConnell},  \href{mailto:mcconnelljh@ornl.gov}{mcconnelljh@ornl.gov}\\
\textbf{Anant Raj},  \href{mailto:raja@ornl.gov}{raja@ornl.gov}
\end{tcolorbox}

\subsubsection{Problem Statement}
\begin{figure}[ht]
        \centering
        \includegraphics[width=0.95\linewidth]{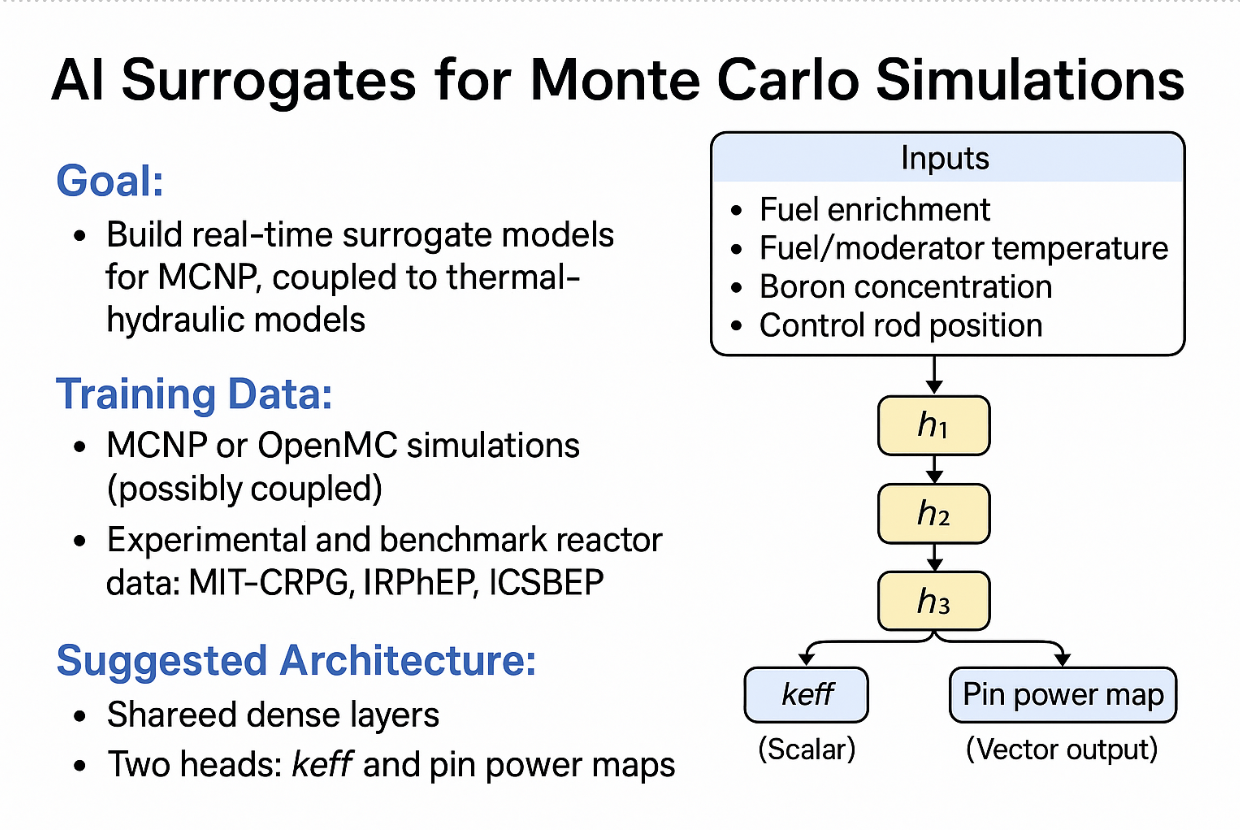}
        \caption{Image generated by ChatGTP to summarize building real-time surrogate models to replicate the accuracy of MCNP simulations (particularly for advanced or transient reactor analysis) using machine learning}
        \label{fig:MCNP_strat}
    \end{figure}

\textbf{Description:} MCNP (Monte Carlo N-Particle) simulations are used to solve complex physical problems involving the transport of neutrons, photons, electrons, and other particles through matter. These simulations aim to model and predict nuclear processes and radiation interactions with materials in a wide range of applications.  Our problem statement was building real-time surrogate models to replicate the accuracy of MCNP simulations (particularly for advanced or transient reactor analysis) using machine learning, with an emphasis on: 
\begin{itemize}
    \item Coupling to thermal-hydraulic (TH) models 
    \item Leveraging real or benchmarked reactor physics data 
    \item Training models that can approximate quantities like keff and pin-level power.
\end{itemize}
Here, keff (k-effective) is a dimensionless parameter that indicates the neutron multiplication factor in a nuclear system. The system is critical for keff = 1,  subcritical when keff < 1.0, and supercritical for keff > 1.0. Pin-level refers to the granular spatial resolution in a nuclear reactor core model, specifically at the level of individual fuel pins (fuel rods) within an assembly. Figure \ref{fig:MCNP_strat} is an image generated by ChatGTP to summarize building real-time surrogate models to replicate the accuracy of MCNP simulations (particularly for advanced or transient reactor analysis) using machine learning.

\textbf{Motivation:} Building surrogates for slow but accurate MCNP simulations enhances computational efficiency, reduces simulation times, and allows for real-time analysis in fusion and fission science. These models enable efficient exploration of design spaces through parameter studies and sensitivity analyses, facilitating optimized configurations. They support multi-physics simulations and integration with other codes, leading to comprehensive system evaluations. Surrogates also aid in validation and benchmarking against experimental data, enhancing confidence in results. Additionally, they serve as educational tools for new researchers, simplifying complex concepts. Overall, surrogates are crucial for advancing nuclear technologies by improving simulation efficiency and supporting safer reactor designs

 \textbf{Team Roles:} The team work interactively on this problem.  Our expertise was split, with have half having a machine learning background and the other half having a nuclear background.  We used each others expertise to judge quality of information being generated by AI.

\subsubsection{Methodology}
We found that questions that were descriptive of what we wanted to solve, including the constraints to problems, were the most effective in getting compelling answers from ChatGPT -03.  Using general questions provided broad general answers, missing the challenges and limitations, and making it hard to later focus on specific aspects of the problem.  In general, we found asking a long specific question provided better results over telescoping.  For example two questions that gave a lot of traction were:

\begin{enumerate}
    \item `Provide an overview of the current challenges in getting data for training models for advanced reactor systems. Focus specifically with respect to (i) neutronics, (ii) thermal hydraulics, (iii) fuel response. On the third one, look at different scales from atomistic response, mesoscale microstructure and grain response and engineering scale thermo mechanical response and fission gas release. Connect it to applications like reactor design optimization, materials design etc. Also, explore development of foundation models in these domains. How can we utilize the data and models from existing reactor systems?'
    \item `We would like to build real-time capable surrogate models for the Monte Carlo code Shift for advanced reactors and for transient simulations. Our specific question is how to gather data for training. Have any surrogate AI models been designed using Shift, even in the steady-state case and/or for standard reactors, like pressurized water reactors (PWR) and boiling water reactors (BWR)? If so, how do such surrogates couple to thermal hydraulics models? Is there experimental data available for validation?'
\end{enumerate}

Shift targets neutron and photon transport physics, with a strong emphasis on reactor core analysis, criticality safety, and high-resolution simulations for advanced nuclear systems \cite{pandya2015shift}. The MOOSE (Multiphysics Object-Oriented Simulation Environment) framework is an open-source, high-performance computational platform developed by Idaho National Laboratory (INL) \cite{icenhour2018multi}. It is designed to solve complex, coupled multiphysics problems, making it especially powerful for nuclear reactor simulations. MOOSE is used to simulate the behavior of nuclear reactor systems by coupling various physics modules (e.g., heat transfer, neutronics, fuel performance, structural mechanics). It enables the development of advanced, multiphysics simulation tools tailored to reactor design, analysis, and safety. 

For the rest of this section we will focus on the teams experience with the second question.  The proposed workflow that ChatGPT-03 offered for the second question was a detailed and reasoned approach and summarized as: 

\begin{tcolorbox}[enhanced,fit to height=5cm,
  colback=green!25!black!10!white,colframe=green!75!black,title=Summary Statement for Surrogate Modeling of Shift,
  drop fuzzy shadow,watermark color=white,watermark text=ChatGPT-03]

  \begin{itemize}

\item \code {No published Shift-based ML surrogate exists yet; you’ll likely be breaking new ground.} 

\item \code {Data gathering will hinge on a well-designed DoE (possibly multifidelity) using Shift (for neutronics) + Griffin (for low-fidelity coupling) runs.} 

\item \code {Coupling uses the same MOOSE MultiApp pattern already exploited by Griffin/Pronghorn: simply substitute your ML surrogate for Griffin in the loop.}

\item \code {Validation should leverage established experimental benchmarks (MSRE, LOFT, BEAVRS/VERA, NEA transients).} 
\end{itemize}

\code {If you’d like pointers to detailed methodologies for DoE sampling, active learning, or embedding in MOOSE, let me know!} 
\end{tcolorbox}

\subsubsection{Key Findings}
Overall, using ChatGPT -03 to develop a research plan to create a surrogate model for the ORNL based exascale code Shift \cite{pandya_implementation_2016} was well structured.  It provided useful workflow and suggestions on how to build the entire framework. It operated at a general knowledge level for the required domains needed to solve this problem.  As described in the limitations and challenges section, there are minimal to no known public databases to train a surrogate model for MCNP. ChatGPT -03 identified this limitation, however was still able to describe the process and write reasonable code that noted if a suitable data-set is generated, how to proceed with all the other steps. 
\subsubsection{Limitations \& Challenges}
Developing accurate surrogate models typically requires a substantial amount of high-quality data from full MCNP simulations. Obtaining this data can be resource-intensive and time-consuming. There are currently no sources available and data generation for these types of challenge could be a value add for the DOE community. In the nuclear field, meeting regulatory standards for safety and reliability can add additional layers of scrutiny. While the application of AI in generating surrogates for MCNP code presents exciting possibilities for improving simulation efficiency, it also brings forth significant challenges that must be addressed. Ensuring high-quality data, model accuracy, interpretability, and successful integration into existing workflows are critical for leveraging the full potential of AI in this domain. 

There was some unique datasets that were discovered, the MIT Reactor Physics Benchmark Models Repository \cite{mit-crpg}, and alternative codes suggested than the more costly models.  However, the main process hurdle was high quality curated data sets ready for AI/ML are hard to come by for nuclear reactors. results.  Large Language Models (LLMs) can significantly assist in tagging metadata by leveraging their ability to understand context, semantics, and patterns in natural language and structured data. LLMs can take previously available data—such as documents, logs, images with descriptions, simulation results, or sensor outputs—and automatically generate or enhance metadata (descriptive labels or annotations). This is valuable for improving dataset quality, discoverability, and readiness for ML tasks.

\subsubsection{Future Directions}
With the strengths of Nuclear science at ORNL and the resources available, creating a high quality surrogate of MCNP is possible. However, the main challenges remain for surrogate modeling of complex physical process.  Specifically, the development of accurate surrogate models necessitates a substantial amount of high-quality data, which can be resource-intensive to obtain. Additionally, the complexity of the physical processes involved can make it difficult for AI to accurately capture all nuances, and the resulting models may suffer from issues with generalization and overfitting. Moreover, the training of complex models may demand significant computational resources, potentially undermining efficiency gains. Integration with existing workflows can also be challenging due to compatibility issues and user adoption hurdles. Additionally, the validation and verification of these surrogate models against traditional MCNP results are crucial yet complex, particularly in ensuring regulatory compliance. Finally, quantifying uncertainties within the surrogate predictions and the need for ongoing updates as new materials and trends emerge further complicate their implementation in the dynamic fields of fusion and fission science.

%\PB{please add at least 5 domain related citations}

% === End LaTeX snippet ===

%% file: team08.tex
\subsection{Investigating Corrosion in Molten Salts}

\begin{tcolorbox}[colback=gray!10, colframe=gray!80, title=Team Members, fonttitle=\bfseries]
\textbf{Rajni Chahal},  \href{mailto:rchahal@tntech.edu}{rchahal@tntech.edu}; chahalr@ornl.gov\\
\textbf{Cihangir Celik},  \href{mailto:celikc@ornl.gov}{celikc@ornl.gov}\\
\textbf{Flavio Dal Forno Chuahy}, dalfornochuf@ornl.gov\\
\textbf{Tyrone Harris}, harristc@ornl.gov
\end{tcolorbox}

\subsubsection{\textbf{Problem description}} Due to high temperatures and radioactive environment in the nuclear reactors, the effect of foreign species, e.g. salt impurities and metallic particles from salt-alloy/metal interface, on the structure and properties of molten salts is difficult to investigate solely using experimental techniques due to difficulties in precise measurements in extreme environments as well as their underlying costs \cite{SRIDHARAN2013241}. Thus, there are gaps where modeling can help: systematically investigating structure-property relations \cite{doi:10.1021/jacsau.2c00526}, predicting solvation structures of ions/complexes that are hard to isolate \cite{D3SC06587A}, predicting solubility limits of corrosion species within reactor temperature ranges \cite{HANSON2022109914}, computing redox potentials \cite{NAM2015224} \cite{D4SC07253G} and activity coefficients for multi-component salts \cite{doi:10.1021/jp075299n}, and simulating electrode processes (nucleation, growth) that experiments cannot easily resolve. 

\textbf{Motivation:} Molten salts are relevant for both fusion (e.g., molten salt blankets) and fission (e.g., coolant and fuel salts) nuclear reactors. In such applications, due to high temperatures and radioactive environment, corrosion in molten salts is accelerated, which in turn can affect the structure and properties of the molten salts. This problem is critical to address for the successful reactor operations and their licensing \cite{osti_4237975}. 

\subsubsection{Methodology}
\begin{itemize}
  \item \textbf{AI Models \& Tools:} OpenAI’s Deep Research was deployed as the reasoning model to investigate: 1) what has been done before to study the problem of corrosion in molten salts, 2) what modeling and simulations efforts can be helpful, 3) what existing and new data will be needed for model development, and 4) what experimental data can be used for validation of models.
Initial prompts for this model were generated using OpenAI’s GPT4.5. 
  \item \textbf{Workflow:} Problem breakdown into small segments, e.g., existing literature on simulations, existing literature on experiments, the gaps in simulations and experiments to further address the problem of corrosion in molten salts
  \begin{itemize}
      \item generate prompts for each problem subsection in OpenAI’s GPT4.5,
      \item feed the prompts to Deep Research,
      \item thorough review of each response by domain expert in the team,
      \item cross-checking the statements made with the reported references, 
      \item checking the validity and trustworthiness of reference sources, 
      \item obtain short summary of each response based on verified content,
      \item LaTeX Integration and Formatting.
  \end{itemize}
  \item \textbf{Team:} Rajni Chahal (expertise in molten salts' modeling and simulations, MLIPs), Flavio Dal Forno Chuahy (expertise in computational fluid dynamics ), Tyrone Harris (expertise in radiation detection), Cihangir Celik ( expertise in modeling and simulation for nuclear criticality safety and radiation shielding)
\end{itemize}

\subsubsection{Key Findings}
\begin{itemize}
  \item \textbf{Results:} We obtained a brief review of existing literature reporting previous efforts done to study the structure of molten salts with corrosion solute species using classical molecular dynamics (MD), \textit{ab initio }MD (AIMD), and MD simulations driven by machine learning interatomic potentials (MLIPs). The limitations of each method was determined correctly but state-of-the-art computational studies using each method were not reported.
Open gaps and opportunities were identified correctly, which include (1) developing accurate MLIPs potentials for actinide/transition metals, (2) bridging atomistic MD with continuum models, (3) understanding cluster formation (e.g. metallic nanoparticle solubility) in salt. In short, while MD (classical or AIMD) has detailed local structure and transport, it often struggles to fully capture redox chemistry and rare events; improved potentials and multiscale methods are needed.
A slightly better review of experimental works was reported listing studies on spectroscopy and scattering, redox potentials, solvation structures of the species, and measured transport properties of molten salts. In this case as well, several important studies were unlisted.
No novel hypothesis was reported by Deep Research model.
  \item \textbf{Insights:} No unexpected observations or new domain-specific learnings were reported in Deep Research model’s responses.
\end{itemize}

\subsubsection{Limitations \& Challenges}
\begin{itemize}
  \item \textbf{Technical Constraints:} The model was not able to report the state-of-the-art computational studies in this area. Also, we traced the references and found that the sources were not necessarily reliable. This could be due to the fact that reliable or journal articles in this area are not open-access. The fact that some statements in model’s response were pulled out of the conference abstracts (generally available openly and free to cost) also supported our hypothesis that the model may not be trained on \textit{closed-access }journal articles. 
A better performance of model was observed in the case of reviewing experimental literature for corrosion in molten salts. This could also be due to the problem of data imbalance. For example, less number of computational studies exist in this area compared to the number of experimental studies.
  \item \textbf{Process Hurdles:} A comparison of Deep Research’s responses with other available models was not performed. However, response time of the Deep Research was considerably slower than a regular inquiry response time which might be a limiting factor for context-based follow up inquires that researchers tend to come up frequently. 
  \item \textbf{Mitigations Tried:} A workaround for using Deep Research effectively would be preparing tailored prompts using regular AI interfaces to ask Deep Research.
\end{itemize}

\subsubsection{Future Directions}
\begin{itemize}
  \item \textbf{Next Steps:} 
1) New AIMD data needs to be generated to study the corrosion solute species in molten salts for the cases where experimental data is available for model’s validation. OLCF computing facility can be utilized for this purpose.\\
2) Machine learning interatomic potentials should be developed on AIMD data using uncertainty quantification approach for efficiently sampling the potential energy surface in order to obtain a trustworthy and robust MLIP model. OLCF computing facility can be utilized for their development.\\
3) Using MLIP models, the quantities of interest, such as solubility limits, redox potentials, activity coefficients of species in molten salts need to be calculated for the temperature range for reactor operations. The effect of solute species and their subsequent aggregation on molten salt’s transport properties needs to be evaluated.\\
4) Existing experimental data \cite{osti_4658668} \cite{1961RuCRv..30...90U} can be used to validate MLIP model’s findings. New experiments need to be proposed through collaboration for further validation of the simulations’ predictions.\\
5) Investments need to be made in both AIMD datasets (for model’s training) and experimental datasets (for model validation).
  \item \textbf{Potential Extensions:} 
Multiscale modeling approaches on “how to link MD with higher length-scale models or continuum models?” can be investigated to gain a better understanding of corrosion of molten salts when interfaced with alloy/metallic structural components in nuclear reactors. This could include feeding-forward the MD-determined diffusion coefficients and other transport properties to higher-scale models allowing investigation of effect of corrosion in molten salts during reactor's life cycle.

\end{itemize}

% === End LaTeX snippet ===

%% file: team10.tex
\subsection{Predictive Modeling of Nuclear Material Degradation}

\begin{tcolorbox}[colback=gray!10, colframe=gray!80, title=Team Members, fonttitle=\bfseries]
\textbf{Samantha Sabatino},  \href{mailto:sabatinosm@ornl.gov}{sabatinosm@ornl.gov}\\
\textbf{Pradeep Ramuhalli},  \href{mailto:ramuhallip@ornl.gov}{ramuhallip@ornl.gov}\\
\textbf{Tom Beck},  \href{mailto:becktl@ornl.gov}{becktl@ornl.gov}
\end{tcolorbox}

\subsubsection{Problem Statement}
\textbf{Description:} Characterizing degradation in nuclear structural materials, including  reinforced concrete and irradiated metals, poses a complex inverse problem. Internal damage (e.g., microcracking from alkali–silica reaction (ASR) or cracking in weldments) is difficult to detect directly during operation, requiring nondestructive examination and inference from sparse measurement data \cite{Ai2021}. Two variations on the science problem were studied: (1) the problem of estimating the modulus loss and crack density in nuclear concrete at 60 years using acoustic and strain measurements; (2) reconstruction of the shape of internal discontinuities (such as internal voids or cracks) from a limited set of acoustic measurements \cite{Kocur2016}.  

For the first case, the expected input data included synthetic acoustic waveform features and strain measurements at mid-life (30 years), with targets being modulus loss and crack density at 60 years. For the second case, expected input data was from a linear acoustic scan of the material under test.

In each case, the objective was to use the AI model to accelerate all aspects of solving the problem, from finding or synthesizing data to generating code that may be used to solve the inverse problem. Overall, the inverse problem is often ill-posed and computationally intensive: small changes in observations can correspond to large differences in internal state, and simulations needed to predict material behavior are demanding \cite{Field2015}. \\

\textbf{Motivation:} This challenge is critical in the context of fission and future fusion reactors, where structural integrity over 80+ years must be ensured under extreme conditions. Traditional inspections often miss early-stage degradation, increasing the need for AI-driven methods that can infer internal damage states from indirect acoustic or strain measurements. Reliable inverse modeling enables better aging management and safety assurance in nuclear infrastructure \cite{Sandhu2023}.

\subsubsection{Methodology}
\begin{itemize}
  \item \textbf{AI Models \& Tools:} We employed OpenAI's ChatGPT (o4-mini or its variants) to identify methods that have been proposed in the literature, with a focus on machine-learning methods. The LLM was further utilized to determine a reasonable workflow for this problem. The scope of the machine learning methods identified for use within the workflow included linear regression, random forests, gradient-boosted trees, Gaussian process regression, and neural networks (MLP and CNNs). 

  \item \textbf{Workflow:}
  \begin{itemize}
      \item  \textbf{ChatGPT}
      \begin{enumerate}
  \item Problem scoping: discussion with ChatGPT to frame the inverse problem and explore solution strategies.
  \item Code generation: ChatGPT produced \texttt{Python} and \texttt{Mathematica} code to simulate, optimize, and visualize data.
  \item Debugging: errors encountered in execution were discussed with ChatGPT, which provided corrections and suggestions.
  \item Documentation: ChatGPT drafted explanatory content, summaries, LaTeX output, and figure descriptions.
\end{enumerate}

\item  \textbf{Inverse Problem}\\
An overview of the following inverse modeling architecture is shown in Figure~\ref{fig:team10_inverse_pipeline}.
      \begin{enumerate}
    \item Synthetic data generation: generate data with probabilistic variations in material, environmental, and geometric parameters \cite{Aliyeva2022}. 
    \item Preprocessing: waveform normalization, noise injection, feature extraction.
    \item Hypothesis generation: selecting strain and acoustic signal features as proxies for damage.
    \item Code development: modular \texttt{Mathematica} or \texttt{Python} framework for training and evaluating AI models.
    \item Evaluation: RMSE, MAE, $R^2$, and depth-dependent prediction plots.
\end{enumerate}
\end{itemize}

  \begin{figure}[h!tbp]
    \centering
    \includegraphics[width=0.85\textwidth]{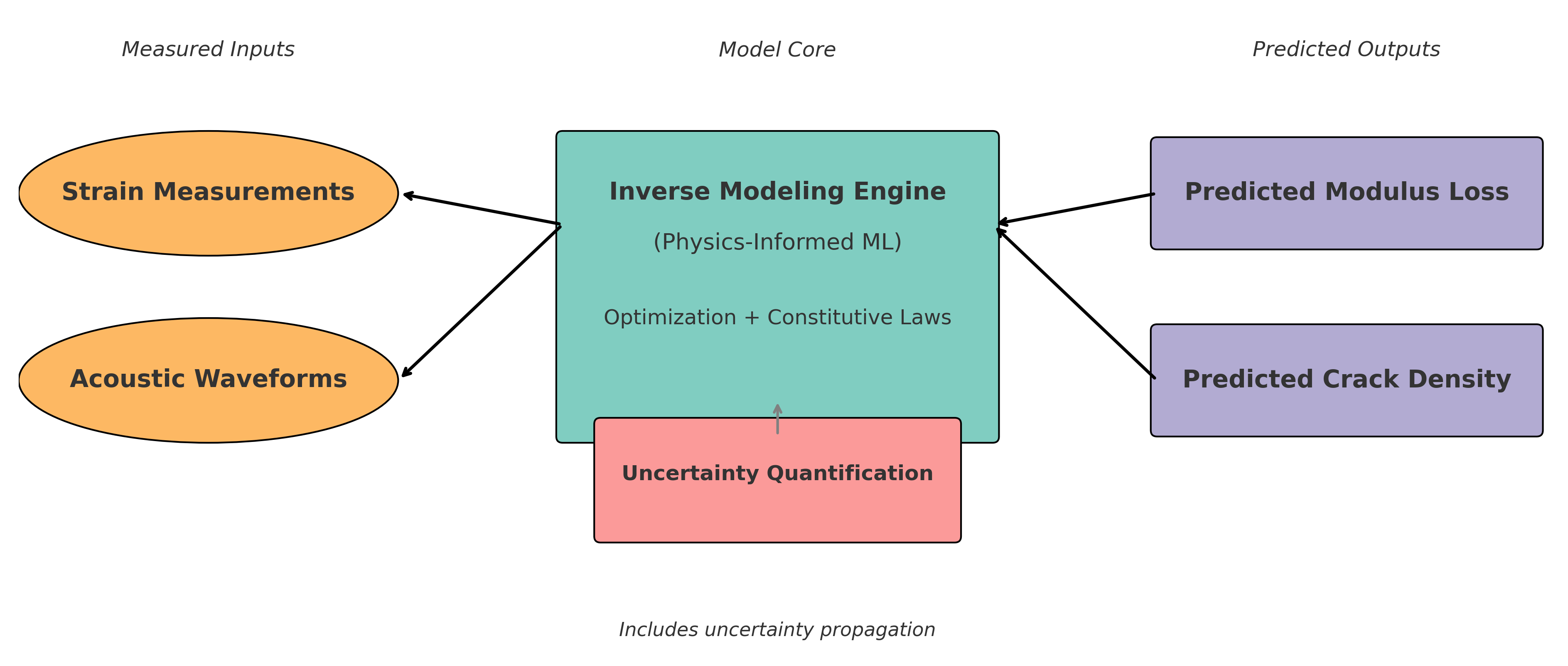}
    \caption{Conceptual diagram of a physics-informed inverse modeling pipeline that maps measured strain and acoustic data to predicted material degradation with uncertainty quantification.}
    \label{fig:team10_inverse_pipeline}
  \end{figure}

  \item \textbf{Team Roles:} Samantha Sabatino provided workflow integration, interfaced with ChatGPT to develop appropriate code to generate synthetic ASR damage data, and debugged \texttt{Mathematica} code that facilitated machine learning. Pradeep Ramuhalli discovered and tested the limitations of ChatGPT, developed and debugged \texttt{Python} code, and interfaced with ChatGPT to provide summary text about the domain problem. Tom Beck provided strategic project oversight, as well as data and workflow management.
\end{itemize}

\subsubsection{Key Findings}
Key findings include both findings relative to the use of ChatGPT as well as from the technical problem selected for this workshop (inverse analysis).
\\

\textbf{ChatGPT}
    \begin{itemize}
    \item \textbf{Results:} ChatGPT provided a reasonable workflow for the problem of interest and was able to provide compilation of information relevant to the problem. Code was generated (\texttt{Python} as well as \texttt{Mathematica}) for solving several components of the workflows. The \texttt{Mathematica} code generated was able to be successfully executed (see \textbf{Inverse Problem: Results} below for a summary of the results from this activity. The \texttt{Python} routines generated focused on the data generation aspect for the generation of acoustic scan data, and were limited in their ability to execute successfully (missing functions, apparent code generation for incorrect domain areas, incorrect results of code execution). 
    \item \textbf{Insights:} ChatGPT appears to be very well suited for information compilation, though the provided information did not seem specific enough but was sufficient for a beginner to understand the problem space and initiate research. While its code generation capabilities seem to be good, greater specificity in queries appears to be needed. It should be noted that the accuracy of data or domain problem results generated using ChatGPT generated code was not evaluated, nor was the accuracy of the provided workflows compared against other, known, solutions that have been proven to work. As a result, at this stage in the evaluation, ChatGPT appears to be a good way to initiate and support research where the researcher needs to conduct due-diligence. Note that other LLM models were initially considered but were not used in this work for a variety of reasons (lack of access, initial inaccurate information generated).
    \end{itemize}

\textbf{Inverse Problem}
  \begin{itemize}
  \item \textbf{Results:} All models achieved strong predictive performance ($R^2 \geq 0.90$) for both modulus loss and crack density. Gradient-boosted trees and linear regression consistently performed best on this dataset. To better understand how predictive accuracy varies with sensor depth, we evaluated RMSE and R² trends across the concrete biological shield (CBS) thickness for five machine learning models trained on synthetic data generated in \texttt{Mathematica}. As shown in Figure~\ref{fig:team10_model_performance}, prediction error (RMSE) increases with depth, while model fit (R²) decreases, reflecting higher uncertainty and signal degradation at deeper regions. These results reinforce the importance of data quality and spatial distribution when designing inverse models for degradation prediction. CNN-based classifiers on waveform data outperformed feature-engineered approaches in ASR detection scenarios \cite{Soltangharaei2020}. The final model framework supports inverse prediction of degradation state using sparse mid-life sensor inputs.
  
  \begin{figure}[h!tbp]
    \centering
    \includegraphics[width=0.95\textwidth]{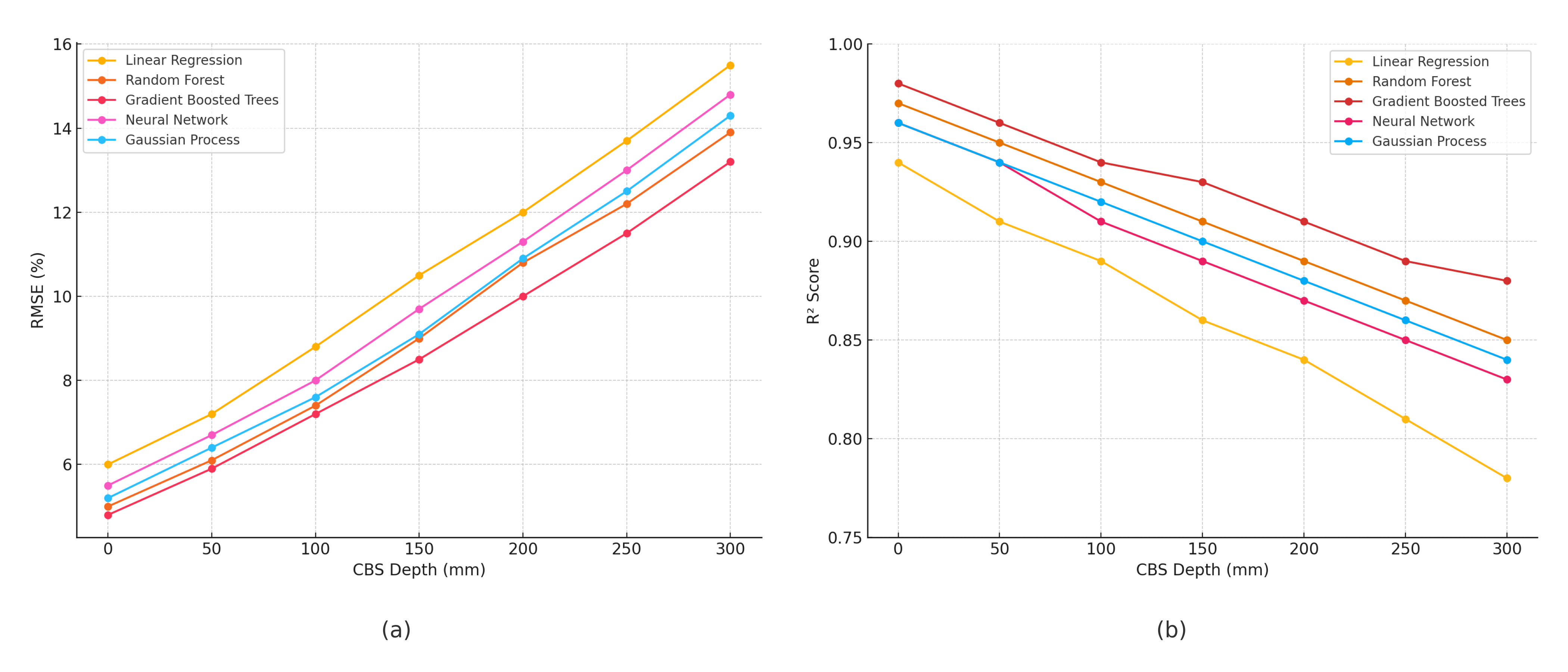}
  \caption{Model performance as a function of CBS depth, showing (a)  prediction error (RMSE) and (b) model fit (R²) across five AI models using synthetic data.}
    \label{fig:team10_model_performance}
  \end{figure}

  \item \textbf{Insights:} Waveform-based features (e.g., normalized velocity) were more predictive than summary statistics alone. Even simple models (linear, RF) captured degradation trends well due to strong signal in features. Combining acoustic and strain data improved accuracy. However, model generalization to real reactor conditions requires further validation with field or high-fidelity experimental data \cite{Zhang2023}.

\end{itemize}

\subsubsection{Limitations \& Challenges}
Again, we include challenges for both the LLMs as well as the domain problem. \\

\textbf{ChatGPT}
    \begin{itemize}
    \item \textbf{Technical Constraints:} 
    \begin{itemize}
  \item Limited memory window resulted in context loss.
  \item No internal code execution; all results had to be externally validated.
  \item Domain-specific assumptions sometimes incorrect.
  \item Limited access to real-world labeled data to provide additional context for ChatGPT. 
\end{itemize}
    \item \textbf{Process Hurdles:} 
    \begin{itemize}
  \item No built-in version control; iterations had to be tracked manually.
  \item Prompt history had to be actively managed.
  \item Multi-turn prompts required careful planning.
\end{itemize}
    \item \textbf{Mitigations Tried:}
    \begin{itemize}
  \item Code was modularized to reduce context load.
  \item Periodic summaries and external notes maintained continuity.
  \item Prompting strategies were refined iteratively.
\end{itemize}
    \item \textbf{Mathematical/Computational Challenges:} Occasional delays in responses from ChatGPT (upwards of an hour or two) were encountered though it is unclear if this is related to technical issues with the servers or computational challenges with the model and the size of the queries/generated results.
\end{itemize}

\textbf{Inverse Problem}
\begin{itemize}
  \item \textbf{Technical Constraints:} There was limited access to real-world labeled data and a heavy reliance on synthetic scenarios. Memory and compute constraints in cloud-based environments limited model complexity and dataset size ($\sim$2,000 samples).
  \item \textbf{Process Hurdles:} Tooling gaps between \texttt{Mathematica} and \texttt{Python}-based ML frameworks slowed integration. Acoustic signal variability due to environmental noise required careful feature design \cite{Carrasco2021}.
  \item \textbf{Mitigations Tried:} We reduced the sample size for tractable training and trained one model at a time using modular loops. Furthermore, simplified feature sets and targeted visualization enabled efficient debugging and validation.
  \item \textbf{Mathematical/computational challenges:} Basic research is needed into how best to incorporate physical constraints in solving inverse problems that utilize limited data to update and improve computational models of stress-strain behavior, fracture, etc \cite{LePape2015}. These improved models can in turn be used for prediction of critical failures in the materials within uncertainty bounds. 
\end{itemize}

\subsubsection{Future Directions}
From the standpoint of the domain problem, next steps and potential extensions are described below. The capabilities and potential limitations of the LLM (ChatGPT) (as encountered during the workshop) are also enabling a review of other potential extensions.
\begin{itemize}
  \item \textbf{Next Steps:} Ideally, we would validate the current models on real ASR-affected structures and irradiated metal data from surveillance programs \cite{Zhuang2025}. Additionally, we would expand the waveform library using laboratory experiments or physics-based signal simulation.
  \item \textbf{Potential Extensions:} \begin{itemize}
   \item Integrate physics-informed neural networks (PINNs) to constrain model predictions. 
   \item Extend to 3D geometry and time-series forecasting for progressive damage.
   \item Apply to fusion materials (e.g., Tungsten, Reduced Activation Ferritic/Martensitic steels) under extreme neutron and thermal conditions.
   \item Incorporate uncertainty quantification for risk-informed decision support.
  \end{itemize}
\end{itemize}

%\PB{please add at least 5 domain related citations}

% === End LaTeX snippet ===

%% file: team11.tex
\subsection{AI-directed Experimental Program for Sodium Fast Reactors and Molten Salt Reactors}

\begin{tcolorbox}[colback=gray!10, colframe=gray!80, title=Team Members, fonttitle=\bfseries]
\textbf{Marie Romedenne},  \href{mailto:romedennem@ornl.gov}{romedennem@ornl.gov}\\
\textbf{Gregory Davidson},  \href{mailto:davidsongg@ornl.gov}{davidsongg@ornl.gov}\\
\textbf{Nathan D. See},  \href{mailto:seend@ornl.gov}{seend@ornl.gov}
\end{tcolorbox}

%choices to fill in gaps in knowledge of Molten Salt and Sodium Fast reactors behavior guided by an AI-generated simulator surrogates}
Our team considered the development of an AI framework that would integrate simulation, uncertainty quantification and experimental design. The framework consists of a physics-based simulator for coolant materials, physical and thermodynamic properties, surrogate and physics-informed neural network (PINN) data pipeline, OpenMC loop modeling, AI-directed experiment and gap detection, high-fidelity corrosion simulation and 40-year lifetime transient extension.

\subsubsection{Problem Statement}
Advanced nuclear reactors, including molten salt and liquid metal-cooled systems, present distinct corrosion challenges owing to their high operational temperatures and the chemically aggressive nature of their coolants (i,e., salts and liquid metals) \cite{PILLAI2021152755,ROMEDENNE2022153919,RomedenneNRC2021}. Current worldwide modeling tools and data integration methodologies are limited in their experimental coverage and responsiveness \cite{PillaiNRC2024}. This research plan aims to address these limitations by integrating surrogate models with physics-informed neural networks (PINNS) that account for uncertainties and incorporate an active learning loop. Unlike traditional frameworks  that rely on one-way validation and model calibration before development, this approach is designed to be more adaptive and data-efficient, enhancing the understanding and management of corrosion phenomena in advanced nuclear reactor environments.  

\textbf{Description:} Our investigation focused on creating a design for an experimental program aimed at generating validation data for a physics-informed neural network (PINN)-based surrogate model, which is trained using a coupled multiphysics simulation capability calibrated through a targeted experimental campaign. The development of this multiphysics simulation capability was guided by an AI algorithm. Additionally, the experimental program was crafted by employing AI to assess and identify existing knowledge gaps in the corrosion data related to molten salts  and liquid metals within the current accessible scientific literature.

\begin{figure}
    \centering
    \includegraphics[width=1\linewidth]{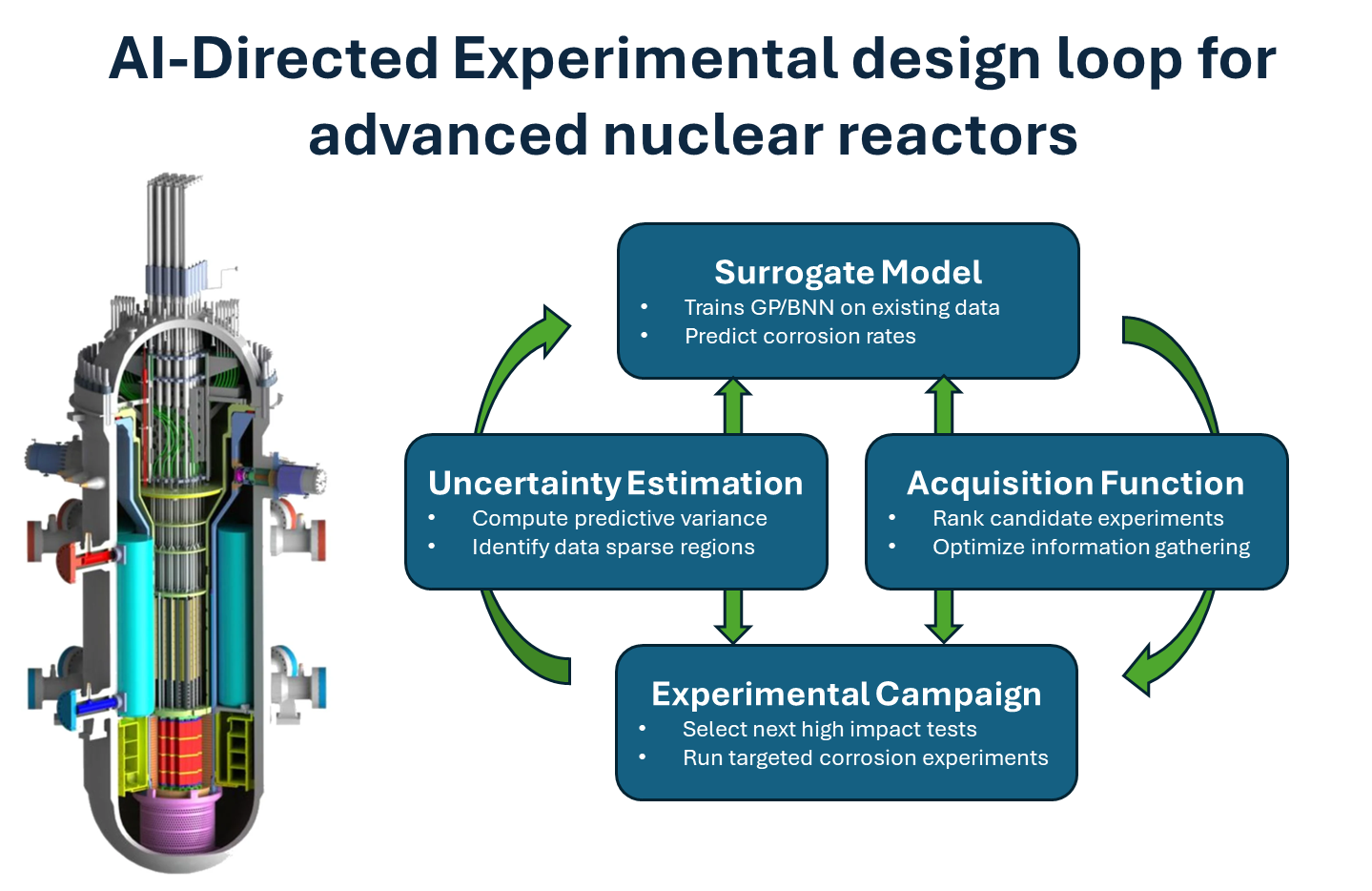}
    \caption{Physics-informed neural network (PINN)-based surrogate model trained using a coupled multiphysics simulation capability calibrated through a targeted experimental campaign}
    \label{fig:enter-label}
\end{figure}

\textbf{Motivation:} Advanced nuclear fission reactors, including molten salt reactors (MSRs) and sodium fast reactors (SFRs), lack the extensive operational data that traditional light water reactors possess, which limits validation opportunities for newm high-fidelity simulation frameworks. To develop a robust simulation framwork that supports the design, licensing, and operation of advanced MSRs and SFRs, it is essential to identify the gaps in existing experimental data. A comprehensive research program must be established to address these gaps and reduce uncertainties in the simulaton models. The present work focused on experimental data related to corrosion and compatibility issues. 

\subsubsection{Methodology}
\begin{itemize}
  \item \textbf{AI Models \& Tools:} Team 11 dedicated its efforts to utilizing various models of ChatGPT, with a preference for ChatGPT 41o due to its advanced reasoning capabilities. In instances where code snippets were required, ChatGPT 4o-mini-high was employed. Importantly, no data was uploaded to ChatGPT other than the prompt queries used throughout the investigation.ChatGPT was used to enhance the writing of this section as well.
  
  \item \textbf{Workflow:} Our workflow involved leveraging ChatGPT to develop an experimental program through a sequence of prompts. Initially, we requested ChatGPT to outline the steps necessary for designing an experimental program to support an AI-generated MSR surrogate simulator. In response, ChatGPT provided a five-step plan that included the creation of a surrogate model using physics-based methods, followed by uncertainty quantification, which would inform the experimental design. 
Next, we prompted ChatGPT to propose an initial coupled multiphysics simulation framework for generating synthetic data to inform the AI-driven surrogate model. ChatGPT provided a plausible Python-based multiphysics simulator that utilized pre-existing solvers for the relevant physics. It accurately identified key components, including neutron transport, thermohydraulics, and corrosion chemistry, and furhter promts led to the inclusion of nonlinear feedback mechanisms among these components using a Gauss-Seidel iteration scheme.
To explore neutron transport, we specifically instructed ChatGPT to use OpenMC as an example and requested a Python script for generating OpenMC inputs tailored for an MSR model. ChatGPT successfully responded with a plausible initial script for this purpose.
Subsequently, we inquired about chemistry codes for simulating corrosion in an MSR. ChatGPT provided a comprehensive table listing various codes along with their licensing information (commercial, open-source, etc.) key features, and a discussion of the pros and cons associated with each software package.
Upon indicating our preference for Thermochimica as the coolant chemistry simulation code, we asked ChatGPT to integrate it to the previously proposed multiphysics framework. This prompted the generation of additional Python code to facilitate the integration.
Finally, we requested a Python code from ChatGPT for traning a PINN using the established multiphysics framework, which resulted in futher code inputs.
After completing the above steps, we expanded our analysis to include sodium-cooled fast reactors (SFRs) to assess ChatGPT's adaptability to novel problems. It successfully identified the relevant corrosion mechanisms for SFRs and how they differ from those in MSRs.

In conclusion, ChatGPT effectively summarized the entire process into a cohesive research program. With additional prompts, it identified gaps in the current corrosion knowledge base, proposed experiments to address these gaps, and located existing experimental facilities capable of conducting the necessary research.

\item \textbf{Team Roles:} 
     Gregory acted as query architect and development subject matter expert (SME) while Marie and Nathan acted as scientific SMEs to review and suggest new pathways.
\end{itemize}

\subsubsection{Key Findings}
\begin{itemize}
  \item \textbf{Results:} 

  ChatGPT was able to propose plausible experiments for filling current experimental gaps in both SFR and MSR corrosion chemistry knowledge.  In Figure~\ref{fig:t11-sfr} we show a screenshow ChatGPT produced for an experimental program to fill knowledge gaps in SFR corrosion.
  \begin{figure}[htbp]
    \centering
    \includegraphics[width=\linewidth]{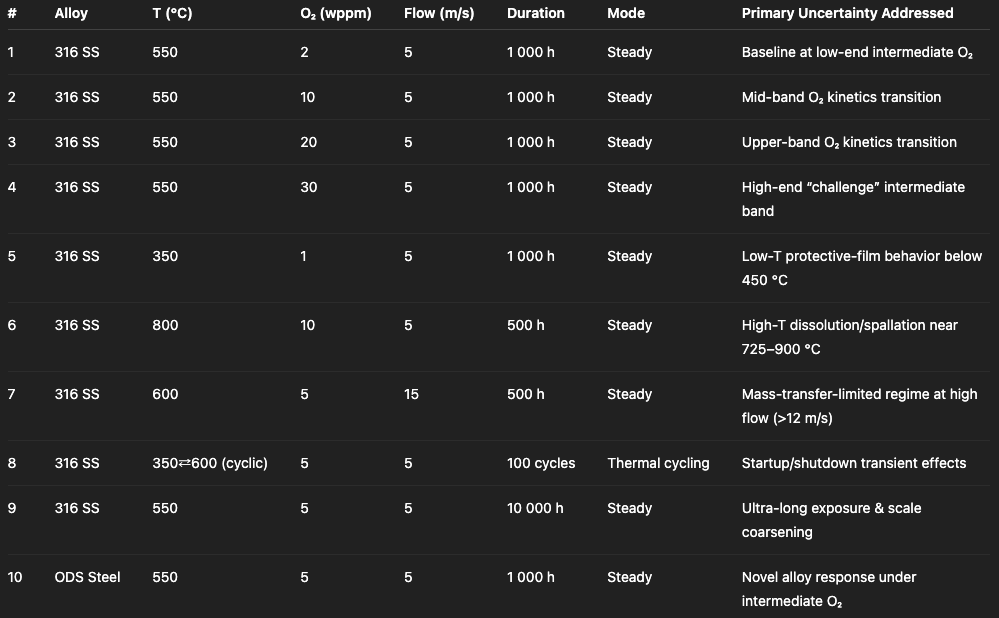}
    \caption{Screenshot from ChatGPT showing a proposed SFR experimental program.}
    \label{fig:t11-sfr}
  \end{figure}

  ChatGPT was also able to provide a proposed budget for the entire research program. This is shown in Figure~\ref{fig:t11-budget}.
  \begin{figure}[htbp]
    \centering
    \includegraphics[width=\linewidth]{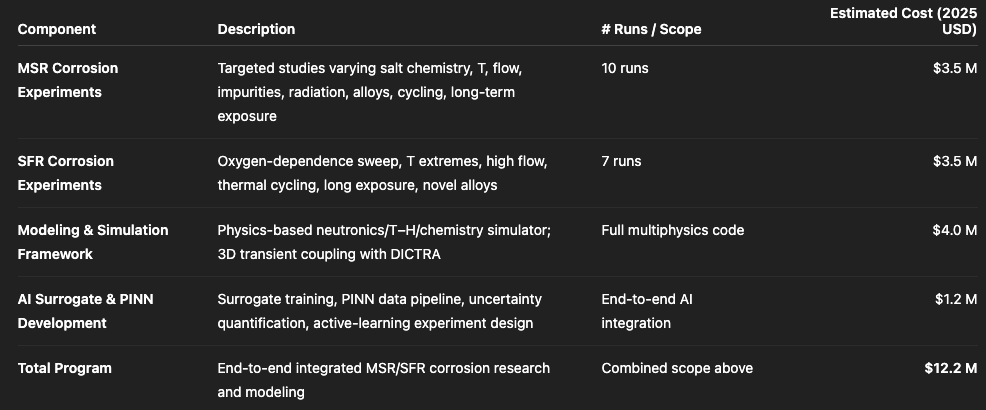}
    \caption{Total research program budget as proposed by ChatGPT.}
    \label{fig:t11-budget}
  \end{figure}
  
  \item \textbf{Insights:} ChatGPT was good at summarizing large amounts of data. It was able to comb through the available literature and find what appeared to be plausible real-world gaps in current experimental knowledge of corrosion chemistry in both SFRs and MSRs. Additionally, it was able to provide a reasonable path-to-solution for a multiphysics modeling capability used to train a PINN-based surrogate model supplemented with experimental data.
\end{itemize}

\subsubsection{Limitations \& Challenges}
Overall, we encountered few technical limitations when using ChatGPT, though we intentionally focused our inquiries on tools and domains with abundant publicly available information online. For example, when exploring neutron transport codes, we specifically asked about OpenMC, an open-source code. Had we instead asked about MCNP or Shift—both of which are export-controlled—the quality and completeness of the responses would likely have been lower due to limited accessible data. We deliberately avoided asking about export-controlled or otherwise sensitive topics.

ChatGPT's access to scientific literature is also inherently limited. It is trained on a combination of licensed and publicly available sources, including open-access journals, preprints (e.g., arXiv), public datasets, codebases, technical documentation, textbooks, white papers, and blogs. However, it does not have access to subscription-based content from publishers such as Elsevier, Springer, or IEEE, unless that content is freely available through open access.

In general, we found that ChatGPT provided helpful high-level frameworks in response to our queries. These outputs could serve as a starting point or outline for a research proposal, with subject matter experts contributing the necessary technical depth. In our examination of surrogate model design, for instance, ChatGPT consistently remained at a high level of abstraction, avoiding detailed technical complications. While it is a useful tool for structuring ideas and workflows, it is unlikely to replace domain experts in developing technically rigorous solutions.

\subsubsection{Future Directions}
\begin{itemize}
  \item \textbf{Next Steps:} 
  It is likely that with further queries more details could be filled-in in all of the areas about which Team 11 asked ChatGPT for output.  The team did not try to coax fine technical detail out of the LLM algorithm, but kept our inquiries high-level and abstract, since that seemed to be the optimal space in which ChatGPT worked. ChatGPT seemed less competent when asked to provide detailed modeling and simulation output, for instance.  It is unknown how much technical expertise and time would be required to engineer the ``correct'' prompts into an LLM to get the desired responses. Also unknown is whether prompt engineering is a better use of subject matter experts than traditional engineering.
  
  \item \textbf{Potential Extensions:} 
  The most obvious potential extensions would be to extend our inquiries into other reactor types (such as small modular reactors), other modeling and simulation toolsets (SCALE, VERA, MOOSE, etc...), and other gaps in experimental knowledge.  This could help both increase knowledge on where further experimental efforts should be directed, as well as explore the limits of LLMs to assist researchers. 
\end{itemize}

%\PB{please add at least 5 domain related citations}

% === End LaTeX snippet ===

%% file: team12.tex
\subsection{Leveraging AI/ML for Atomic Structure Modeling and S/TEM Image Simulation}

\begin{tcolorbox}[colback=gray!10, colframe=gray!80, title=Team Members, fonttitle=\bfseries]
\textbf{Yan-Ru Lin},  \href{mailto:liny@ornl.gov}{liny@ornl.gov}\\
\textbf{Arindam Chowdhury},  \href{mailto:chowdhurya1@ornl.gov}{chowdhurya1@ornl.gov}\\
\textbf{Calvin Lear},  \href{mailto:learcr@ornl.gov}{learcr@ornl.gov}\\
\textbf{Soyoung Kang},  \href{mailto:kangs1@ornl.gov}{kangs1@ornl.gov}
\end{tcolorbox}

Our team had four members: three from the Radiation Effects and Microstructural Analysis Group, MSTD (Yan-Ru Lin, Soyoung Lang, and Calvin Lear), and one from the Computational Coupled Physics Group, CSED (Arindam Chowdhury). During the workshop, our discussions focused on applying AI to materials characterization data and image analysis—for example, Electron Backscatter Diffraction (EBSD) data from Scanning Electron Microscopy (SEM) and various imaging techniques from Transmission Electron Microscopy (TEM)—for nuclear materials.

\subsubsection{Problem Statement}
\textbf{Description:} We formulated the problem of S/TEM image analysis and atomic-scale 3D reconstruction for identifying irradiation-induced defects—a grand challenge beyond the reach of traditional 2D projected S/TEM images. Despite the complexity, advancing S/TEM simulation, atomic modeling, and AI/ML integration is essential for bridging experiments and simulations in the study of complex atomic structures, with broad implications for fundamental science .

\textbf{Motivation:} In 1959, Richard Feynman remarked that “it would be very easy to analyze any chemical substance; all one would have to do is look at it and see where the atoms are.”\cite{Feynman1959} However, more than sixty years later, scientists have yet to achieve the ability to image and determine the precise 3D positions of individual atoms in a material. Neutron radiation in fission and fusion reactors displaces atoms in structural materials, creating crystalline defects that degrade properties and threaten nuclear safety\cite{Zinkle2013}. Understanding the atomic structure of nanoscale and point defects (often involving just one to a few interstitial/vacancy atoms) is critical for predicting defect evolution and designing radiation-resistant materials. From a fundamental science standpoint, the complex local atomic rearrangements caused by neutron irradiation present both a significant challenge and a unique opportunity for 3D atomic structure reconstruction\cite{Lin2017}\cite{Lin2024}. To meet this challenge, we seek to apply AI/ML in atomic structure modeling and S/TEM image simulation\cite{Cheng2022}.

\subsubsection{Methodology}
Our approach consisted of two primary phases: (1) problem formulation and exploratory model development using GPT-4o and (2) targeted generation of 3D atomic structures and S/TEM image simulations using prompt engineering and deep research tools. Together, these efforts aimed to evaluate the feasibility of AI-driven reconstruction and analysis of defect structures in materials imaged by S/TEM.
\begin{itemize}
  \item \textbf{AI Models \& Tools:} We leveraged the capabilities of GPT-4o and Deep Research to explore both conceptual and generative aspects of the atomic-scale reconstruction task:
\begin{enumerate}
\item \textbf{Problem Formulation:}
We used GPT-4o to frame the challenge of 3D reconstruction and defect identification from S/TEM images. The model provided a structured workflow for automating S/TEM image simulation and suggested a classical (non-ML) pipeline for 3D reconstruction involving image alignment, tomography, and atom position extraction.
\item \textbf{STEM Image Mode Comparison:}
We prompted GPT-4o to compare different STEM imaging modes (e.g., high-angle annular dark-field (HAADF) and annular bright-field (ABF)) in terms of their effectiveness for defect visualization. This analysis informed our selection of HAADF-STEM due to its superior Z-contrast, which enhances the identification of atomic-level disruptions.
\item \textbf{Graph-ML Exploration:}
We explored the use of graph-based machine learning models to represent and analyze atomic structures, particularly for identifying and classifying defect clusters. GPT-4o proposed methods to encode atomic environments as graphs and use Graph Neural Networks (GNNs) for defect prediction.
\item \textbf{Deep Research Integration:}
In the second phase, we employed the Deep Research tool to conduct a literature review on interstitial dislocation loops and ML-based STEM image simulation models. While it returned comprehensive references and contextual understanding, it did not generate structural visualizations directly.
\end{enumerate}

\item \textbf{Workflow:} The following steps summarize the workflow we designed and implemented:
\begin{enumerate}
\item \textbf{3D Defect Structure Generation:} We used prompt engineering with GPT-4o to generate lattice structures containing voids and interstitial dislocation loop defects.
\item \textbf{Literature-Guided Structure Synthesis:} Using Deep Research, we conducted an in-depth review of interstitial dislocation loop formation and behavior. While the tool synthesized knowledge effectively, it could not produce structural models or images.
\item \textbf{Python Code for Visualization:} We wrote Python scripts to visualize the generated 3D atomic structures, enabling manual verification and comparison with literature-based defect morphologies.
\item \textbf{S/TEM Image Simulation:} We then prompted GPT-4o and Deep Research to simulate S/TEM images from the 3D structures. After directing Deep Research to literature on ML models for image simulation (e.g., multislice-based approaches and generative networks), it successfully produced image outputs closely aligned with expected features.
\end{enumerate}
\item \textbf{Team Roles:} \textbf{Y. Lin:} Conceptualization, Writing - Original Draft, Formal analysis, Investigation. \textbf{A. Chowdhury:} Methodology, Data Curation, Writing - Review \& Editing, Formal analysis, Investigation. \textbf{S. Kang:} Writing - Review \& Editing, Investigation. \textbf{C. Lear:} Writing - Review \& Editing, Investigation.
\end{itemize}
  
\subsubsection{Key Findings}
\begin{itemize}
  \item \textbf{Results:} 
In the first half of the work, we defined the problem of S/TEM image analysis and 3D reconstruction for identifying defect clusters in atomic scale. Using the GPT-4o model, we constructed a prompt-driven workflow for automating S/TEM image simulation tasks. This involved guiding the model through each stage of the image generation process, laying the foundation for an automated pipeline. To establish a baseline, we also asked the model to outline a classical, non-ML-based pipeline for reconstructing 3D atomic structures from S/TEM images. This comparison helped us evaluate the potential advantages of AI-enhanced methodologies.

We further explored the effectiveness of different S/TEM imaging modes in capturing defect structures. By querying the model for mode-specific characteristics, we gained insights into the relative strengths of techniques like HAADF and ABF in resolving local structural disruptions. Additionally, we asked the model to propose a graph-based machine learning framework, where atomic configurations are represented as graphs consisting of nodes (atoms) and edges (interatomic distances or bonding). This representation enables the use of Graph Neural Networks to predict 3D structures from 2D image data, offering a promising direction for inverse imaging problems.

In the second half, we focused on two main experimental tasks using GPT-4o and Deep Research. First, we used prompt engineering to generate 3D atomic lattice structures containing both void defects and interstitial dislocation loops. While voids were easily generated (Figure. \ref{fig:12-1}), producing interstitial loops proved more challenging. The model required detailed structural definitions to yield outputs with even partial physical realism. Despite iterative improvements, the generated dislocation loop structures remained inaccurate, with interstitial atoms appearing only along the loop’s circumference rather than forming a proper two-dimensional planar defect. To overcome this limitation, we used Deep Research to conduct a literature review on interstitial dislocation loops. Although it provided a comprehensive synthesis of the relevant materials science literature, it was unable to produce atomic visualizations or models based on the reviewed content.
\begin{figure}
    \centering
    \includegraphics[width=1\linewidth]{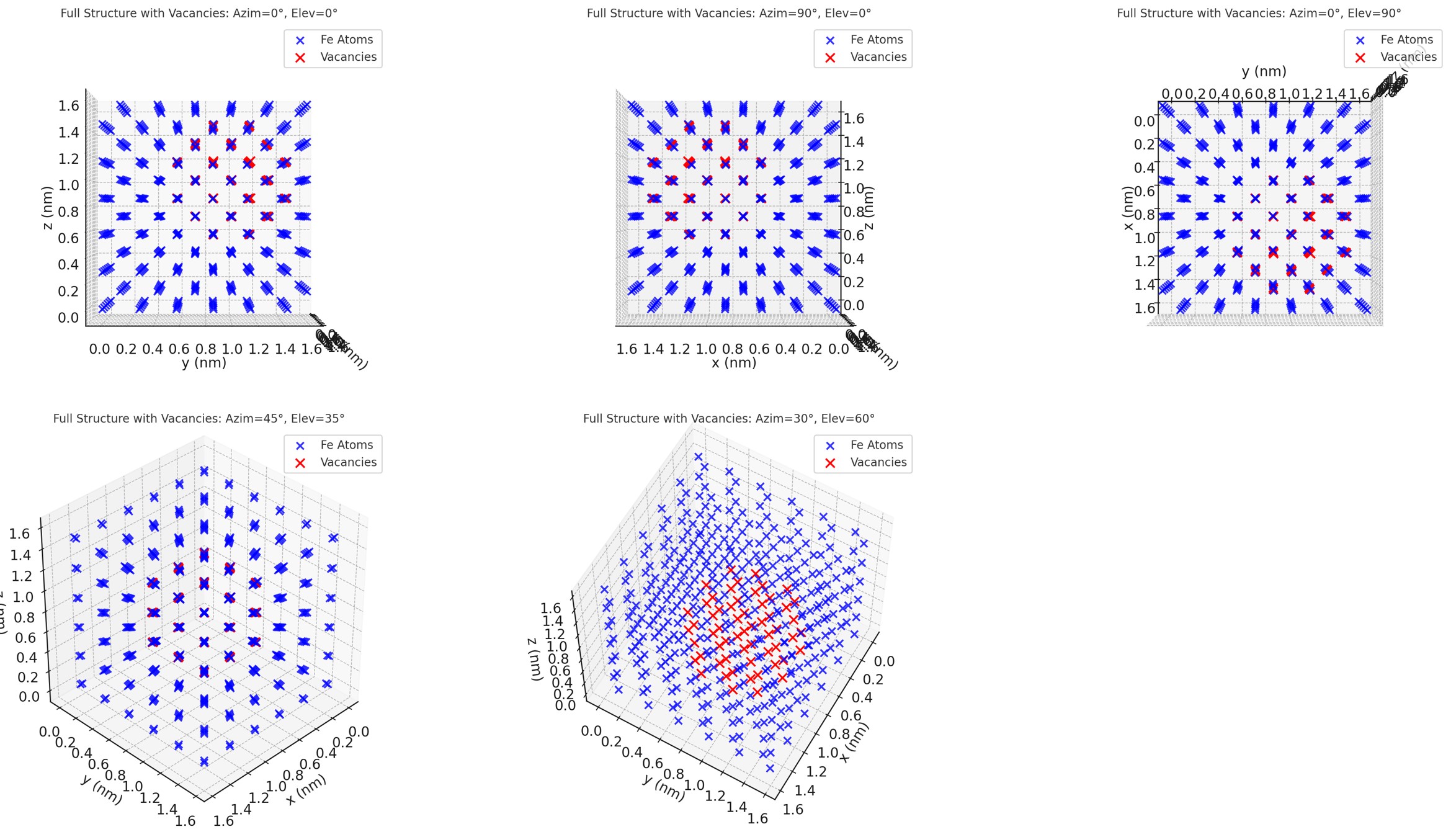}
    \caption{Various perspectives of the AI-generated atomic model depicting a 2 nm diameter void in body-centered cubic (BCC) structured iron.}
    \label{fig:12-1}
\end{figure}

We then extended prompt engineering to simulate S/TEM images from the generated 3D structures (Figure \ref{fig:12-2}). After guiding Deep Research to literature on multislice and ML-based image simulation methods, the model produced synthetic S/TEM images that closely approximated realistic experimental outputs. These results demonstrate the potential of combining large language models with targeted domain knowledge to support both atomic structure modeling and image simulation, although further refinement is needed to ensure physical accuracy and completeness.
\begin{figure}
    \centering
    \includegraphics[width=1\linewidth]{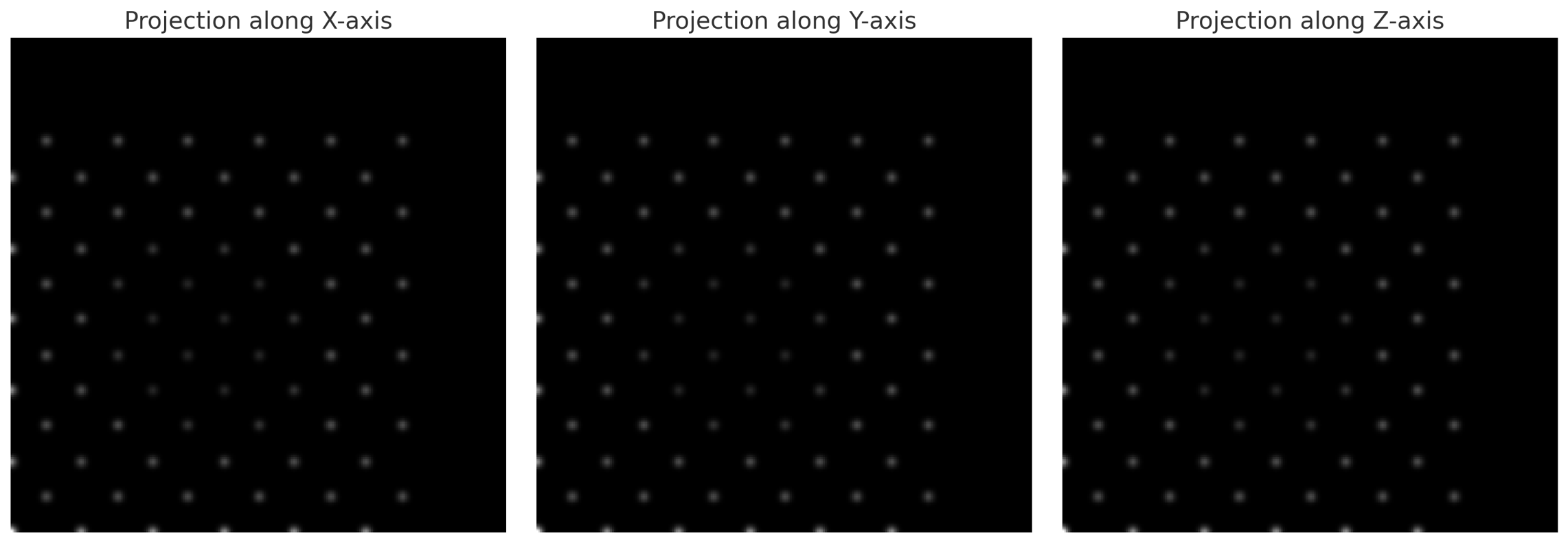}
    \caption{Simulated STEM HAADF image of a 2 nm diameter void in BCC iron.}
    \label{fig:12-2}
\end{figure}

  \item \textbf{Insights:} The preliminary works revealed several promising and unexpected capabilities of AI in advancing defect studies at the atomic scale. Notably, after reviewing selected journal articles on S/TEM image simulation, the AI demonstrated an emergent ability to infer and request essential microscopy parameters—an indication of contextual understanding beyond simple text generation. This suggests that AI can engage in domain-informed reasoning, opening new possibilities for autonomous assistance in microstructure characterization.

A key insight is the dual role AI/ML can play in both generating atomic structures for image simulations and predicting 3D atomic configurations directly from experimental S/TEM images. This integration could significantly improve the accuracy of simulation studies (such as strain mapping and defect bias analysis) by anchoring them in experimentally derived structures rather than assumed models. Such a shift would enable more realistic and reliable predictions in materials science.

However, current limitations remain. For instance, ChatGPT could only construct atomic structures within a limited spatial volume (less than 10 × 10 × 10 nm), which restricts its applicability for larger or more complex systems.  Moreover, any structure it proposes still requires careful expert review, because the model may closely imitate published descriptions yet inadvertently generate physically impossible configurations—such as the overlapping atoms mentioned above and shown in Figure 4. Overcoming this constraint is crucial for modeling extended defect networks or mesoscale phenomena.

Despite these challenges, the results reinforce the transformative potential of AI/ML in bridging experimental imaging and atomistic modeling. Although meaningful breakthroughs will require advances in S/TEM simulation fidelity, atomic modeling, and AI/ML integration, this direction represents a critical and timely step forward in irradiation-induced defect research, with far-reaching implications for materials science and nuclear engineering.
\end{itemize}

\subsubsection{Limitations \& Challenges} Generating the true 3D structure of defect clusters from S/TEM images is a multi-step and technically demanding process involving image acquisition, preprocessing, 3D reconstruction, atomistic modeling, and simulation-based validation. The key challenges encountered in this workflow can be categorized as follows:
\begin{itemize}
  \item \textbf{Technical Constraints:}
  \begin{enumerate}
\item \textbf{Data Availability:} High-quality tilt-series datasets are difficult to obtain due to limited tilt ranges, beam damage, drift, and sample stability issues—especially for radiation-sensitive materials.
\item \textbf{Model Performance Ceilings:} Traditional reconstruction methods such as Filtered Back Projection (FBP) are highly sensitive to noise, while iterative methods like SIRT offer improved robustness but still fall short in resolution. Even emerging deep learning-based reconstructions often require extensive training data and do not generalize well across material systems.
\item \textbf{Computational Resource Demand:} The full workflow—including 3D reconstruction, atom position extraction, atomistic simulation (e.g., MD or DFT), and multislice image simulation—requires significant computational power and specialized software, limiting throughput and scalability.
\end{enumerate}
  \item \textbf{Process Hurdles:} 
  \begin{enumerate}
\item \textbf{Collaboration Bottlenecks:} Effective execution requires seamless collaboration across microscopy, image analysis, machine learning, and computational materials science. Coordination delays and gaps in shared domain knowledge can slow progress.
\item \textbf{Tooling and Integration Gaps:} Most available software tools are optimized for specific steps but lack interoperability, requiring manual handoffs between platforms for alignment, reconstruction, atomistic modeling, and validation.
\item \textbf{Sensitivity and Noise Propagation:} Noise during imaging and misalignments in tilt-series propagate through the entire reconstruction pipeline, introducing uncertainty in atom localization and defect identification—especially for point defects like a single interstitial.
\end{enumerate}
  \item \textbf{Mitigations Tried:} 
  \begin{enumerate}
\item \textbf{Enhanced Imaging Practices:} Utilized HAADF-STEM for improved Z-contrast and prioritized fine-angle tilt acquisition in tomography to improve input quality.
\item \textbf{Advanced Preprocessing Techniques:} Implemented fiducial marker-based alignment and state-of-the-art denoising algorithms (e.g., non-local means, total variation) to reduce noise while preserving atomic details.
\item \textbf{Algorithmic Improvements:} Explored compressed sensing and machine learning-based reconstruction techniques to reduce the number of tilt images required and improve reconstruction quality from incomplete data.
\item \textbf{Post-Reconstruction Refinement:} Applied MD and DFT simulations to refine atomic coordinates and validate structural plausibility. Introduced feedback loops by comparing simulated and experimental images using multislice simulation methods.
\item \textbf{Interdisciplinary Coordination:} Established cross-functional collaboration between S/TEM experimentalists, image analysts, and computational modelers to ensure consistency and feedback across each step of the workflow.
\end{enumerate}
\end{itemize}

\subsubsection{Future Directions} To advance our goal of generating realistic STEM images of defect clusters and reconstructing their corresponding 3D atomic configurations using Graph Machine Learning (Graph ML), we propose the following future research directions.
\begin{itemize}
  \item \textbf{Next Steps:} Building on the current pipeline, the immediate next steps include experimental validation, model refinement, and structured dataset development:
   \begin{enumerate}
\item \textbf{Benchmarking with Experimental STEM Data:} Compare simulated STEM images (generated from atomic graphs using multislice tools such as PyMultislice or Dr. Probe) with high-resolution experimental data to validate simulation fidelity and assess noise robustness.
\item \textbf{Inverse Model Development:} Implement and train encoder-decoder models combining CNNs or vision transformers with GNN decoders to predict 3D atomic graphs directly from 2D STEM images. Begin with simple defect configurations (e.g., voids, single interstitials) to test predictive reliability.
\item \textbf{Defect Cluster Validation via Atomistic Simulations:} Refine and energetically validate the reconstructed atomic configurations using molecular dynamics (MD) or density functional theory (DFT) simulations to ensure physical plausibility.
\item \textbf{Integrated Dataset Construction:} Assemble a standardized training and evaluation dataset of paired STEM images and labeled atomic structures—including both experimental and simulated data—annotated for defect type, position, and energy state.
\end{enumerate}

\item \textbf{Potential Extensions:} Several promising interdisciplinary and technical directions could enhance the scalability, accuracy, and utility of the pipeline:
\begin{enumerate}
\item \textbf{Generative Graph Models for Defect Discovery:} Apply graph generative models (e.g., GraphVAE, GraphGAN, or diffusion-based methods) to explore complex atomic configurations of defect clusters, and validate their thermodynamic stability using DFT or empirical potentials.
\item \textbf{Neural Volume Rendering for 3D Reconstruction:} Explore neural radiance field (NeRF) or geometric deep learning approaches to reconstruct 3D atomic structures directly from 2D STEM tilt-series, potentially bypassing explicit tomography.
\item \textbf{Multimodal Fusion for Enhanced Prediction:} Combine multiple imaging modalities (e.g., HAADF, ABF, and EELS) within a multimodal ML framework to improve defect visibility and reduce prediction uncertainty.
\item \textbf{Synthetic Data Augmentation for Robust Training:} Leverage the GNN-to-image generative models to create large-scale synthetic datasets of labeled defect configurations, improving model generalization under limited experimental data scenarios.
\item \textbf{Unsupervised Defect Motif Learning:} Employ unsupervised graph clustering or contrastive learning to mine STEM image datasets for recurring defect motifs, potentially revealing previously unknown structural anomalies.
\item \textbf{Toolchain Integration and Open Benchmarking:} Develop a modular and reproducible workflow using PyTorch Geometric, DGL, abTEM
\end{enumerate}
\end{itemize}

%\PB{please add at least 5 domain related citations}

% === End LaTeX snippet ===

%% file: team13.tex
\subsection{Pressurized water reactor power uprate analysis}

% \PB{please add at least 1 figure and 5 domain related citations}
% \

% \PB{Pradeep will review} \PR{[Done]}

\begin{tcolorbox}[colback=gray!10, colframe=gray!80, title=Team Members, fonttitle=\bfseries]
\textbf{Lisa Fassino},  \href{mailto:reedke@ornl.gov}{reedke@ornl.gov}\\
\textbf{Vineet Kumar},  \href{mailto:kumarv@ornl.gov}{kumarv@ornl.gov}\\
\textbf{Mohammed Alnaggar},  \href{mailto:alnaggarmg@ornl.gov}{alnaggarmg@ornl.gov}\\
\textbf{Rike Bostelmann},  \href{mailto:bostelmannf@ornl.gov}{bostelmannf@ornl.gov}
\end{tcolorbox}

\subsubsection{Problem Statement}
\textbf{Description:}  
The technical challenge was to find ways to support the input definition and component integration of a robust multiphysics strategy to evaluate a power uprate in a Pressurized Water Reactor (PWR). Reactor power uprates are important as they improve the economics of the light water reactor fleet in the US and have been an ongoing effort for the past two decades. PWRs have higher heating rates compared to BWRs, which present additional challenges for power uprates. In addition to the power uprate at current fuel enrichment, if a higher fuel enrichment (upto 7 percent) is also considered, it is critical that both the regulator and the operator rely on advanced high-fidelity modeling and simulation tools to assess reactor safety margins for new operational regimes. Some of the ongoing efforts on the development of a high-fidelity multiphysics tools for LWR simulations using the VERA and MOOSE frameworks are discussed in \cite{gamble2019, Avramova2021, Yu2022}. The goal of this effort was in using LLMs to support the input definition and model development for a multiphysics framework which includes neutronics, thermal-hydraulics, fuel performance, and structural mechanics codes. Furthermore, the goal was to automate the process of input creation using fully open-source or a combination of open-source and export-controlled software, and setting up the multi-way coupling framework. 
% \PR{Provide a bit more detail on the problem statement for readers who may not be familiar with it. Inputs to what? Components = PWR components or multiphysics components or something else? etc.}
% \VK{Is this sufficient?}

\begin{figure}
    \centering
    \includegraphics[width=0.9\linewidth]{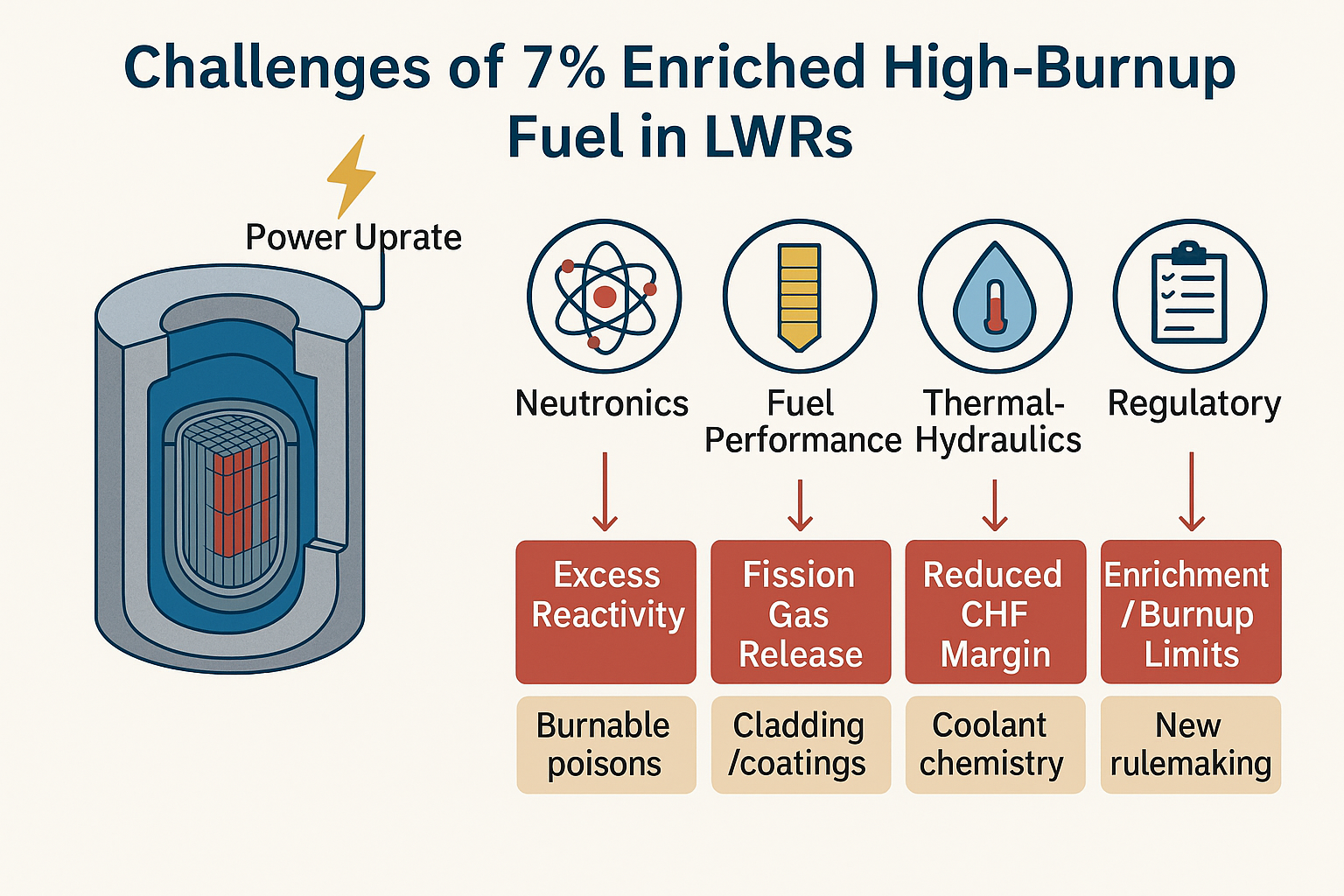}
    \caption{Multi-domain challenges for higher enrichment and higher burnup fuel generated by ChatGPT-4o.}
    \label{fig:Team13_workflow}
\end{figure}

\textbf{Motivation:}  
Power uprates extend the lifetime and efficiency of existing nuclear infrastructure, but impose higher thermal and mechanical loads on various reactor components such as, the fuel, the reactor pressure vessel (RPV), the steam generator and other balance of plant components \cite{NunezCarrera2017}. As an example, the higher power increases the neutron flux in the core and the surrounding structure, which can accelerate the irradiation embrittlement of the RPV, potentially reducing the plant's operational life . The development of such coupled models requires deep knowledge of each multiphysics aspect specifically as well as breadth across the disciplines to ensure accuracy in the system at large, as summarized in Fig. \ref{fig:Team13_workflow} for higher enrichment and higher burnup fuel. 
The use of LLMs to enhance the input development process and multiphysics pipeline would prove immensely beneficial, allowing for more time to fine tune, test, and analyze the models.

\subsubsection{Methodology}
\begin{itemize}
  \item \textbf{AI Models \& Tools:}  
  ChatGPT (GPT-4) was used as a scientific reasoning tool to support development of a multiphysics workflow with specific support for the input generation at each stage. Based on various prompts regarding the potential tools, inputs to those tools, and workflows between them, ChatGPT would provide literature-based recommendations for each of these components. Potential routes to supplement validation of inputs was also explored, including creation of custom-tailored GPTs which would focus specifically on their topics of "expertise" to support the multiphysics pipeline.

  \item \textbf{Workflow:}
  \begin{enumerate}
      \item Establish the problem scope to perform a quasi steady-state power uprate simulation of a PWR with the modeling of thermal-hydraulics (subchannel for the core + system for balance of plant), core neutronics, fuel performance, and RPV structural analysis. 
      \item Identify the relevant tools to perform such a simulation using a combination of ChatGPT suggestions and domain expertise judgement. List down the pros and cons of each tool. 
      \item Setup the multiphysics framework and clearly state the data streams in and out of each tool. Identify the coupling strategy for both domain overlapping as well as boundary exchange coupling, depending on the problem domain and the tools being used.
      \item Generate input decks for the identified tools based on the multiphysics framework and compile available validation data.
      \item Identify mitigation strategies for export-controlled tools and for lack of available validation data. 
  \end{enumerate}
  % \PR{Expand and provide a bit more detail on each step. how did you approach the problem?}
  % \VK{Is this sufficient?}

  \item \textbf{Team Roles:}  
  M. Alnagger developed a potential problem where all team members' areas of expertise could be leveraged to gauge suitability of ChatGPT for addressing this problem. R. Bostelmann and L. Fassino investigated specifically ways to improve ChatGPT's ability to support input creation for the SCALE software suite in a variety of capacities. V. Kumar focused on the use of ChatGPT to support coupling with COBRA-TF (CTF). M. Alnagger documented the project plan throughout the way, and it was summarized and expanded upon by L. Fassino.
\end{itemize}

\subsubsection{Key Findings}
\begin{itemize}
  \item \textbf{Results:}  
  The GPT model delineated various multiphysics coupling strategies and a selection of tools, categorized by domain (e.g., PARCS for neutronics, CTF for thermal hydraulics, FRAPCON for fuel performance, TRACE for system thermal-hydraulics, Grizzly for RPV degradation and aging, etc.), \cite{Avramova2021, Spencer2012} and provided a potential workflow for integrating the different data streams and time scales. It also provided example inputs for the chosen tools and further advice on setting up models to get necessary data. ChatGPT emphasized pros and cons to each strategy, including potential bottlenecks. It also provided integrated solutions with tools like MOOSE. In particular, ChatGPT was able to provide a workflow for how the multiphysics framework could be used for the uprate scenario. To briefly summarize, the simulation setup would first model a baseline model of a PWR reactor and validate it with operating data of temperatures, pressures, neutron fluxes etc. The reactor would then slowly be stepped up in power with the corresponding neutronic calculation performed at the new core power to determine the updated flux distribution, along with the thermal-hydraulic subchannel and system calculations at the new operating conditions. The updated thermal and radiation loads are then fed to the structural model of the RPV, and the material embrittlement assessed by the fracture toughness model using the updated neutron fluence. This gives a flavor of how the LLM was able to present a robust strategy for the power uprate simulation. 
  % \PR{Tie back to the problem statement - How did it help address the question/problem you stated at the beginning re: inputs and component integration for multiphysics power uprate evaluation?}

  % \VK{Is this sufficient?}

\end{itemize}

\subsubsection{Limitations \& Challenges}

\begin{itemize}
  \item \textbf{Technical Constraints:}  
  The LLM provided a thorough summary of the current state of the art multiphysics code coupling strategies and was able to tie it back to our study. It also provided key scenario parameters such as the updated neutron fluence (with some assumptions) and how that would affect material embrittlement of the RPV. However, it failed at its task when taking a step further to actually setup the problem in terms of using the available open-source tools and supporting input deck creation. It must be stated that not all the available tools for supporting such a simulation are open-source and therefore, it is expected that the LLM cannot effectively provide a framework to setup the problem. Furthermore, when prompting the LLM to use available literature to support input deck creation for export-control software, it failed to create a reasonable input deck, even with excessive supplemental information provided in terms of user manuals and input decks of simple benchmark problems. 
  
  % \PR{The first part of this is a general statement. Can you narrow down your focus to the specific problem you tackled based on the results you obtained?}

  % \VK{Is this sufficient?}

  \item \textbf{Process Hurdles:}  
  One of the key process hurdles as mentioned previously, is the lack of state-of-the-art open-source tools which are required across all the relevant multiphysics domains to perform such a simulation, with the majority of the tools falling under export-controlled software. Additionally, the data that would be required to validate such a study is not available in the public domain. To perform a limited validation study, data available from relevant studies in the literature in combination with synthetic data could be a viable path forward. 
  
  % \PR{are you making a general statement about connectivity of models or is this specific to a GPT-based solution to the problem of model connectivity?} 

  % \VK{Is this sufficient?}

  \item \textbf{Mitigations Tried:}  
  To solve a problem with this strategy, it would be essential to use open-source tools with expansive documentation and example inputs. The produced inputs for tools like SCALE \cite{scale_manual} and COBRA-TF \cite{salko2020} were very far from functional. ChatGPT provides the option to create specialized GPTs, as seen in Figure~\ref{fig:t13-custom-gpt}, to support input creation, but even that approach proved challenging. Certain tools could be developed however to counter this. For example, a SCALE API which provides input validation, could support an iterative input development loop where enforced input validation provides at least a minimal, working model. 
  
  % \PR{This is good and points to potential future directions to try. Curious if you tried any of these or other mitigation approaches during the workshop or after and if so, can you describe them briefly?}

  % Response - We did not follow through with this but hope to do so in the future.

  \begin{figure}
  \centering
  \includegraphics[width=0.9\linewidth]{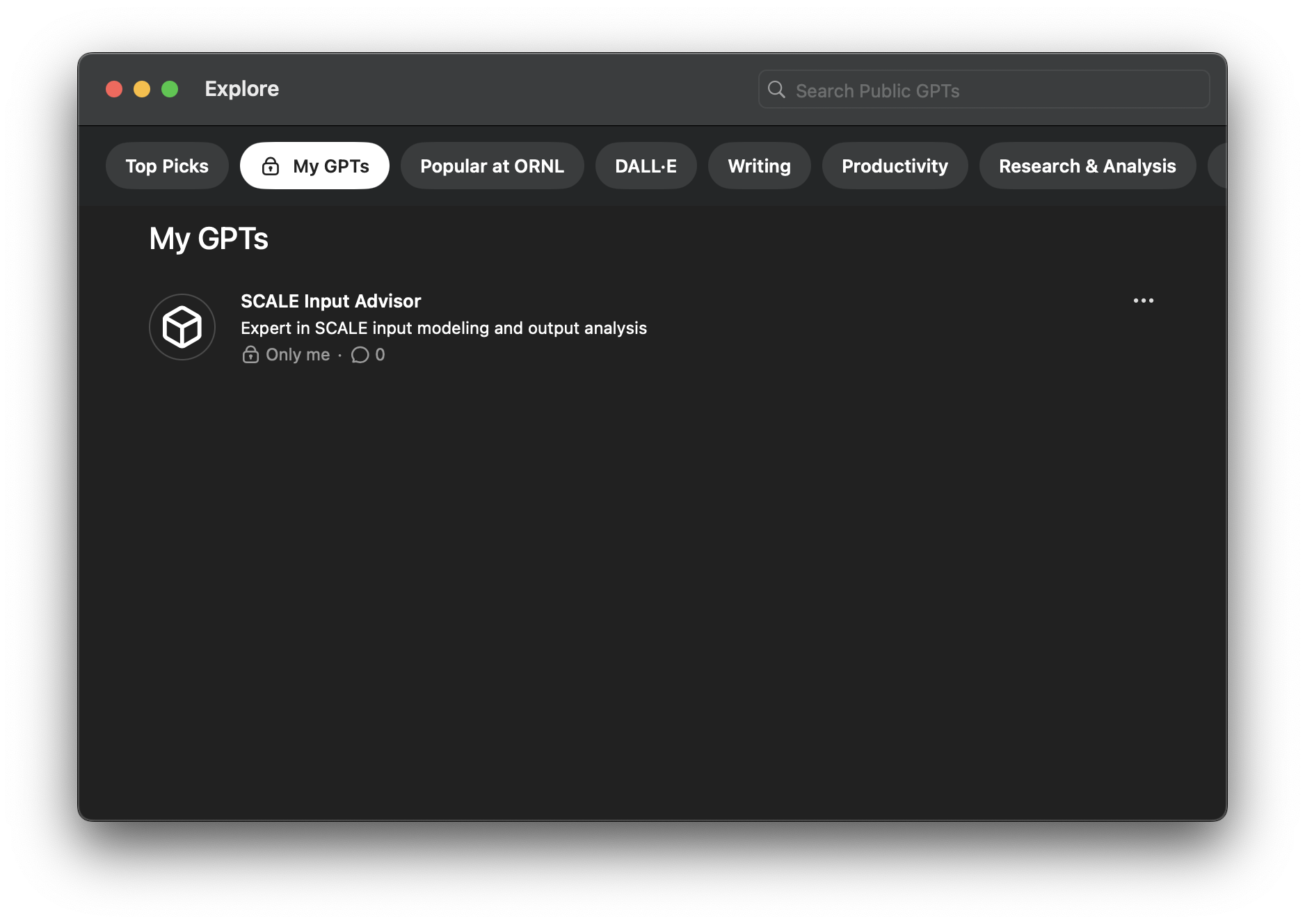}
  \caption{Example of potential GPT to support input development.}
  \label{fig:t13-custom-gpt}
  \end{figure}

\end{itemize}

\subsubsection{Future Directions}
\begin{itemize}
  \item \textbf{Next Steps:}  
  Implement and test AI-generated multiphysics coupling in a reduced-order PWR model using open-source tools.
  % \PR{you also have some good stuff in the mitigations section. Consider adding them here.}
  \item \textbf{Potential Extensions:}  
  Training models specifically for export-controlled tools using secure means to better support tools which are not open source.
\end{itemize}

%\PB{please add at least 5 domain related citations}

%% file: team16.tex
\subsection{Combining LLM and Domain-driven Prompt Engineering for Efficient Microscopy Image Simulation of Materials Defects}
%For each team, use the following headings:

\begin{tcolorbox}[colback=gray!10, colframe=gray!80, title=Team Members, fonttitle=\bfseries]
\textbf{Arpan Biswas},  \href{mailto:biswasa@ornl.gov}{biswasa@ornl.gov} \& \href{mailto:abiswas5@utk.edu}{abiswas5@utk.edu}\\
\textbf{José Salcedo-Pérez},  \href{mailto:salcedoperjl@ornl.gov}{salcedoperjl@ornl.gov}
\end{tcolorbox}

\subsubsection{Problem Statement}
\textbf{Description:} This team explored the capabilities of LLMs on simulating microscopy images of  defects of choice in  SS304 material. SS304, while a common and versatile stainless steel, is susceptible to certain material defects in nuclear applications due to the harsh environments it's exposed to. In this project, we wanted to validate the workflows of LLMs in providing relevant simulated images and whether the results can be improved with domain-knowledge specific prompt engineering. 

\textbf{Motivation:} The broad goal of this project is the scope of future developments of AI tools for predictive maintenance of nuclear reactors, and thereby motivated to find better materials. Designing a ML model to predict different defects in real-time that can arise over time in a nuclear defects and do preventive measures before catastrophic failure, saving life and cost. The broader goal is to accelerate the generation of AI-produced data and, where possible, augment it with small amounts of experimental data to increase the overall training set—otherwise too expensive and nearly infeasible. In order to aim this broader goals, the first step is to visualize and understand different defects at the microscopic levels, that can arise in materials for nuclear applications. In general, in-situ experiments for material analysis is an extensive process\cite{Harris2024PLD}. Thus, conducting different microscopy experiments to learn several pattern of defects is a time and cost-consuming process which need extensive labor, facility, sample costs. Thus, leveraging LLMs we can conduct a early interpretation of these defects before the need to go for expensive experiments. On the other hand, meaningful simulation is also necessary to improve the required learning. In recent years, for improving microscopy experiments and analysis, domain-knowledge has been integrated with data-driven ML methods, to focus on maximizing extraction of physical insights from low volume of data\cite{Biswas2024pVAE, Biswas2024BOARS, Kalinin2024}. Inspiring from similar idea, we aim to define domain-specific prompt workflows to systematically generate meaningful microscopy images for various material defects. We finally compare the study with and without systematic prompt engineering to highlight the current achievements and limitations of these LLM models, and how we can accelerate research in the area of nuclear engineering with AI and domain knowledge integration.

\subsubsection{Methodology}
\begin{itemize}
  \item \textbf{AI Models \& Tools:} For this study, we considered Gemini AI, Gemini Imagen 3 and ChatGPT. Gemini AI were used for extracting domain-specific textual information where Imagen 3 is used for fine tuning model based on the domain-specific prompts, and finally to generate simulated images. Regarding ChatGPT, we also used deep research to improve microscopic images, as specified in the prompt. Here, the aim was to compare the results across the different LLM models we had. We used the Gemini AI to first tune the model defined with domain-specific prompt (See Prompt 3 in \textit{Workflow}). We conduct the same prompt workflow for ChatGPT separately. For image generated from Imagen 3, Gemini AI was prompted to tune. For image generated from ChatGPT, ChatGPT was prompted to tune. Each tools we used produced different quality images once we optimized the prompts.%\textbf{JSP: the aim was to compare the results across the different AI models and tools we had. Each tool we used produced quality images once we optimized the prompts.} \textbf{AB: No, we use the Gemini AI to first tune the model defined with domain-specific prompt (See Prompt 3 below). We conduct the same prompt workflow for ChatGPT and Gemini AI separately. For image generated from Imagen 3, Gemini AI was prompted to tune. For image generated from ChatGPT, ChatGPT was prompted to tune. We got better results however from Gemini AI which we have mentioned later.}
  \item \textbf{Workflow:} The design of the prompt workflows is stated as, \textit{Prompt 1:} “What are the potential defects of SS304 in nuclear applications?”
 $\rightarrow$ Selection of a specific defect (Eg. Sensitization) $\rightarrow$ \textit{Prompt 2:} “What is the best microscope to characterize the sensitization defect of SS304?”
 $\rightarrow$ Selection of a microscope, relevant to this study (eg. Electrochemical Scanning Tunneling Microscope) $\rightarrow$ \textit{Prompt 3:} “Tune the model to find Electrochemical-STM Microscopy images of sensitized defects of SS304” $\rightarrow$ \textit{Prompt 4:} “Supply training samples of Electrochemical-STM Microscopy images of sensitized defects of SS304” $\rightarrow$ Validation of the simulated images with domain knowledge and open source microscopy images. 
  \item \textbf{Team Roles:} A.Biswas conduct the microscopy image generation and analysis over Gemini AI and J. Salcedo-Pérez conduct the microscopy image generation and analysis over ChatGPT. Both of them conceived the research problem and design the prompt workflows. 
\end{itemize}

\subsubsection{Key Findings}
\begin{figure}
    \centering
    \includegraphics[width=1\linewidth]{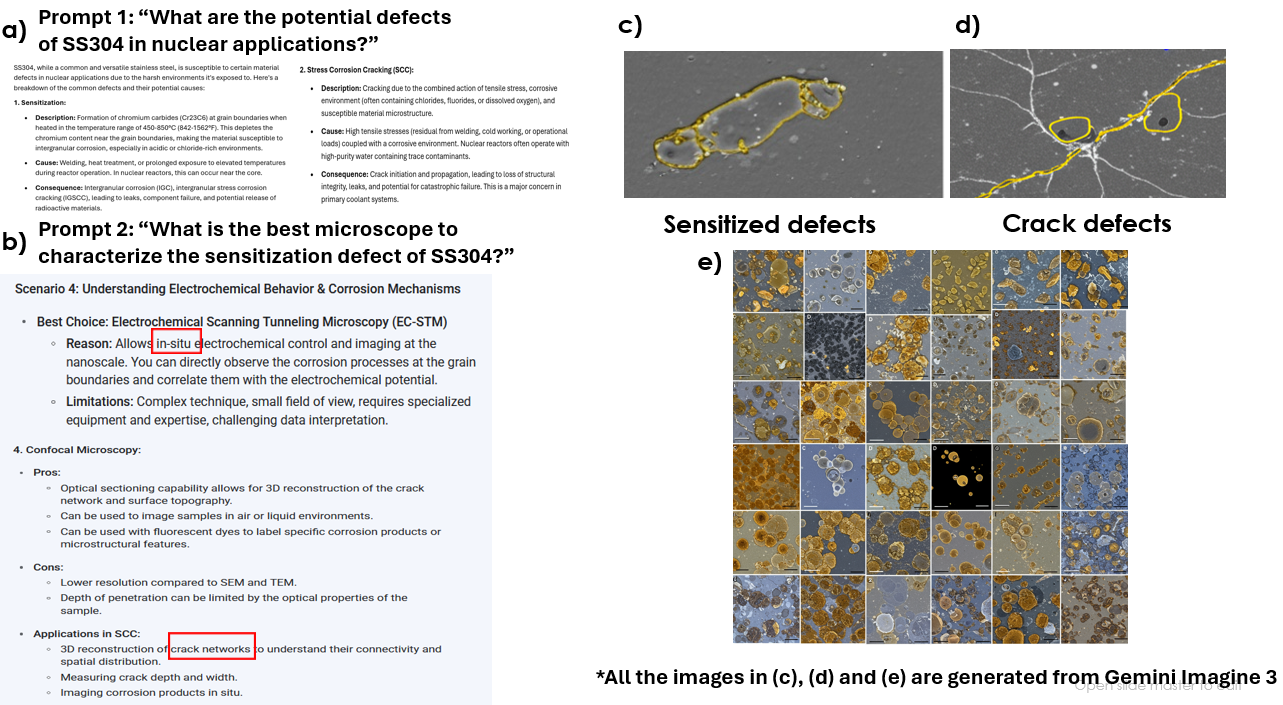}
    \caption{Prompt engineering and LLM assisted generation of simulated microscopy defect images. The screenshots in (a) and (b) are generated from Gemini AI and the images in (c)-(e) are generated from Gemini Imagine 3.}
    \label{fig:Team16_workflow}
\end{figure}

\begin{itemize}
  \item \textbf{Results:} For this study, our choice of materials was SS304, and we provide and example of the results (Fig. \ref{fig:Team16_workflow}) generated from \textbf{Gemini AI} and \textbf{Imagine 3} via the domain specific prompt engineering workflow (as mentioned in above section). From the first prompt we find a list of material defects due to nuclear application (Fig. \ref{fig:Team16_workflow}a). We considered 4 primary defects such as 1)\textbf{Sensitization}\cite{Rodriguez2013Sensitize}, 2)\textbf{Stress Corrosion Cracking}\cite{Guo2020Crack}, 3)\textbf{Irradiation Embrittlement}\cite{Kumar2022Radiation} and 4)\textbf{Weld Defects}. Then through the second prompt, we find a broad review of each microscopic images with pros and cons (Fig. \ref{fig:Team16_workflow}b). We select the suitable microscope for generating high quality/informative images on each defects based on insitu imaging capabilities, learning dynamic defect (Eg. crack) propagation etc, which is critical for developing AI tools to gain a broader physical insights of various defects for real-time predictive maintenance. For eg. based on these selective criteria, we found the best microscopy imaging would be \textbf{Electrochemical-STM} for \textbf{Sensitization}, \textbf{Confocal Microscopy} for \textbf{Stress Corrosion Cracking} and \textbf{Transmission Electron Microscopy} for \textbf{Irradiation Embrittlement}. Figures \ref{fig:Team16_workflow}c, \ref{fig:Team16_workflow}d shows examples of the defect specific simulated images, generated from Gemini AI (Imagine 3). 
  \item \textbf{Insights:} 
  \begin{itemize}
      \item We compared the AI generated sensitized and crack defects, as in \ref{fig:Team16_workflow}c-d, with the literature and see good agreement among them\cite{Rodriguez2013Sensitize, Guo2020Crack}.
      \item   To validate the prompt engineering, we input a very generalized non-systematic and non-domain specific prompt such as \textit{Prompt: Supply training microscopy image samples of SS304 defects}. From fig. \ref{fig:Team16_workflow}e., we can clearly see there are no diversity of the defects that we learn from the systematic prompts (eg. no simulated images of crack defects) and the simulated images is very hard to interpret the label of these defects and the microscopy imaging. This shows the infusion of the domain knowledge via prompt engineering improve the AI alignment with intended search. 
  \end{itemize}
\end{itemize}

\subsubsection{Limitations \& Challenges}
\begin{itemize}
  \item \textbf{Technical Constraints:}
  \begin{itemize}
        \item We could not find any relevant images for the \textbf{Welding} defects in ChatGPT, even we considered the deep research option. 
        \item Overall Gemini AI performs much better than ChatGPT, as we considered similar prompt workflows.
        \item As stated earlier, the basic prompts did not provide relevant results, lacks deep domain knowledge.
  \end{itemize}

  \item \textbf{Process Hurdles:} 
  \begin{itemize}
      \item As this study was based on the image generation, one key aspect is the non-transparency of the source of these images. For example, one major drawback of Gemini AI is not providing information of any cited paper from where the knowledge of the image simulation have been trained. This could be key addition in the capability of LLM models to gain more usability of the data via proper citation and permissions if required.
  \end{itemize}
  \item \textbf{Mitigations Tried:} 
  \begin{itemize}
      \item We considered the deep research option to search for Weld defects in ChatGPT
      \item As stated earlier, we designed the domain-specific systematic prompts to improve the interpretation of the simulated images.
      \item We partially tried to mitigate this by manually search for microscopy images which matches exactly with the generated images in Gemini AI, if any, to map the cited paper. However, we did not find anything like that to our knowledge. Therefore, to fully mitigate the usability challenges, we need deeper understanding of the workflow of image generation of Gemini AI.
  \end{itemize}
  
\end{itemize}

\subsubsection{Future Directions}
\begin{itemize}
  \item \textbf{Next Steps:} 
    \begin{itemize}
        \item A more deeper evaluation of workflow is necessary to understand the training of the image generation, and how we can apply in future research.
        \item A more deeper exploration of domain-specific prompt engineering to further improve the results. Though the defect images generated are meaningful, however, it is not clear and not always provided the specific material SS304 images that we prompted for. It will require for deeper search to pin point with specific material simulated images. 
        \item To design a AI tools where we can train by augmenting high volume of relevant simulated images with low-volume of real-time experimental images, thus increasing the size and diversity of the training data. 
    \end{itemize}
  \item \textbf{Potential Extensions:} In order to suitably train a AI model, it need to be trained with all the pattern of defects. In other word, if the AI is only trained with sensitization defects, it will likely fail to detect any cracking defect in high-precision level, even if the training data-volume is high. This is due to the low-quality of training data. On the other hand, it is very expensive to capture all the defects in different microscopy imaging. Therefore, with mitigating the stated challenges of these LLM models, we can easily consider a future extension of an efficient augmentation of AI generated and experimental data to better train the AI model for predictive maintenance design of nuclear reactors. 
\end{itemize}

% === End LaTeX snippet ===

%% file: team17.tex
\subsection{Evaluating LLMs for Applications in Nuclear Isotope Studies and MCNP Geometry Automation}

\begin{tcolorbox}[colback=gray!10, colframe=gray!80, title=Team Members, fonttitle=\bfseries]
\textbf{Scott Greenwood},  \href{mailto:greenwoodms@ornl.gov}{greenwoodms@ornl.gov}\\
\textbf{Christopher Greulich},  \href{mailto:greulichcr@ornl.gov}{greulichcr@ornl.gov}\\
\textbf{Punam Thakur},  \href{mailto:thakurp@ornl.gov}{thakurp@ornl.gov}\\
\textbf{Nageswara Rao},  \href{mailto:raons@ornl.gov}{raons@ornl.gov}
\end{tcolorbox}

\subsubsection{Problem Statement}
\textbf{Description:} This team explored the application of LLMs to explore two topical areas. The first area was assessing LLM accuracy for nuclear medical isotope research, specifically identification and ranking of important medical isotopes and their ability to be created at the High Flux Isotope Reactor (HFIR) and processed at the Radiochemical Engineering Development Center (REDC) based on specific criteria based on difficulty to produce and economic value. The second area was understanding the challenges of generating watertight geometries from LiDAR to use in MCNP simulations and developing a preliminary coding framework.

\textbf{Motivation:} 
For topic area 1, the accurate and efficient retrieval and synthesis of scientific information are fundamental to accelerating research and development in fusion and fission energy. If LLMs can reliably perform complex scientific literature reviews and data analysis in specialized fields like nuclear isotope production, they could serve as powerful tools for researchers, potentially speeding up the identification of promising isotopes for applications ranging from medical treatments to advanced nuclear fuel cycles and fundamental physics studies or at least help to accelerate learning and help provide an initial hypotheses and evaluation of new ideas. Validating their accuracy is a critical step towards their wider adoption and impact in various nuclear engineering and science problems related to HIFR and REDC.

Accurate geometric modeling is essential for obtaining reliable results from nuclear simulations used in the design, safety analysis, and optimization of fusion and fission systems. Real-world nuclear facilities and experimental setups often have complex geometries that are time-consuming and difficult to manually model in simulation codes. Automating the creation of watertight geometries from LiDAR data can improve the fidelity of simulation inputs, leading to more accurate predictions of neutron and photon transport, radiation shielding requirements, and material activation, ultimately contributing to safer and more effective systems.

\subsubsection{Methodology}
\begin{itemize}
  \item \textbf{AI Models \& Tools:} ChatGPT, Gemini, and Grok were each used to explore the topical areas. For each of these LLMs, reasoning models and deep research were both employed. Reasoning models were used to improve and clarify prompts and to generate coding frameworks. For this study only text prompts were used to iterate on the creation of prompts for deep research. The results of the deep research were then provided back to reasoning models for creation of an outline and implementation of an initial coding framework for task two.
  %\item \textbf{Workflow:} Prompt Engineering $\rightarrow$ Deep Research $\rightarrow$ Response Validation $\rightarrow$ Reasoning Model for Outline/Coding $\rightarrow$ Review $\rightarrow$ Discussion, Code Development, and Iteration. \PB{expand and provdie more details; see the other sections for examples}
  \item \textbf{Team Roles:} P. Thakur reviewed the responses for topic area 1 (isotopes) not only on identified medical isotopes but the claims AI made on the ranking and HFIR's ability to generate those isotopes. C. Greulich reviewed topic area 2 (LiDAR) responses for completeness and proper challenge identification. S. Greenwood guided the workflow of using the LLM including prompt crafting, discussed topical areas with the team members, and evaluated the generated coding framework. N.~Rao contributed to post-analysis of the findings and general applicability of LLMs to nuclear engineering problems, including the sensor drift estimation which is joint work with Greulich \cite{Raoetal2019ans}.
\end{itemize}

\subsubsection{Key Findings}
\begin{itemize}
  \item \textbf{Results:}
  \begin{itemize}
    \item LLMs provided generally accurate information, when publicly available online, regarding isotopes creatable at HFIR and summaries of the LiDAR-to-MCNP geometry challenge. The citations included previous work done by C. Greulich on the topic \cite{greulich2022method} which was summarized adequately. 
    \item Reasoning model seemed to generate a plausible coding outline for the LiDAR task, indicating points for tool/code insertion, that was informed by a deep research inquiry. The outline while plausible had been previously discarded as it fails on non-public domain datasets. 
  \end{itemize}
  \item \textbf{Insights:}
  \begin{itemize}
    \item Base LLMs often miss the latest research and specific methods in deep scientific domains, especially domains that have not yet widely adopted open access and preprint publishing, potentially providing an incomplete or outdated understanding (e.g., overstating exclusivity of certain isotope production methods at HFIR).
    \item LLMs may default to high-level summaries or outlines rather than suggesting specific technical tools, libraries, or code implementations for complex engineering tasks like geometry processing. Follow on questioning may improve this response but is not guaranteed.
    \item This limitation in providing current, detailed, actionable scientific/engineering information from general models is concerning for researchers relying on them for state-of-the-art insights.
  \end{itemize}
\end{itemize}

\subsubsection{Limitations \& Challenges}
\begin{itemize}
  \item \textbf{Technical Constraints:}
  \begin{itemize}
    \item Base LLM training data lacks comprehensive, up-to-date coverage of niche, deep scientific fields.
    \item The performance ceiling of general models for highly specialized, task-specific outputs (like suggesting precise coding tools or methods) was observed.
    \item The general responses lacked understanding of the users specific technical background and interests as well as the implicit (i.e. non prompted) challenges of the targeted application. While the users had previous experience with LLMs the accounts were recently created to give access to more advance models and features and therefore the lacked a significant chat history. 
  \end{itemize}
  \item \textbf{Process Hurdles:}
  \begin{itemize}
    \item Significant hurdles exist in easily setting up workflows for domain scientists (non-AI/coding experts) to supply LLMs with proprietary or specialized technical data for fine-tuning or contextual grounding.
    \item Clear and accessible methods for validating the accuracy and reliability of fine-tuned AI models in scientific contexts are lacking.
    \item LLMs exhibited a tendency to provide general information rather than the specific tools or code required to solve a defined technical problem. 
    \item The tendency towards generality also manifests in specific domain terminology used in prompts that were repeated in responses but when the responses elaborated on the terminology it failed to understand the specific context. 
  \end{itemize}
  \item \textbf{Mitigations Tried:}
  \begin{itemize}
    \item Switched to a reasoning model after a deep research model failed to provide concrete solutions for the LiDAR coding task, resulting in a basic structural outline but still lacking specific technical details. The reasoning model selected depended on the employed LLM. Additional, focused prompting for areas is required for further development.
    \item The lack of complete contextual understanding was partially mitigated with repeated clarification prompts, more detailed initial prompts, and the use of reasoning and deep research models. 
  \end{itemize}
\end{itemize}

\subsubsection{Future Directions}
\begin{itemize}
  \item \textbf{Next Steps:}
  \begin{itemize}
    \item Develop generalized methods to create domain specialized LLMs, either via fined tuning, retrieval augmented generation (rag), or another method. Specialized models which can access domain-specific nuclear science data, technical publications, and laboratory procedures.
    \item Investigate and establish robust validation methodologies for assessing the scientific accuracy and reliability of fine-tuned AI outputs in specialized technical fields.
   \end{itemize}
  \item \textbf{Potential Extensions:}
  \begin{itemize}
    \item Explore the creation of secure, institution-specific AI platforms designed for handling proprietary technical data along with user-friendly interfaces to enable non-AI/coding experts to easily provide technical data and documents to LLMs for improved accuracy and context.
    \item Compare LLMs with existing ML methods in the area of nuclear science and engineering, including the sensor drift estimation in reactor systems \cite{Raoetal2020ned} where ML methods are shown to achieve better accuracy compared to existing auto associative kernel regression (AAKR) methods \cite{MaJiang2011}. Furthermore, principles of information fusion and thermal hydraulics  are critical to establishing the correctness and quantified generalization property of this method; a potential extension is to investigate these properties of LLMs.
  \end{itemize}
\end{itemize}

%\PB{please add at least 5 domain related citations}

% === End LaTeX snippet ===